\documentclass[12pt,preprint,onecolumn]{emulateapj}
\usepackage[american]{babel}
\usepackage{natbib}
\usepackage{amsmath}
\usepackage{graphicx}
\usepackage{multirow}
\usepackage{rotating}
\usepackage{mathptmx}
\usepackage[unicode,pageanchor,colorlinks]{hyperref}
\hypersetup{
  pdftitle={Multi-Scale CLEAN: A comparison of its performance against
    classical CLEAN using THINGS galaxies},
  pdfauthor={Rich, J. W., de Blok, W. J. G., Cornwell, T. J., Brinks,
    E., Walter, F., Bagetakos, I., Kennicutt, R. C. Jr.},
  pdfsubject={techniques: image processing --- galaxies:
    individual(NGC~2403, Holmberg~II, IC~2574) --- ISM:
    structure},
  citecolor=blue, 
  linkcolor=blue,
  urlcolor=red}
\usepackage[T1]{fontenc}
\usepackage{microtype}
\usepackage[]{caption}
\pdfoutput=1

\newcommand{\msclean}{{\scshape msclean}}
\newcommand{\bmsclean}{{\bfseries\small MSCLEAN}}
\newcommand{\mrclean}{AIPS Multi-Resolution {\scshape clean}}

\newcommand{\wclean}{windowed-\clean}
\newcommand{\cwclean}{Windowed-\clean}

\newcommand{\clean}{{\scshape clean}}
\newcommand{\bclean}{{\bfseries\small CLEAN}}
\newcommand{\eg}{{\em e.g.,}}
\newcommand{\ie}{{\em i.e.,}}
\newcommand{\etc}{{\em etc.}}

\newcommand{\HI}{\hbox{\rmfamily H\,{\scshape i}}}
\newcommand{\HIfat}{\hbox{\rmfamily\bfseries H\,{\small I}}}

\newcommand{\kms}{\hbox{km s$^{-1}$}}
\newcommand{\mjbeam}{\hbox{mJy beam$^{-1}$}}

\newcommand{\jkms}{\hbox{Jy km s$^{-1}$}}
\newcommand{\coldensity}{\hbox{cm$^{-2}$}}
\journalinfo{Accepted in AJ, 2008-10-12}
\slugcomment{For a high-resolution version visit:
  \url{http://www.mpia.de/THINGS/Publications.html}}

\begin{document}

%%%%####################################################################

\title{Multi-Scale \bclean: A comparison of its performance
 against classical \bclean\ on galaxies using THINGS}
\author{Rich, J. W.\altaffilmark{1}, de Blok,
 W. J. G.\altaffilmark{2}, Cornwell, T. J.\altaffilmark{3}, Brinks,
 E.\altaffilmark{4}, Walter, F.\altaffilmark{5}, Bagetakos,
 I.\altaffilmark{4}, Kennicutt, R. C. Jr.\altaffilmark{6}}

\email{joshua@mso.anu.edu.au}
\email{edeblok@ast.uct.ac.za}
\email{tim.cornwell@csiro.au}
\email{E.Brinks@herts.ac.uk}
\email{walter@mpia.de}
\email{I.Bagetakos@herts.ac.uk}
\email{robk@ast.cam.ac.uk}

\altaffiltext{1}{Research School of Astronomy and Astrophysics,
 Australian National University, Cotter Road, Weston Creek, ACT 2611,
 Australia}
\altaffiltext{2}{Department of Astronomy, University of Cape Town,
 Private Bag X3, Rondebosch 7701, South Africa}
\altaffiltext{3}{Australia Telescope National Facility, PO Box 76,
 Epping NSW 1710, Australia}
\altaffiltext{4}{Center for Astrophysics Research, University of
 Hertfordshire,College Lane Hatfield, AL10~9AB, United Kingdom}
\altaffiltext{5}{Max-Planck-Institut f\"ur Astronomie, K\"onigstuhl
 17, 69117
 Heidelberg, Germany}
\altaffiltext{6}{Institute of Astronomy, University of Cambridge,
 Madingley Road, Cambridge CB3~0HA, United Kingdom}

%%%%####################################################################

\begin{abstract}

A practical evaluation of the Multi-Scale \clean\ algorithm is
presented. The data used in the comparisons are taken from The \HI\
Nearby Galaxy Survey (THINGS). The implementation of Multi-Scale
\clean\ in the CASA software package is used, although comparisons are
made against the very similar Multi-Resolution \clean\ algorithm
implemented in AIPS. Both are compared against the classical \clean\
algorithm (as implemented in AIPS). The results of this comparison
show that several of the well-known characteristics and issues of
using classical \clean\ are significantly lessened (or eliminated
completely) when using the Multi-Scale \clean\ algorithm.
Importantly, Multi-Scale \clean\ reduces significantly the effects of
the clean `bowl' caused by missing short-spacings, and the `pedestal'
of low-level un-cleaned flux (which affects flux scales and
resolution). Multi-Scale \clean\ can clean down to the noise level
without the divergence suffered by classical \clean. We discuss
practical applications of the added contrast provided by Multi-Scale
\clean\ using two selected astronomical examples: \HI\ holes in the
interstellar medium and anomalous gas structures outside the main
galactic disk.

\end{abstract}

\keywords{techniques: image processing --- galaxies: individual(NGC~2403,
IC~2574, Holmberg~II) --- ISM: structure}

%%%%####################################################################

\section{Introduction}
\label{sec:intro}

The art of astronomical observations lies in reconstructing the
highest fidelity representation of the sky, given the inherent
limitations of the data, which are always to some degree affected by
atmosphere, instrument, technique or telescope. Some kind of
processing always needs to be applied to minimize these issues and
produce the most accurate representation of the astronomical source of
interest. This processing is particularly important for observations
obtained with radio synthesis telescopes, \eg\ observations of the
$21$cm line of atomic hydrogen. Even though, in this domain,
instrumental artifacts are generally small and the atmosphere is
well-behaved, the noise and instrumental response (or Point Spread
Function, PSF) are still a major problem in generating a true
representation of a source. For radio interferometers, the full width
at half maximum (FWHM) of the PSF to first order depends on the length
of the longest baseline. The precise shape of the PSF and the surface
brightness sensitivity depend, however, on the configuration of the
array and its filling factor. The incompletely sampled size-scale
distribution inherent to an unfilled aperture instrument leads to a
complicated PSF and consequently to an image with significant
artifacts (we refer the reader to \citealp{tms_radioint} for more
details).

Deconvolution is the technique of choice for attempting to remove
these artifacts.  It does so by interpolating and extrapolating the
incompletely sampled \textit{uv} plane, thus attempting to correct for
the effects of the complex PSF. Other factors that influence the
observations, such as noise and atmospheric effects, cannot be
corrected by deconvolution.  Deconvolution can be thought of as a
simple equation: $(F+G)*H~=~I$, where $(F+G)*H$ indicates the
convolution of $F$ plus $G$ with $H$. $F$ is the true image, $G$ is
the noise, $H$ is the PSF and $I$ represents the raw measurements or
observations. Deconvolution is the process of solving for $F$, given
$G$, $H$ and $I$. A measure for the performance of any deconvolution
algorithm thus lies in its effectiveness of reducing the effect of a
complex PSF ($H$) on the image ($F$). Data that are not (or cannot be)
recorded at the telescope, due to, for example, incomplete sampling of
the \textit{uv}-plane are not recovered through deconvolution, but
have to be interpolated. This interpolation is at the heart of some of
the complications with deconvolution described in this paper.

The `de-facto' standard of deconvolution algorithms for radio
astronomy is the \clean\ algorithm \citep{hogbom_clean}. This
classical \clean\ is an iterative procedure. It assumes that the sky
is only sparsely filled with a limited number of point sources
superimposed on an otherwise empty background. It finds the location
and strength of all these sources, and uses them to reconstruct a
representation of the true image using an ideal PSF. Over the years
several variants \citep[\eg][]{schwab_clean,clark_clean} have been
proposed to overcome various limitations and nuances that are
intrinsic to \clean\ \citep[see
also][]{schwarz_cleananal,bs_m31}. However, there still remains a
basic assumption that the sky is composed of point sources, which is
adequate if the source being observed has limited extent. Extended
structure is modeled by \clean\ as a large number of point sources and
the iterative procedure becomes a time-consuming process involving a
constantly increasing number of point sources. To partly compensate
for this problem, \clean\ can be restricted to only work in regions
where the source of interest is known to exist, through the use of
\clean\ `windows'.  Such windows require prior knowledge of the source
structure and extent, but are flexible and multiple windows can be
used to bound a complicated source.

An obvious enhancement to \clean\ would be to make it more efficient
at modeling extended structures. Many such `scale-sensitive'
algorithms have been proposed.  These methods generally assume that
the sky is not composed of only point-sources as \clean\ does, but of
sources of many different sizes and scales.  One of the first such
methods, multi-resolution \clean, tackles the problem by smoothing and
decimating the dirty image and PSF to emphasize the extended emission.
The image resulting from cleaning this dirty image is then used as an
initial model for cleaning the next higher resolution image, \etc\
\citep{ws_mrclean}. Wavelet \clean\ uses the wavelet transform in the
\clean\ process but basically operates in a similar manner to
multi-resolution \clean\ \citep{sblp_wavelet}.  The more recently
developed Adaptive Scale Pixel \clean\ uses scale sizes that can
`adapt' their size during processing \citep{bc_ssd}. There is a wealth
of literature explaining the inner workings of multi-resolution
\clean\ and the other algorithms mentioned above in detail
\citep[see][in addition to the references
above]{bc_msc,deconv_review}, including a considerable amount of
modeling as well as some practical tests.

Multi-scale \clean\ (hereafter referred to as \msclean) is a
straight-forward extension of the classical \clean\ algorithm for
handling extended sources \citep{msclean}. Just like the
multi-resolution \clean\ defined by \cite{ws_mrclean}, it works by
assuming the sky is composed of emission at different spatial scales.
However, whilst the \citeauthor{ws_mrclean} multi-resolution \clean\
works on each scale sequentially, \msclean\ works simultaneously on
all scales that are being considered.  This prevents potential errors
made at a previous larger scale from being `frozen in' and requiring
many iterations at smaller scales to correct.  \msclean\ has been
implemented in both CASA\footnote{Common Astronomy Software
Applications, \url{http://casa.nrao.edu/}} (formerly AIPS++) and
classical AIPS\footnote{Astronomical Image Processing System,
\url{http://www.aips.nrao.edu/}} (as part of the multi-resolution
options of the IMAGR task). In this paper, we compare the operation of
\msclean\ against classical \clean\ both with and without the use of
\clean\ windows, using some real-world data and astronomical problems
derived from high quality HI data obtained through the THINGS project.
We refer to \cite{msclean} for a more technical description of
\msclean, as well as a description of tests involving artificial data.

THINGS, The \HI\ Nearby Galaxy Survey, consists of a sample of 34
nearby galaxies observed with the NRAO Very Large Array\footnote{The
National Radio Astronomy Observatory is a facility of the National
Science Foundation operated under cooperative agreement by Associated
Universities, Inc.} (VLA) B,C and D arrays \citep{THINGS}. The high
resolution ($\sim6\arcsec$, $\sim5$~\kms) achieved with THINGS pushes
the VLA to the limits of its current performance for observations of a
significant sample of galaxies. The THINGS observations, being
combined multi-array data with a large range of spatial scales
therefore present one of the most challenging data-sets with which to
compare the efficiency of deconvolution algorithms. We emphasize that
classical \clean\ works very well on the THINGS data sets, and that
with a full knowledge of its intricacies excellent results can be
achieved. However, \clean\ is not perfect, and our goal in this paper
is to describe some practical applications of the \msclean\
algorithm. \msclean\ seems to have fewer limitations than \clean\ does
when it comes to handling data of extended sources. It is promising
enough that it might point the way to an efficient exploration of the
high-quality data, similar to the THINGS data set, that will be
routinely produced by the next-generation radio and millimeter
facilities such as ALMA\footnote{\url{http://www.alma.info/}},
EVLA\footnote{\url{http://www.aoc.nrao.edu/evla/}},
LOFAR\footnote{\url{http://www.lofar.org/}} and
SKA\footnote{\url{http://www.skatelescope.org/}}

This paper begins with a brief description of the \clean\ and
\msclean\ algorithms in Section~\ref{sec:deconv-desc}.  In addition to
summarizing the procedure each algorithm uses in Sections
\ref{sec:clean-proc} and \ref{sec:msclean-proc}, the paper discusses
the limitations of the \clean\ algorithm and the advantages that
\msclean\ provides (Section~\ref{sec:clean-msclean-issues}).  In
Section~\ref{sec:clean-msclean}, a detailed comparison of the
performance of \clean\ and \msclean\ on a sample of three galaxies
from THINGS is presented.  Section~\ref{sec:ms-mr-w-c} then presents a
practical comparison between \msclean\ as implemented in CASA, the
version implemented in AIPS and classical \clean\ as well as
windowed-\clean. Having compared \msclean\ to a number of related
algorithms, the paper then takes this comparison a step further by
describing some astrophysical applications of \msclean, namely the
detection of structures within the \HI\ disks of galaxies but also the
detection of anomalous gas structures outside the disk, in
Section~\ref{sec:applications}. The paper ends with a short summary of
the key results of the tests and comparisons performed
(Section~\ref{sec:summary}).

%%%%####################################################################

\section{Deconvolution with \texorpdfstring{\bclean}{CLEAN} and
  \texorpdfstring{\bmsclean}{MSCLEAN}}
\label{sec:deconv-desc}

Excellent detailed descriptions of the \clean\ algorithm can be found
in \cite{hogbom_clean}, \cite{schwarz_cleananal} and
\cite{cornwell_deconv}. For \msclean, \cite{msclean} contains a
detailed description of this algorithm.  For completeness, and to
introduce some nomenclature, we briefly describe the classical
\clean\, as well as the \msclean\ procedure here.

Deconvolution algorithms attempt to create a model of the true `sky brightness
distribution' (the object) from the `dirty map' (the observed image)
using the `dirty beam' (the observed PSF). This model is called the
`\clean\ map' (or `restored map') and is created using the `\clean\
beam', which represents an ideal PSF. It is typically a Gaussian
function of the same FWHM as the central component of the dirty beam
\citep{schwarz_cleananal}. 

\subsection{The \texorpdfstring{\bclean}{CLEAN} Procedure}
\label{sec:clean-proc}

The \clean\ algorithm is an iterative procedure operating on the dirty
map. \clean\ can be described as follows:

\begin{enumerate}
 \item Find the location of the maximum absolute brightness point source in
  the `dirty map' and optionally within a user-defined window in the image.
 \item Multiply the strength of this point source by a gain factor
  (usually $10$\%) to generate a `\clean\
  component' at this location.
 \item Convolve the \clean\ component with the `dirty beam', and subtract
  this from the dirty map, recording the position and
  strength of the \clean\ component subtracted.
 \item Repeat the above three steps on the dirty map, until all
  emission is found, or a certain flux threshold is reached, or a number
  of iterations has been achieved.
\end{enumerate}

After the procedure has stopped by meeting one of the conditions in
Step 4, the result is a list of \clean\ components, and a residual
image with all flux cleaned away to the specified conditions. The
final image (the \clean\ map or restored image) is created by adding
all of the \clean\ components convolved with the \clean\ beam onto the
residual map.  Adding the residual image left after the iterative
subtraction procedure is in principle optional, but is usually always
done in order to retain information on the noise and any remaining
residual flux \citep{schwarz_cleananal}.  We defer to
Section~\ref{sec:clean-msclean-issues} a discussion on the potential
problem of combining cleaned and uncleaned data.

\subsection{The Multi-Scale \texorpdfstring{\bclean}{CLEAN} Procedure}
\label{sec:msclean-proc}

Multi-Scale \clean\ (in both its CASA and AIPS implementations) is a
modification of the classical \clean\ algorithm.  Instead of
representing the sky as empty and containing a limited number of point
sources as is done by \clean, \msclean\ presumes that sources in the
sky are actually extended structures of different scales (which can
include point sources).

An important constraint is that the function used to define the shape
of the scales has a finite extent, so that, for example, a \clean\
window can be applied. As such, a Gaussian function is not a good
choice, unless some kind of truncating function is applied. Generally,
a paraboloid with an appropriate tapering function is chosen, which
becomes a delta-function in the limit of zero scale-size
\citep{msclean}. By pre-computing the convolution of the dirty beam
with each scale size, one can proceed with a parallel \clean-like
procedure on a set of images generated for each of the scales. It is
important to include a scale-size zero delta-function which enables
proper fitting of point sources. The process is then as follows:

\begin{enumerate}
\item Convolve the dirty map with each scale
  size to create a set of convolved images.
\item Find the global peak among these images, \ie\ the scale that contains
  the maximum total flux and
  record the position, flux and scale size for the image in which this occurs.
\item Subtract the pre-computed scale (of the same size in which the
  peak was found) convolved by the dirty beam, multiplied by some
  gain factor, from all the images made in the first step.
\item Store the subtracted component and the scale size in a table.
\item Repeat the above steps on the current convolved images until all
  emission has been removed, a flux threshold is reached or an
  iteration limit has passed.
\end{enumerate}

In other words, where \clean\ operates on a single residual image,
\msclean\ keeps a set of residual images, one image per scale size
defined. The peak subtraction is performed on all of these images, but
only the one subtracted component and its scale size are stored in the
\clean\ component table. The restoration is then an addition of the
appropriately scaled, positioned and convolved components subtracted at
each iteration on top of the final residual image.  We refer to
\cite{msclean} for a technical description of the algorithm.

\subsection{The Limitations of \texorpdfstring{\bclean}{CLEAN} and
  \texorpdfstring{\bmsclean}{MSCLEAN}'s Advantages}
\label{sec:clean-msclean-issues}

For purely practical reasons, any implementation of \clean\ will have
an iteration limit built in \citep[\eg][]{cornwell_deconv}. The design
of the algorithm is to iteratively find and remove the strongest point
sources; the number of iterations is therefore one of the factors that
defines how deep \clean\ will go in terms of the flux
level. Eventually, noise peaks will start to be cleaned away as well,
as the algorithm cannot distinguish them from faint real signals. The
number of components will therefore start to increase
dramatically. This can easily spiral out of control as \clean\
approaches the noise limit and leads to a diverging total flux
\citep{cornwell_deconv}. All of this illustrates the necessity of
imposing a hard limit on the number of iterations, either directly or
through a flux threshold.

However, despite its usefulness in placing a practical constraint on
\clean, the iteration limit is not without its disadvantages. It is a
compromise between cleaning as deeply as possible and avoiding most
of the noise. As mentioned in the introduction, this is a particularly
important point for extended sources (such as the THINGS galaxies),
which \clean\ will attempt to model with a large number of point
sources.  In contrast, practical results show that \msclean\ removes
large-scale structure before finer details \citep[][and see
Section~\ref{subsec:mscleanscale}]{msclean}. This provides a useful
advantage over \clean\ in most cases: the prior removal of underlying
extended emission will reduce the strength of small-scale emission
peaks that will remain. This means that when \msclean\ begins to
remove the small-scale structure, it will require fewer iterations to
do so.  The lack of this scale-size advantage in classical \clean\
means it is required to slowly cut down sources. 
\msclean, with its convergence from large to small scales, does not spend
its final cycles slowly removing a large amount of extended emission
in small increments.

Classical \clean\ must also use low loop gains (typically $10$\% or
less) to improve the reconstruction of extended emission.  Using a
high loop gain would be more efficient, especially for extended
sources, but can easily lead to instabilities \citep{cornwell_deconv}.
Furthermore the gain is one parameter of \clean\ that has a large
effect on the final \clean\ solution
\citep{schwarz_cleananal,tan_cleananal}.  \msclean\ is much less
dependent on the gain and is able to use much higher values
\citep{msclean}.  This means that each \msclean\ scale has the ability
to remove a much larger portion of the flux in each individual
iteration, again reducing the overall number of iterations required
compared to \clean.

The problem whereby \clean\ poorly models extended emission is further
compounded by the interpolation for the missing spacing information.
Generally, in interferometric observations, the extremely short or
zero spacings, which measure the largest structure on the sky, are
missing. In other words, the innermost part of the \textit{uv}-plane
is not (well) sampled. In these cases the total flux of an
\textit{extended} source (\ie\ containing structures at scales larger
than sampled by the shortest baseline) cannot be recovered.  Together
with the strong side-lobes present in the dirty beam and the pedestal
of uncleaned flux, this leads to a \clean\ `bowl': the source sits in
a region with a negative background. This \clean\ bowl will cause
skewed noise and flux estimates.  The higher efficiency of \msclean\
in removing extended structures means it can approach the noise limit
more easily than \clean, with the result that the presence of the
\clean\ bowl is also greatly reduced yielding a noise-like residual
background.

A subtle, but important limitation of \clean\ lies in the
reconstruction of the \clean\ or restored image from the final
residual image and the \clean\ components. The restored image is an
addition of two separate flux scales, a residual map with the units
\mbox{Jy per dirty beam} and a \clean\ component map with units
\mbox{Jy per clean beam}. As the extent of the dirty beam is always
larger than that of the best-fitting \clean\ beam, using the \clean\
beam to determine the flux of the residuals will always lead to an
overestimate of this flux. As such a correction factor needs to be
applied to determine the flux in regions with signal
\citep{jvm_residflux}. This is described in \cite{THINGS} for the
THINGS galaxies. This scaling of the residuals, and therefore the
noise, yields correct fluxes in areas with signal, however the noise
properties are no longer representative.  Residual-scaled cubes should
thus be only used to measure fluxes in areas with significant signal.
Any other applications which critically depend on noise properties
(such as profile fitting) should use the unscaled cubes
\citep[see][for a detailed discussion]{THINGS}.

In order to illustrate the limitations of \clean\ and potential
advantages that \msclean\ could provide, the remainder of this paper
is devoted to a practical comparison between \clean\ and \msclean\
(Section~\ref{sec:clean-msclean}).  We also compare the implementation
of \msclean\ in the CASA and AIPS software packages to classical
\clean\ with the use of \clean\ windows (Section~\ref{sec:ms-mr-w-c}).
We use a small sample of galaxies from THINGS to compare the results.

%%%%####################################################################

\section{A Detailed \texorpdfstring{\bclean}{CLEAN} and
  \texorpdfstring{\bmsclean}{MSCLEAN} Comparison}
\label{sec:clean-msclean}

\subsection{Data Processing}
\label{sec:dataproc-clean-msclean}

We chose three galaxies from the THINGS survey to use for our
comparison of \msclean\ and \clean. These galaxies are
\object{NGC~2403}, \object{Holmberg~II} and \object{IC~2574} and they
cover a range of \HI\ masses and morphologies \citep{THINGS}. Each
galaxy was observed with the VLA in B, C and D configurations. This
study uses the calibrated and combined \textit{uv} data-set from all
arrays for each galaxy. More details on the observations and
generation of the \textit{uv} data-sets can be found in
\cite{THINGS}. The \msclean\ implementation in the CASA software
package and the \clean\ implementation in the AIPS software package
\citep[based on the Clark \clean, see][]{clark_clean} was used on the
\textit{uv} data-set for this comparison. For each galaxy, two data
cubes were generated with two different weightings; a `natural'
weighting and a `robust' weighting with a robustness parameter of
$0.5$ \citep{briggs_robust}. These weightings were the same as used in
the creation of the original THINGS data cubes.

The data cubes were created with the same spatial and velocity pixel
sizes as the original THINGS cubes \citep[for details see][]{THINGS}.
The cubes were then `{\scshape msclean}-ed' down to $2.5\sigma$ which
was also the flux threshold used in THINGS. As mentioned in
Section~\ref{sec:clean-msclean-issues}, \msclean\ can usually use much
higher gain factors than classical \clean. In this study, the
\msclean\ gain factor was set at $0.7$ for all three galaxies
\citep[see][for a discussion of the \msclean\ loop gain]{msclean}. No
\clean\ windows were used. A maximum number of \msclean\ iterations of
$1200$ (NGC 2403), $700$ (Holmberg~II) and $1000$ (IC~2574) were
chosen for each galaxy per channel. In all cases the flux limit was
reached well before the iteration limit.  Unless mentioned otherwise
below, the cubes were also corrected for primary beam attenuation
using the \textsc{linmos} task of the MIRIAD\footnote{Multichannel
Image Reconstruction, Image Analysis and Display,
\url{http://www.atnf.csiro.au/computing/software/miriad/}} software
package. Residual cubes were also created for the \clean\ and
\msclean\ data.

\subsection{\texorpdfstring{\bmsclean}{MSCLEAN} Scale Choice}
\label{subsec:mscleanscale}

As discussed briefly in Section~\ref{sec:msclean-proc}, CASA (and
AIPS++) use paraboloids for the shape of the scale components.
Extensive experimentation shows that the results did not depend
significantly on the number of scales chosen nor their
distribution. The most important choice is that of the largest size in
the distribution. We found that optimum results were obtained when
this was chosen to correspond roughly to the size of the largest
coherent structures visible in individual channel maps. This choice
does not have to be exact. We chose a total of six scales distributed
where each scale is three times larger than the preceding
scale. Again, this choice was not critical, but we found that this
distribution was more efficient (in terms of number of iterations
required) than a linear distribution. The largest scale for each
galaxy was $130\arcsec$ (NGC~2403), $270\arcsec$ (Holmberg~II) and
$160\arcsec$ (IC~2574). 

\begin{figure}[htbp]
  \centering
  \includegraphics[angle=270,width=0.5\textwidth]{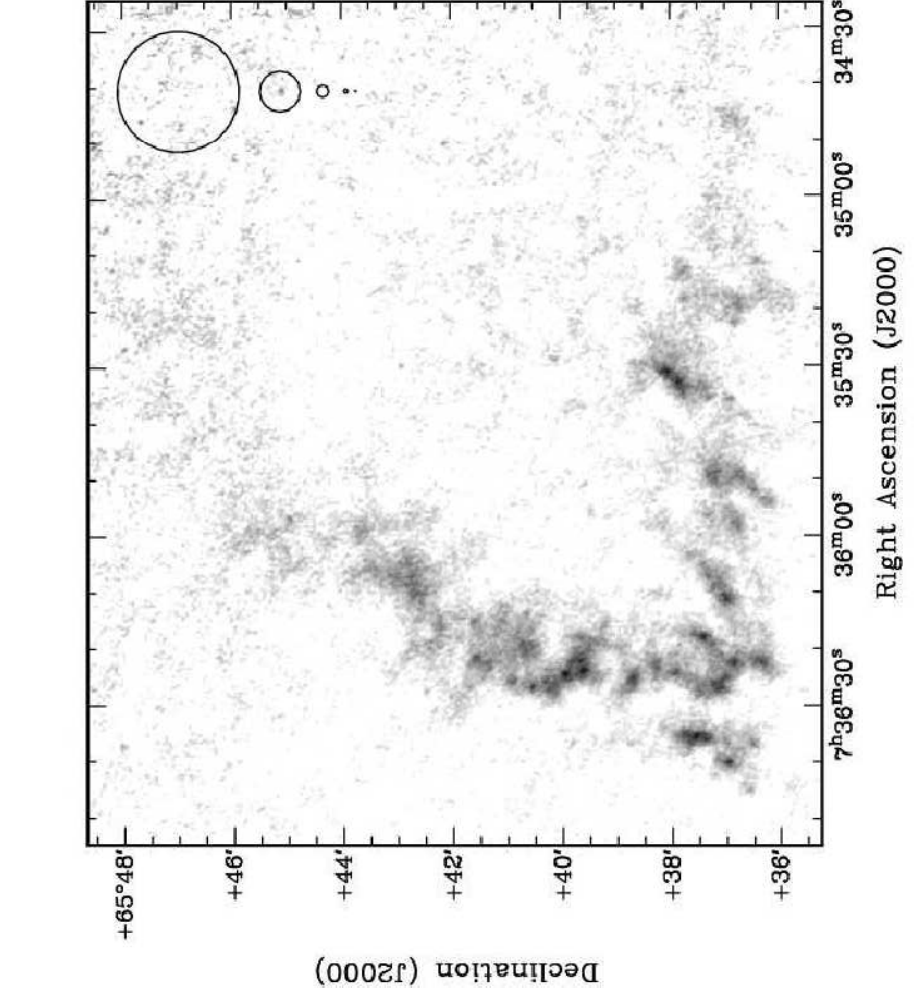}%
  \caption{The relationship between the \msclean\ scale sizes and
    structure in a single velocity channel of NGC~2403.  The upper right
    corner shows the \msclean\ scale sizes used in processing this galaxy,
    from largest to smallest down the figure. All scales excluding the
    smallest scale (a delta peak) are shown. The diameters of the scales
    are $130\arcsec$, $44\arcsec$, $13\arcsec$, $4\arcsec$ and
    $1.3\arcsec$.  The largest scale size ($130\arcsec$) roughly
    corresponds to the largest coherent structure visible in the channel
    map.  This relation was used as the basis for choosing the \msclean\
    scales to be used for each galaxy, as discussed in
    Sec.~\ref{subsec:mscleanscale}.\label{fig:ngc2403-scale2struct-comp}}
\end{figure}

For comparison, Figure~\ref{fig:ngc2403-scale2struct-comp} shows how
the range of scales chosen and structure size for a single channel map
are related in the galaxy NGC~2403. It can be seen that for the
largest scale size, the largest coherent structure visible in the
channel map fits within its diameter. Choosing an even larger scale
would have no effect, as the largest structure is already optimally
contained within the scale distribution shown, and \msclean\ would
simply not choose these even larger scales. Choosing a smaller,
largest scale would simply increase the number of iterations.

%\placefigure{fig:ngc2403-scale2struct-comp}

\begin{figure}[htbp]
  \centering
  \includegraphics[angle=270,width=0.45\textwidth]{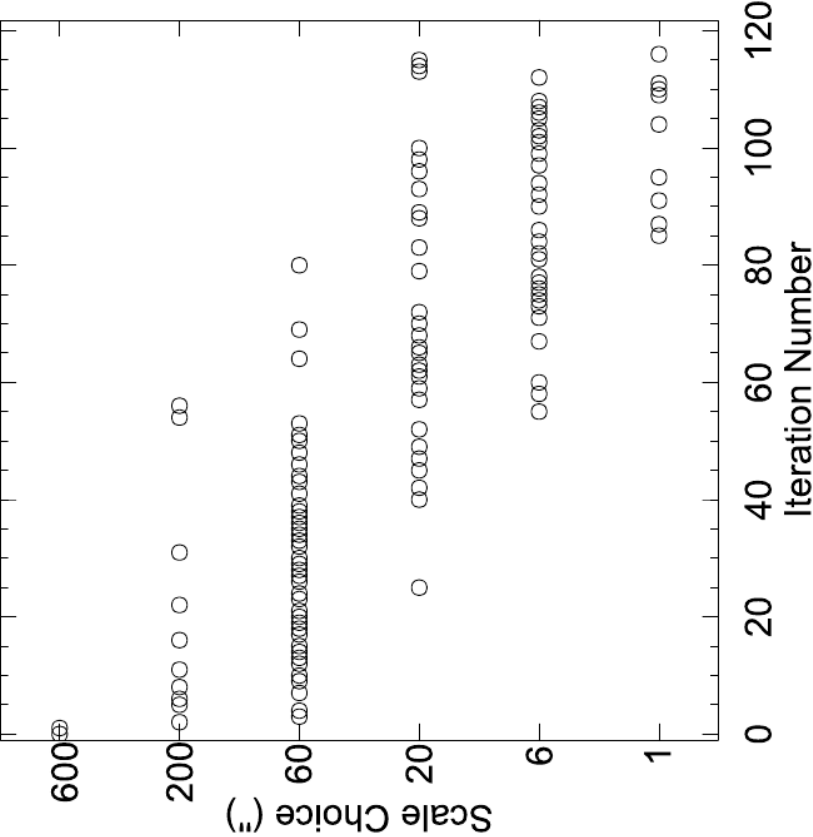}%
  \caption{The choice of scale size with iteration number for a single
    velocity channel from Holmberg~II in \msclean\ .  The x-axis shows the
    iteration number and the y-axis shows the scales (in arc-seconds). As
    shown, \msclean\ uses the largest scales first, which corresponds to
    extended emission, before switching to smaller scales and cleaning
    finer structure.  \label{fig:msclean-scalechoice}}
\end{figure}

Figure~\ref{fig:msclean-scalechoice} shows the choice of scales the
algorithm made with iteration number for a single velocity channel of
Holmberg~II. In other words, the figure shows the \msclean\ scale
component used as the algorithm progressed to lower flux levels. A
similar trend is observed across all channels of all three galaxies.
It shows that \msclean\ indeed removes emission at larger scales
before smaller scales (as explained in
Section~\ref{sec:clean-msclean-issues}).

% \placefigure{fig:msclean-scalechoice}

\subsection{Beams}
\label{sec:comp-beams}

In Table~\ref{tab:beam-size-comp} we compare the sizes of the
restoring beams of the AIPS \clean\ and the CASA/AIPS++
\msclean. These are both determined by fitting a Gaussian function to
the central component of the respective dirty beams. As their dirty
beams are identical, the small differences seen in
Table~\ref{tab:beam-size-comp} are due to the different fitting
procedures for a Gaussian function used in the restore processes of
both software packages.

\begin{deluxetable}{lcccc}
  \tablecaption{FWHM Beam Size Comparison for \clean\ and \msclean\
    data.\label{tab:beam-size-comp}}
  \tablewidth{0pt}
  \tablecolumns{5}
  \tablehead{
    \colhead{\multirow{2}{*}{Source}}& 
    \colhead{\multirow{2}{*}{Weighting}}&
    \multicolumn{3}{c}{Beam Size ($\arcsec$)}\\
    \colhead{} & \colhead{} &
    \colhead{NGC~2403} & \colhead{Holmberg~II} & \colhead{IC~2574}}
  \startdata
  \multirow{2}{*}{\clean}%
    & Natural & $8.8\times7.7$ & $13.7\times12.6$ & $12.8\times11.9$\\
    & Robust  & $6.0\times5.2$ & $7.0\times6.1$ & $5.9\times5.5$\\
  \multirow{2}{*}{\msclean}%
    & Natural & $8.0\times7.1$ & $11.4\times10.9$ & $10.8\times10.3$\\
    & Robust  & $5.7\times4.9$ & $6.6\times5.8$ & $6.3\times5.9$\\
  \enddata
\end{deluxetable}
% \placetable{tab:beam-size-comp}

\begin{figure}[htbp]
  \centering
  \begin{tabular}{cc}
    \includegraphics[width=0.4\textwidth,angle=270]{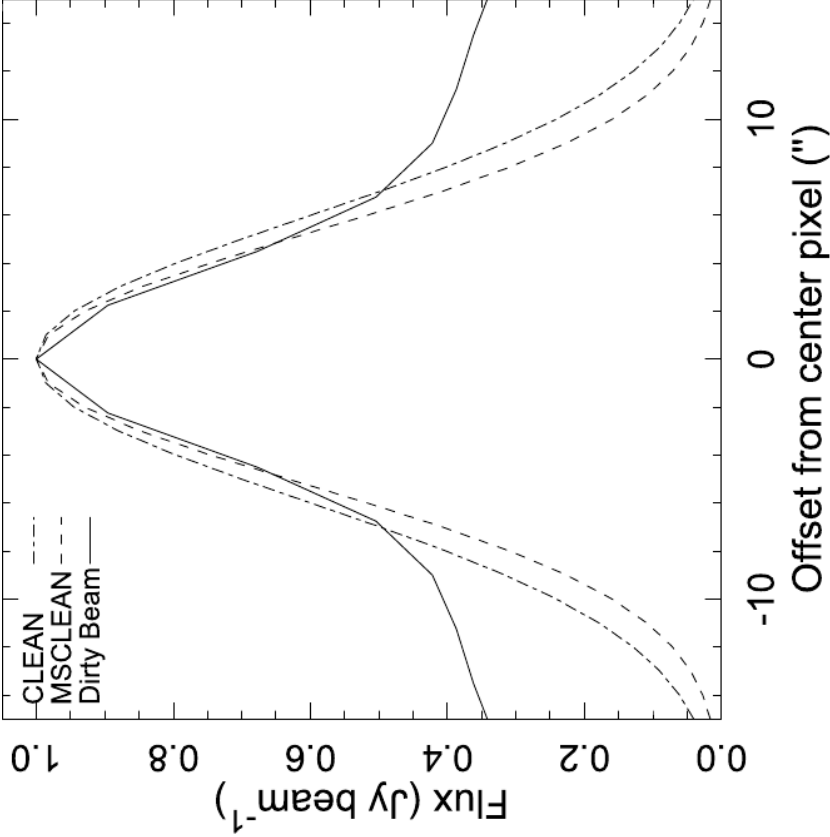}
    \includegraphics[width=0.4\textwidth,angle=270]{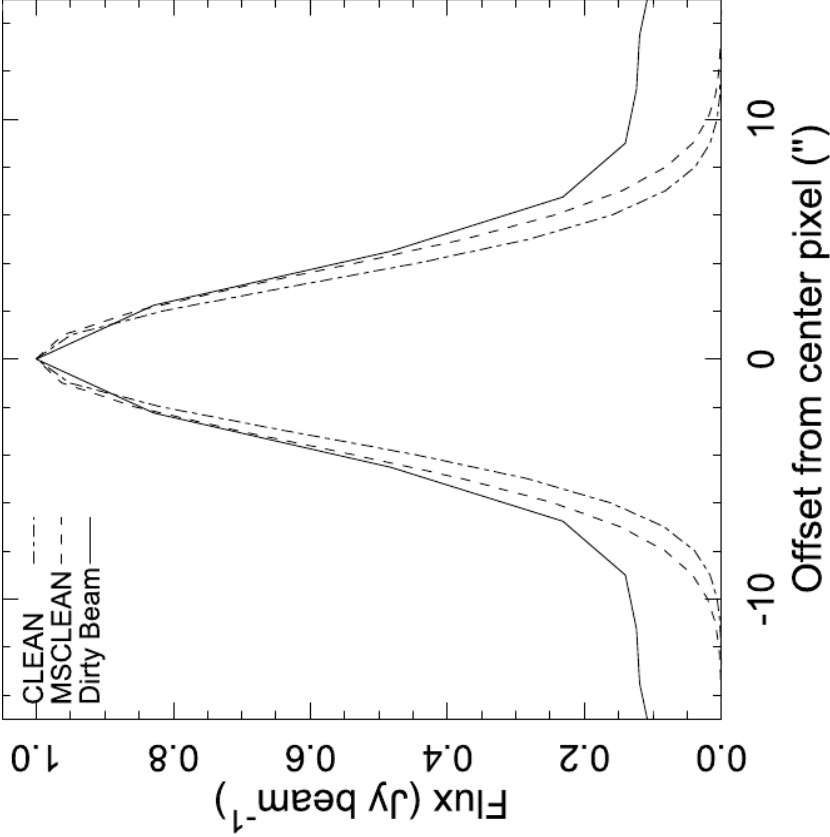}
  \end{tabular}
  \caption{Profiles of the dirty beam (solid lines) and restoring or
    \clean\ beams for both \clean\ (dot-dash lines) and \msclean\ (dashed
    lines) for the natural weighted (left) and robust weighted (with a
    robustness parameter of $0.5$, right) data of the galaxy IC~2574.  The
    dirty beam for the natural weighted data on the left shows the
    significant wings which make it difficult to fit a Gaussian function
    to it.  Nevertheless, the restoring beams of \clean\ and \msclean\ are
    very similar for this weighting.  On the other hand, The restoring
    beams for the robust weighted data are very similar and provide a good
    fit to the dirty beam, which is approximated well by a Gaussian
    function. \label{fig:ic2574-beam-comp}}
\end{figure}

In Figure~\ref{fig:ic2574-beam-comp} profiles of the restoring beams
for both weightings and both data-sets are plotted against the
respective dirty beam. It should be noted that natural weighting
results in a beam that cannot be well approximated by a Gaussian
function as shown in these figures. The natural dirty beam has
extended `wings', while for a robust dirty beam these wings are much
less pronounced and a Gaussian function is a much better fit.

% \placefigure{fig:ic2574-beam-comp}

Natural weighting is more sensitive to diffuse, extended
emission. This is where \clean\ is at its weakest.  The wings observed
in the natural beam (Figure~\ref{fig:ic2574-beam-comp}, left image)
result in a large difference of the beam integral between the natural
dirty and restoring beams. Robust weighting
(Figure~\ref{fig:ic2574-beam-comp}, right image) on the other hand is
a much better fit to the actual dirty beam. So residual scaling for
robust weighting is less important as there is less emphasis on
extended structure. Therefore \msclean\ and \clean\ results are much
more similar for robust weighting in the case of the THINGS data. In
the remainder of this comparison, the paper focuses on the analysis
based on the natural weighted data.

\subsection{Channel Maps}
\label{sec:comp-chanmaps}

\begin{figure}[htbp]
  \centering
  \begin{tabular}{cc}
    \includegraphics[angle=270,width=0.275\textwidth]{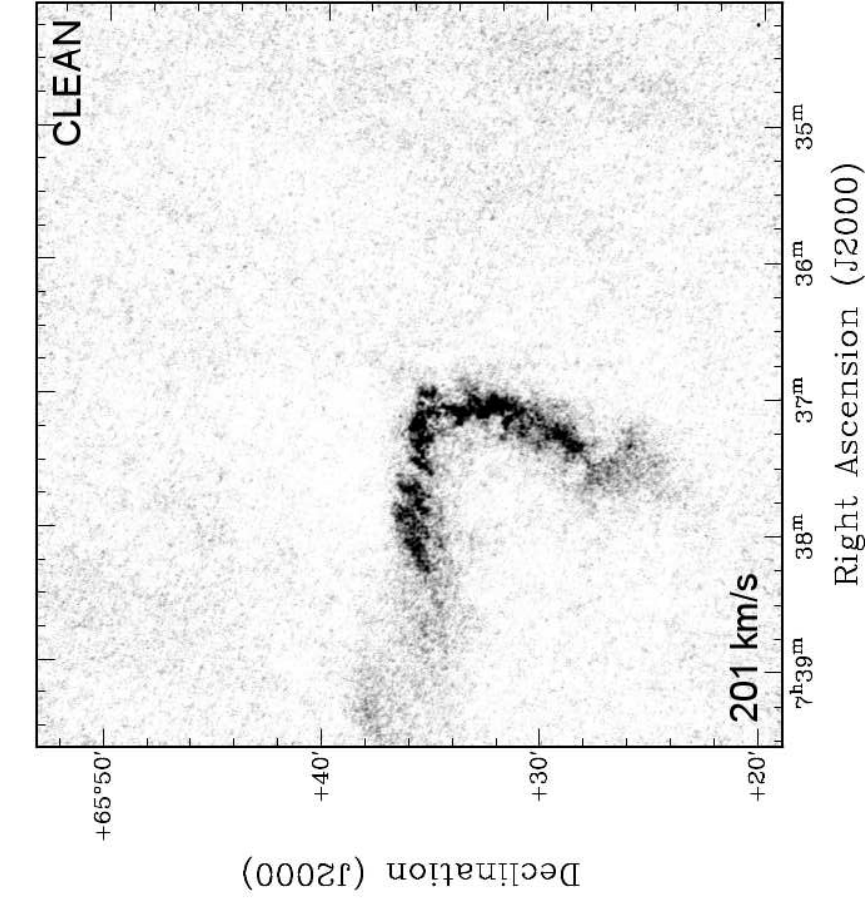}&
    \includegraphics[angle=270,width=0.275\textwidth]{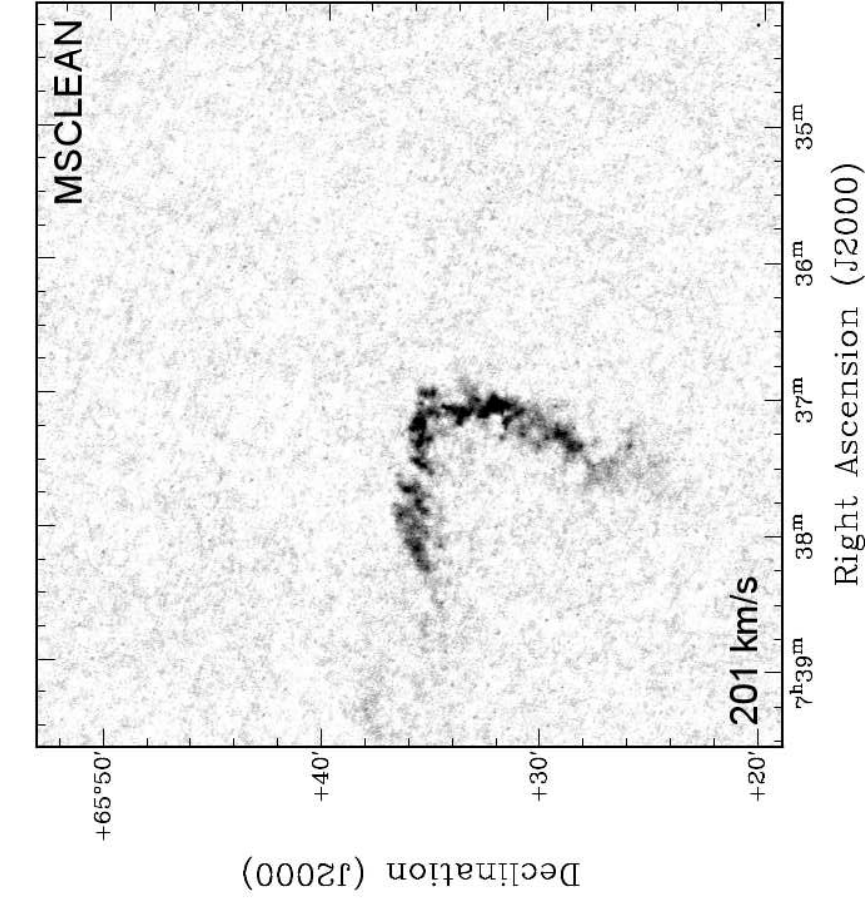}\\
    \includegraphics[angle=270,width=0.275\textwidth]{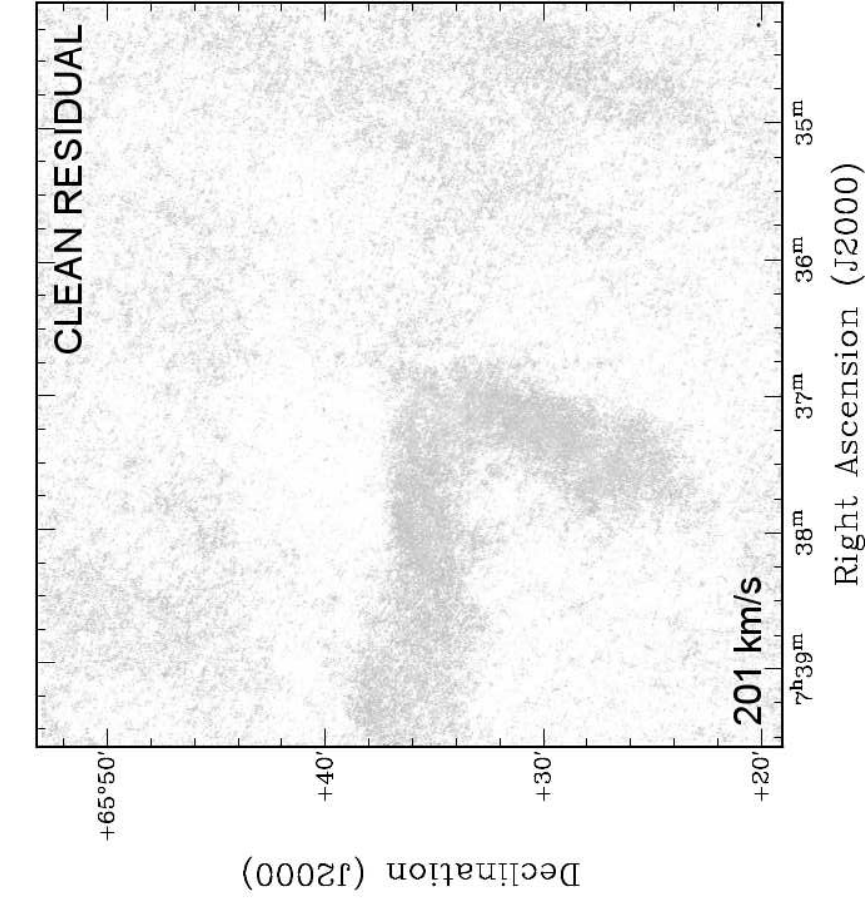}&
    \includegraphics[angle=270,width=0.275\textwidth]{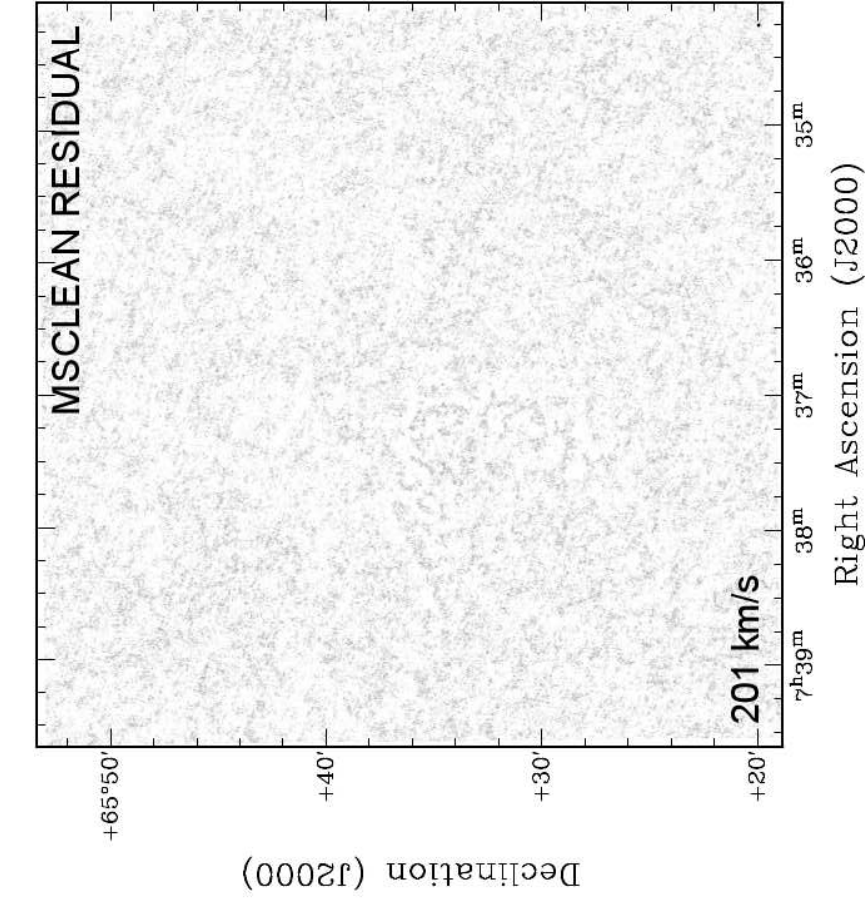}\\
  \end{tabular}
  \caption{Comparison of a single channel map of NGC~2403 for the
    \clean\ (left) and \msclean\ (right) data (top panels).  Also shown is the
    respective residual channel map for the \clean\ (left) and \msclean\
    (right) data (bottom panels), which is the \clean\ or \msclean\ 
    channel without adding in the convolved \clean\ components.  The
    channel maps were extracted from the natural-weighted, unmasked cubes.  No
    primary beam correction has been applied.  The \clean\ channel
    maps have not been residual corrected.  The gray-scale levels run
    from $0$ to $4$ \mjbeam.\label{fig:ngc2403-chanmaps}}
\end{figure}

\begin{figure}[htbp]
  \centering
  \begin{tabular}{cc}
    \includegraphics[angle=270,width=0.275\textwidth]{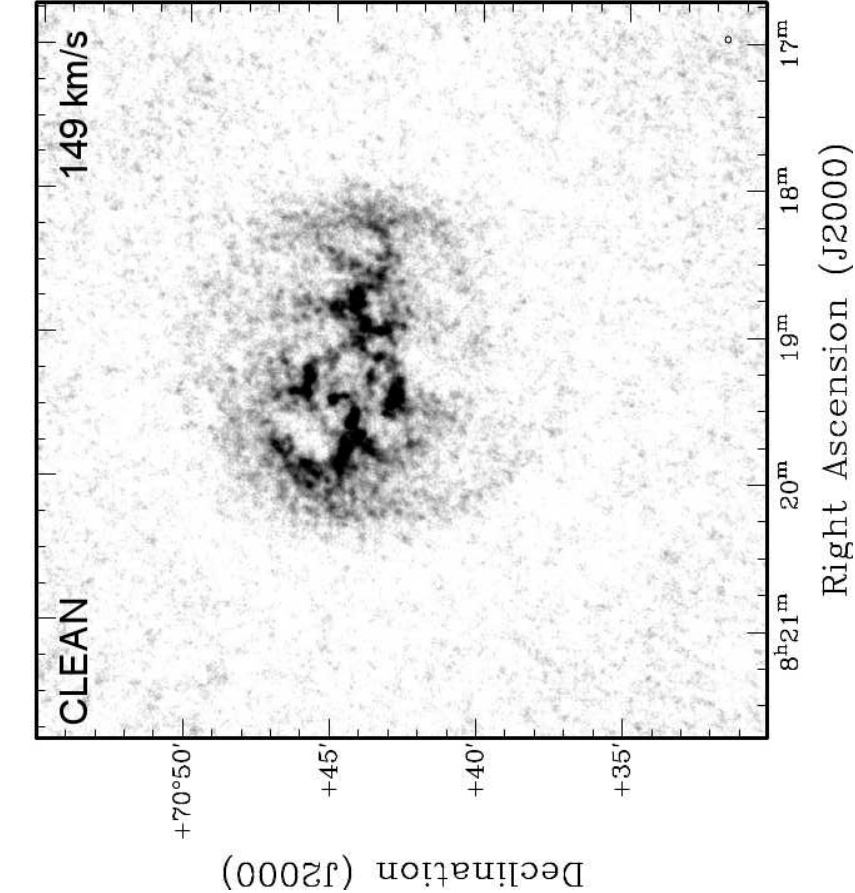}&
    \includegraphics[angle=270,width=0.275\textwidth]{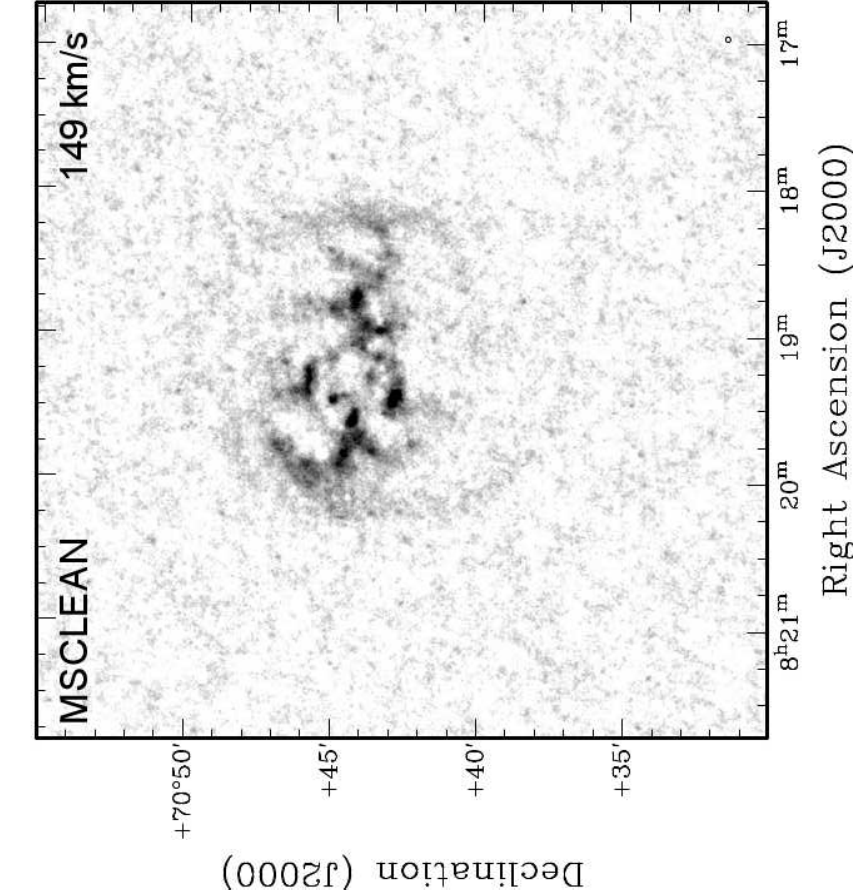}\\
    \includegraphics[angle=270,width=0.275\textwidth]{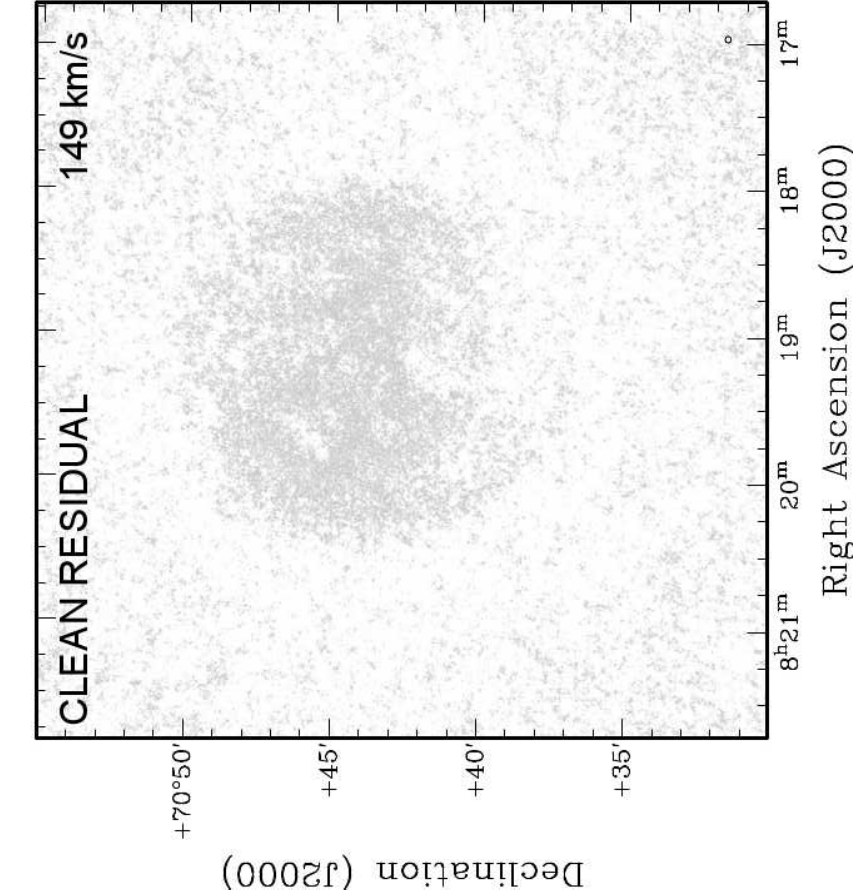}&
    \includegraphics[angle=270,width=0.275\textwidth]{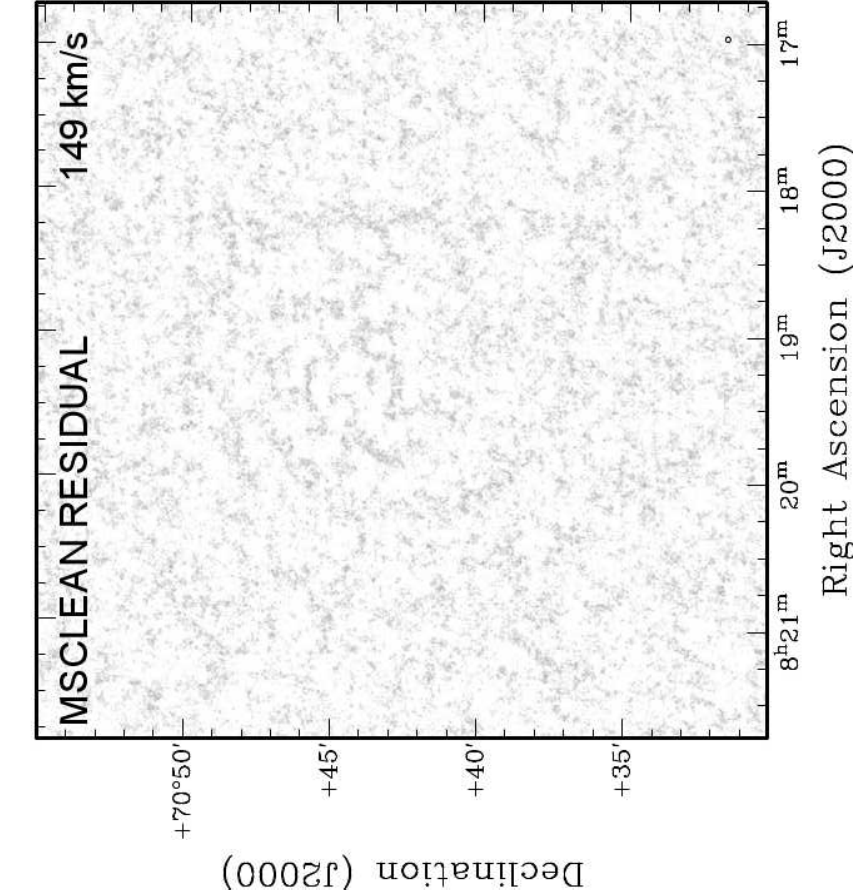}\\
  \end{tabular}
  \caption{Comparison of a single channel map of Holmberg~II for the
    \clean\ (left) and \msclean\ (right) data (top panels).  Also shown is
    the respective residual channel map for the \clean\ (left) and
    \msclean\ (right) data (bottom panels).  The channel maps were
    extracted from the natural-weighted, unmasked cubes.  No primary beam
    correction has been applied. The \clean\ channel maps have not been
    residual corrected. The gray-scale levels run from $0$ to $10$
    \mjbeam.\label{fig:hol2-chanmaps}}
\end{figure}

\begin{figure}[htbp]
  \centering
  \begin{tabular}{cc}
    \includegraphics[angle=270,width=0.275\textwidth]{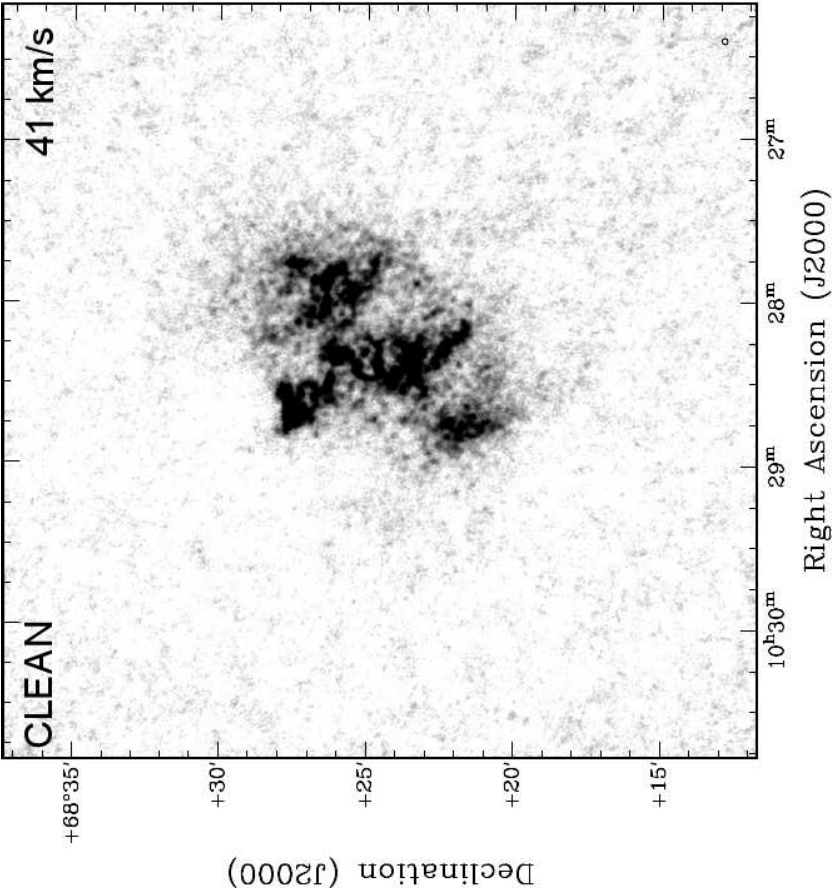}&
    \includegraphics[angle=270,width=0.275\textwidth]{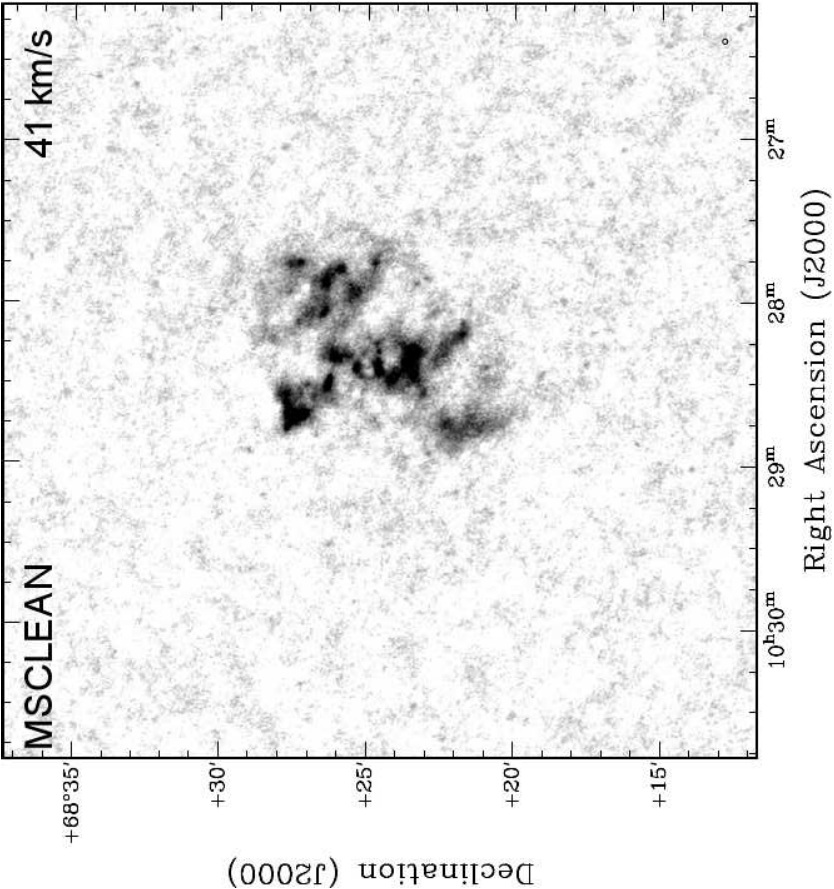}\\
    \includegraphics[angle=270,width=0.275\textwidth]{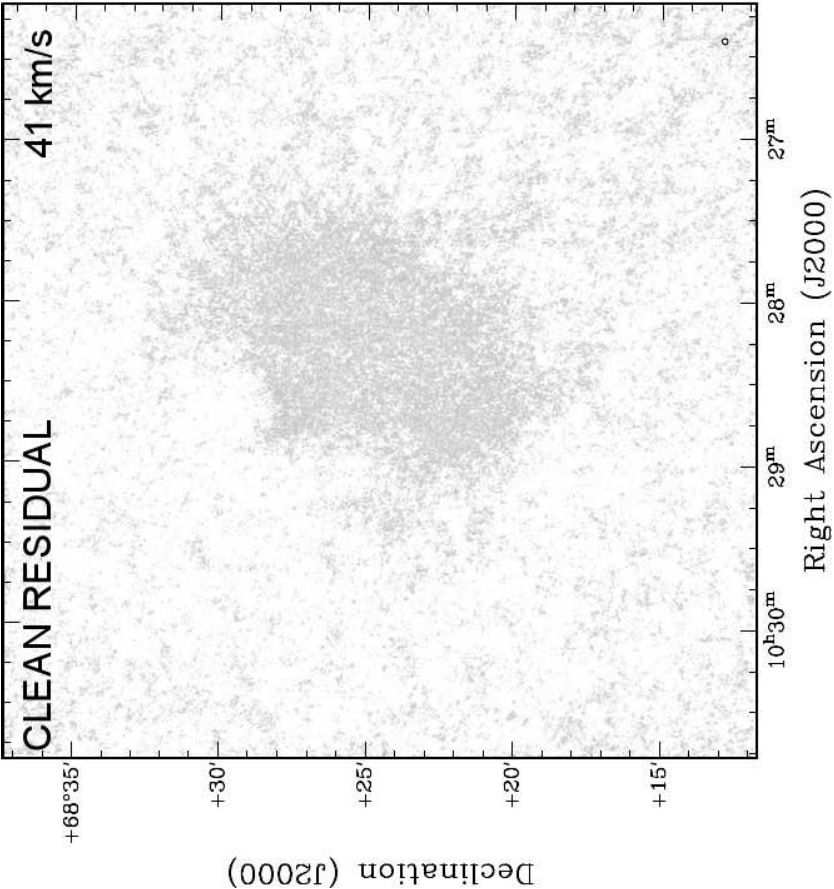}&
    \includegraphics[angle=270,width=0.275\textwidth]{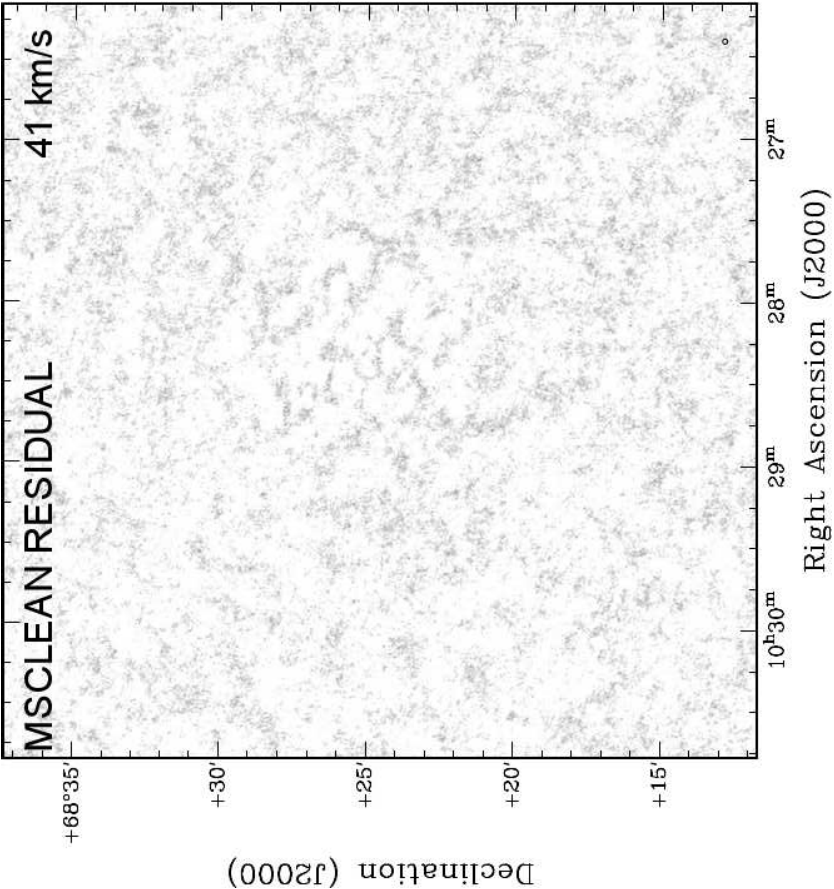}\\
  \end{tabular}
  \caption{Comparison of a single channel map of IC~2574 for the
    \clean\ (left) and \msclean\ (right) data (top panels).  Also shown is
    the respective residual channel map for the \clean\ (left) and
    \msclean\ (right) data (bottom panels).  The channel maps were
    extracted from the natural-weighted, unmasked cubes.  No primary beam
    correction has been applied. The \clean\ channel maps have not been
    residual corrected.The gray-scale levels run from $0$ to $6$
    \mjbeam.\label{fig:ic2574-chanmaps}} % -0.5 6.0
\end{figure}

A comparison of selected single channel maps for both the \clean\ and
\msclean\ natural-weighted cubes, including the respective residual
channel maps, are shown in Figures \ref{fig:ngc2403-chanmaps} to
\ref{fig:ic2574-chanmaps}. The channel maps shown in these figures are
from the non-residual scaled, unmasked THINGS cubes before correcting
for primary-beam attenuation.

The \clean\ bowl is most noticeable in the NGC~2403 channel map
(Figure~\ref{fig:ngc2403-chanmaps}), obscuring the noise in the inner
region of the image. The \clean\ residuals also show a significant
pedestal. This contributes to the difference in contrast between
regions of low and high flux across the \clean\ and \msclean\
data. For example, the extent of the high-contrast regions (black) are
much larger in the channels from the \clean\ data. This pedestal also
covers up the `holes' in the flux in the \clean\ images, where in the
\msclean\ images one can see right down to the noise levels in these
regions, as most readily seen in the channel map for Holmberg~II
(Figure~\ref{fig:hol2-chanmaps}).

% \placefigure{fig:ngc2403-chanmaps}
% \placefigure{fig:hol2-chanmaps}
% \placefigure{fig:ic2574-chanmaps}

\subsection{Integrated \texorpdfstring{\HIfat}{HI} Maps}
\label{sec:comp-mommaps}

\begin{figure}[htbp]
  \centering
  \begin{tabular}{cc}
    \includegraphics[angle=270,width=0.275\textwidth]{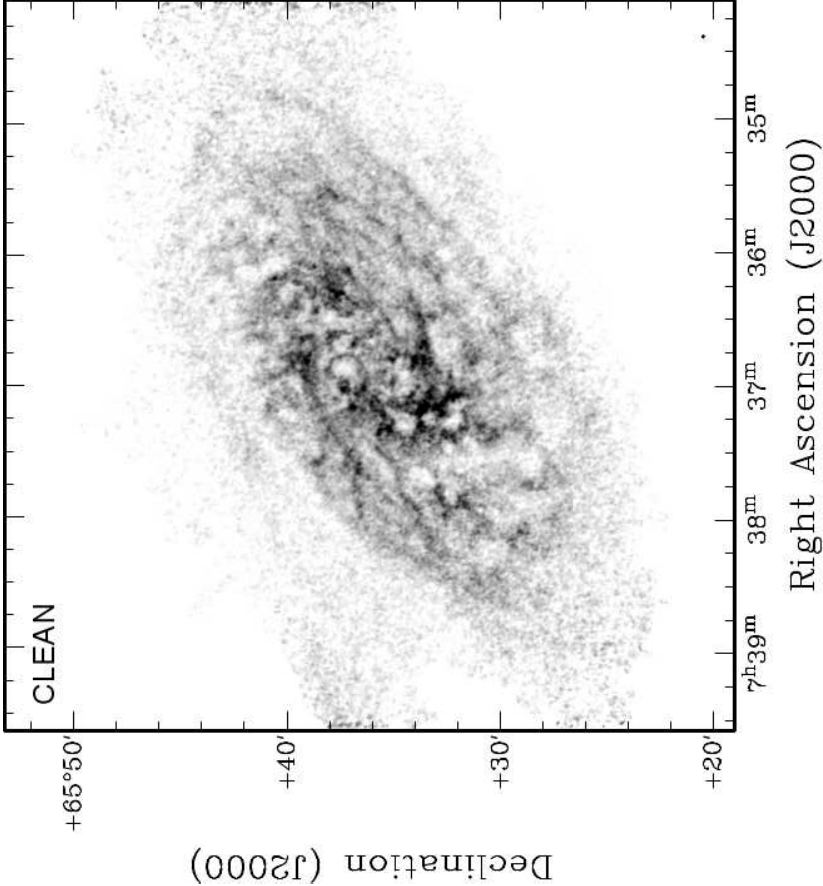}&
    \includegraphics[angle=270,width=0.275\textwidth]{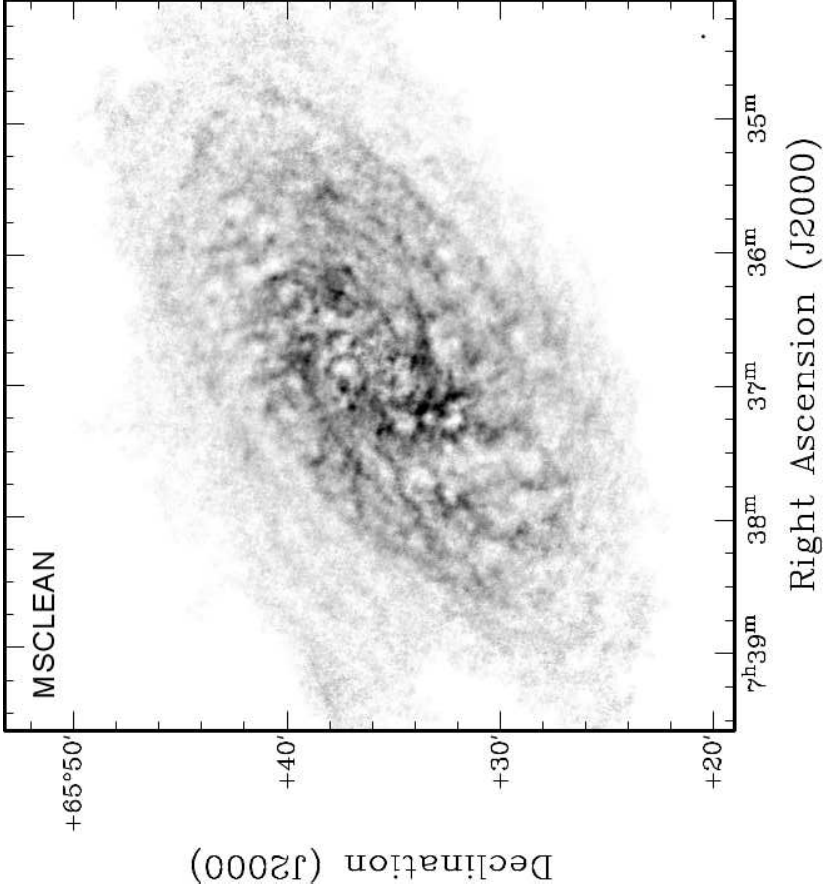}\\
    \includegraphics[angle=270,width=0.275\textwidth]{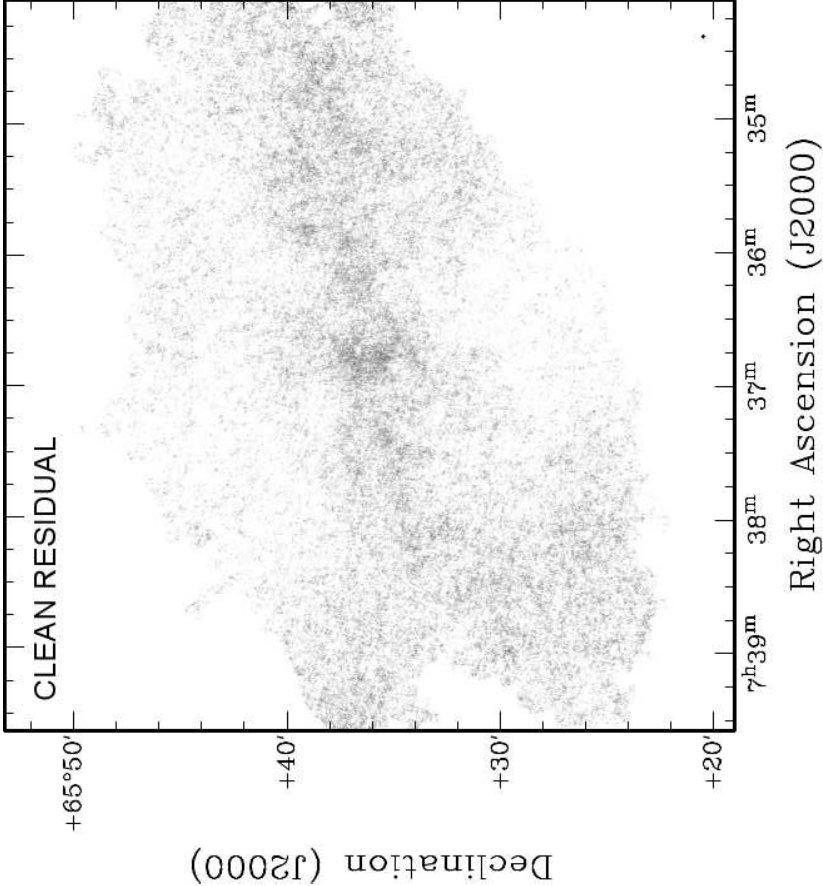}&
    \includegraphics[angle=270,width=0.275\textwidth]{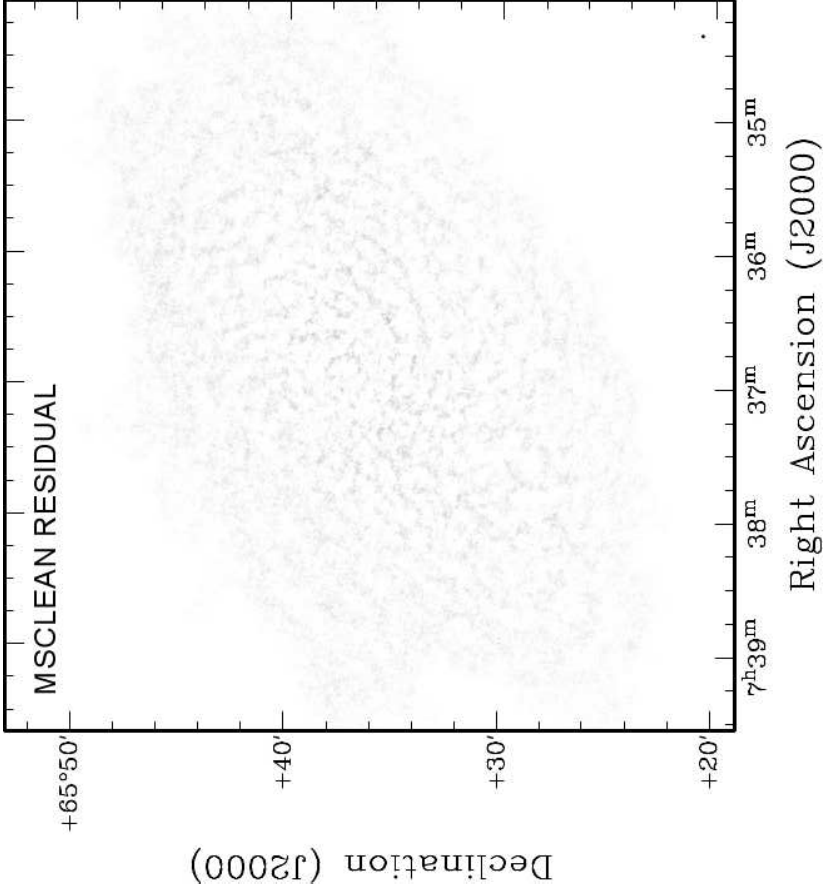}\\
  \end{tabular}
  \caption{\clean\ (left) and \msclean\ (right) integrated \HI\ maps
    (top row) for NGC~2403. The maps were generated from the
    natural-weighted, masked, primary-beam corrected and for \clean,
    residual flux corrected cubes. The equivalent residual integrated \HI\
    maps are also shown (bottom row), generated as described in
    Sec.~\ref{sec:comp-mommaps}.  No primary-beam corrections and for
    \clean, no residual flux corrections have been applied to these
    images. The gray-scale levels run from $0$ to
    $200$~\mjbeam~\kms. Beams are marked in bottom right corner of
    images.\label{fig:ngc2403-msclean-mom0}}
\end{figure}

\begin{figure}[htbp]
  \centering
  \begin{tabular}{cc}
    \includegraphics[angle=270,width=0.275\textwidth]{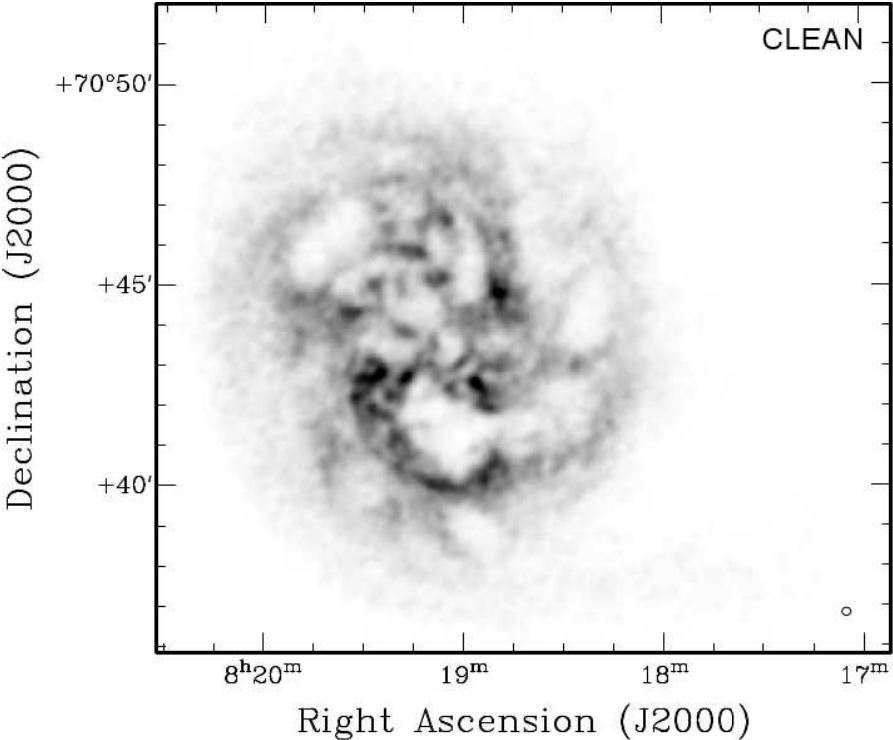}&
    \includegraphics[angle=270,width=0.275\textwidth]{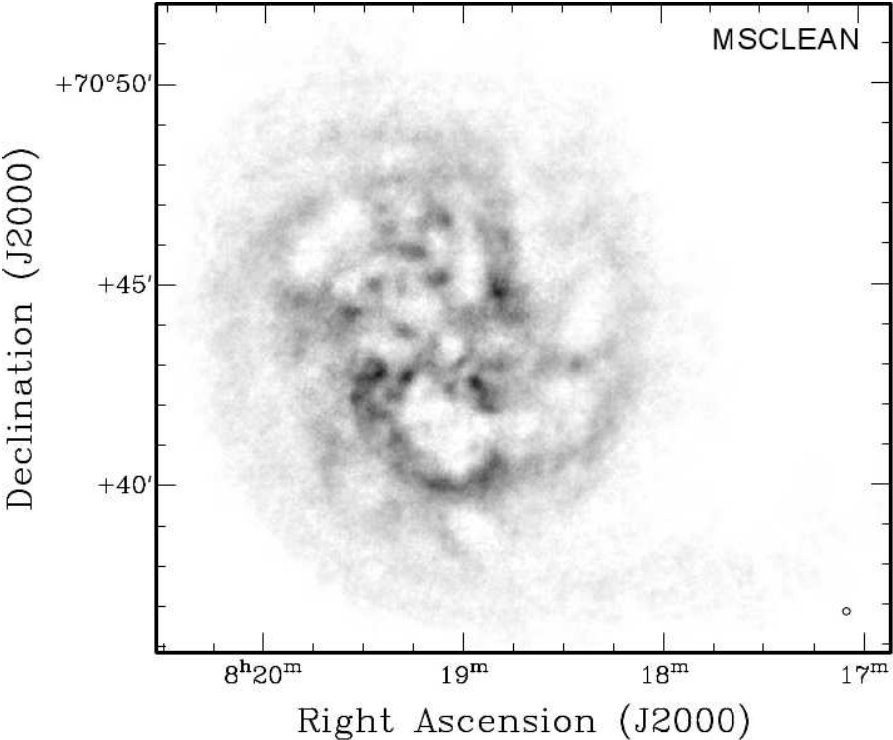}\\
    \includegraphics[angle=270,width=0.275\textwidth]{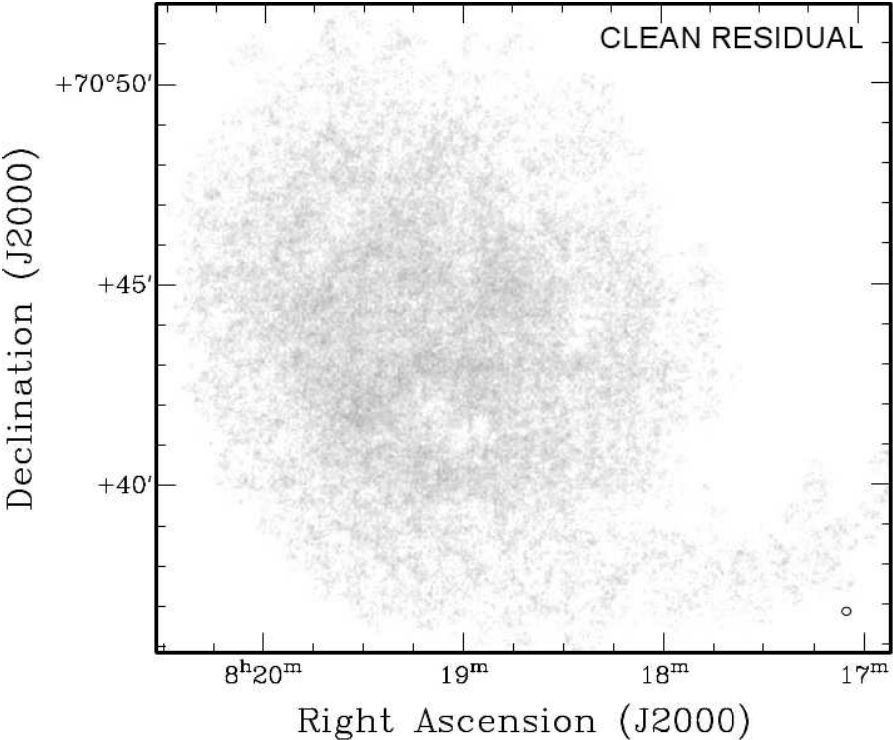}&
    \includegraphics[angle=270,width=0.275\textwidth]{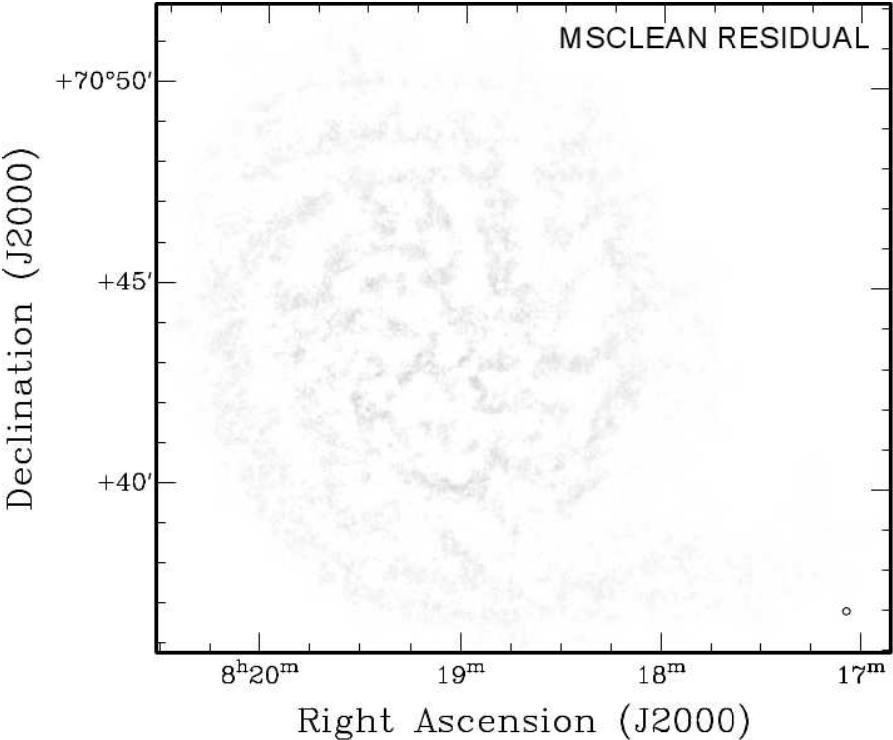}\\
  \end{tabular}
  \caption{\clean\ (left) and \msclean\ (right) integrated \HI\ maps
    (top row) for Holmberg~II. The maps were generated from the
    natural-weighted, masked, primary-beam corrected and for \clean,
    residual flux corrected cubes. The equivalent residual integrated \HI\
    maps are also shown (bottom row).  No primary-beam corrections and for
    \clean, no residual flux corrections have been applied. The gray-scale
    levels run from $0$ to $500$~\mjbeam~\kms. Beams are marked in bottom
    right corner of images.\label{fig:hol2-msclean-mom0}} % min -17
\end{figure}

\begin{figure}[htbp]
  \centering
  \begin{tabular}{cc}
    \includegraphics[angle=270,width=0.275\textwidth]{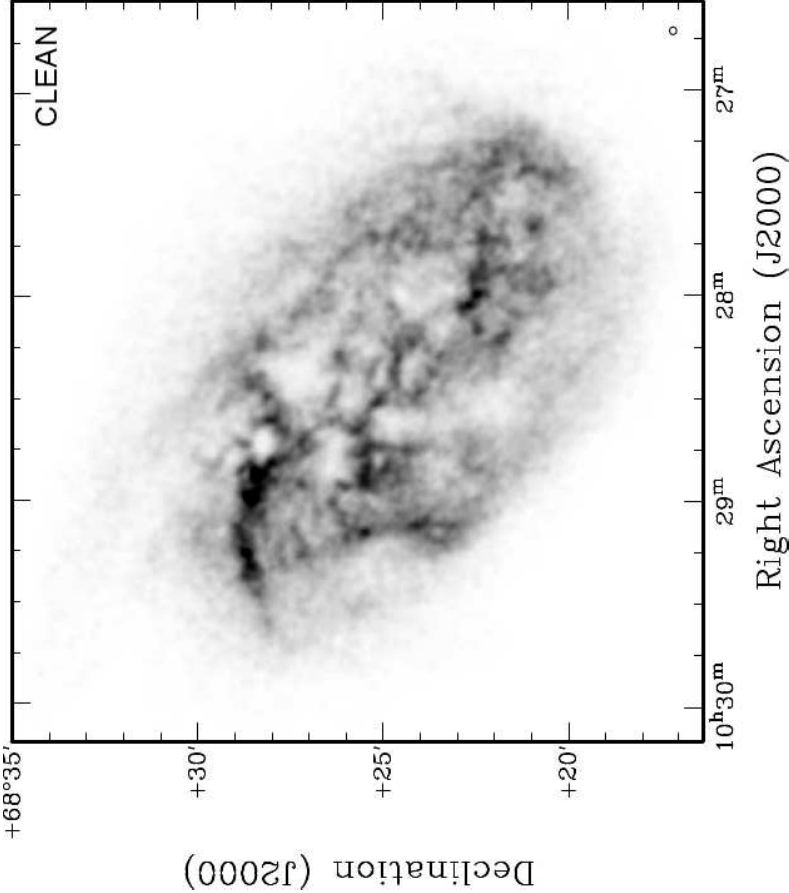}&
    \includegraphics[angle=270,width=0.275\textwidth]{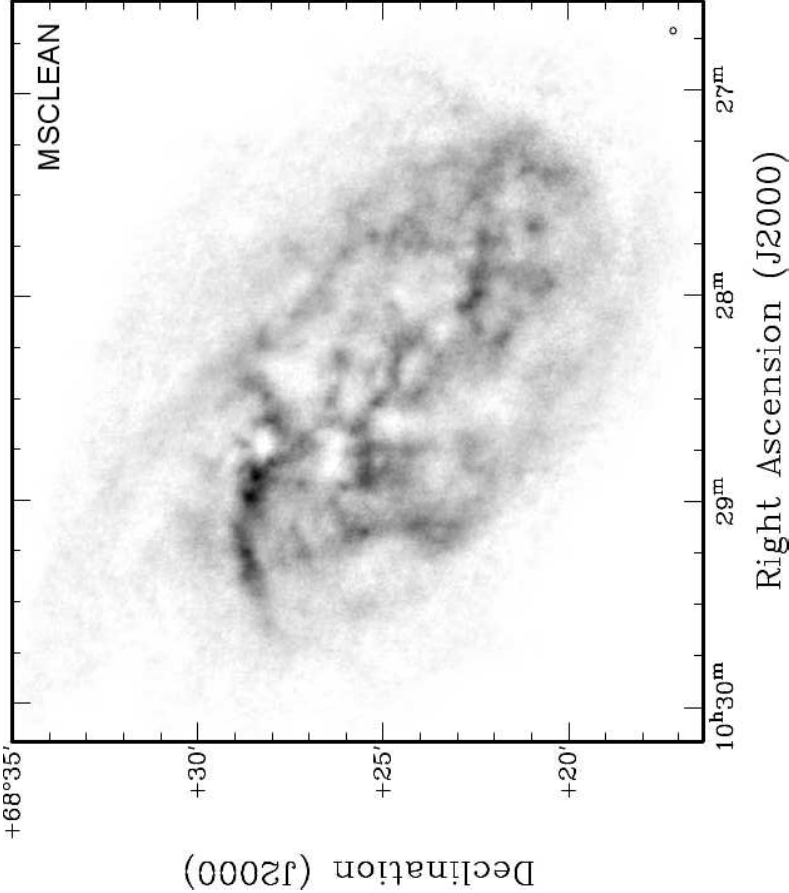}\\
    \includegraphics[angle=270,width=0.275\textwidth]{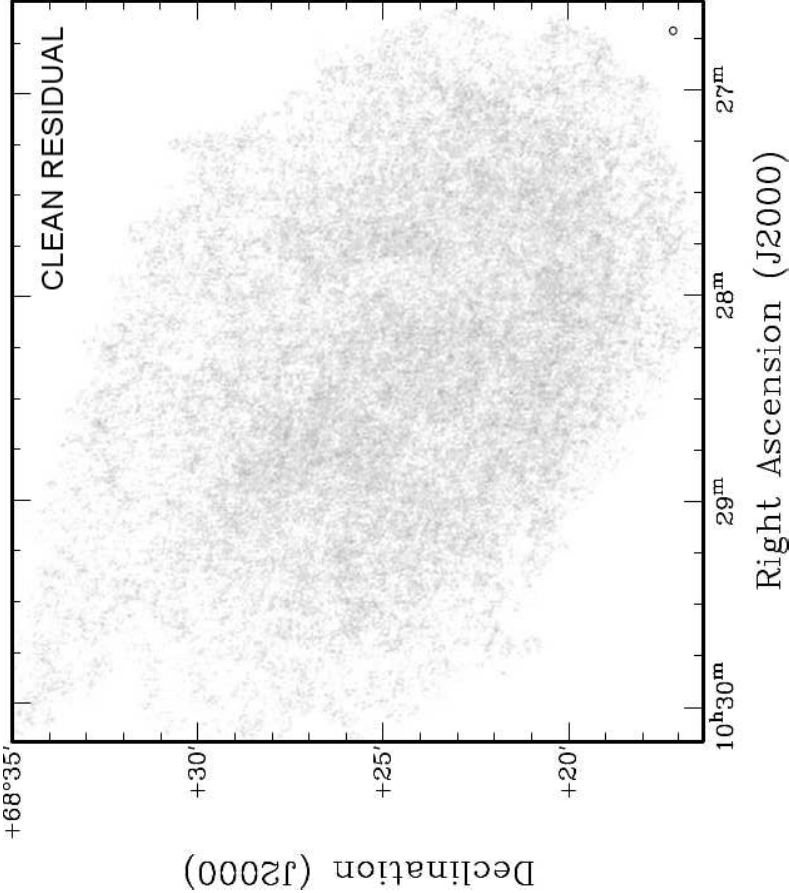}&
    \includegraphics[angle=270,width=0.275\textwidth]{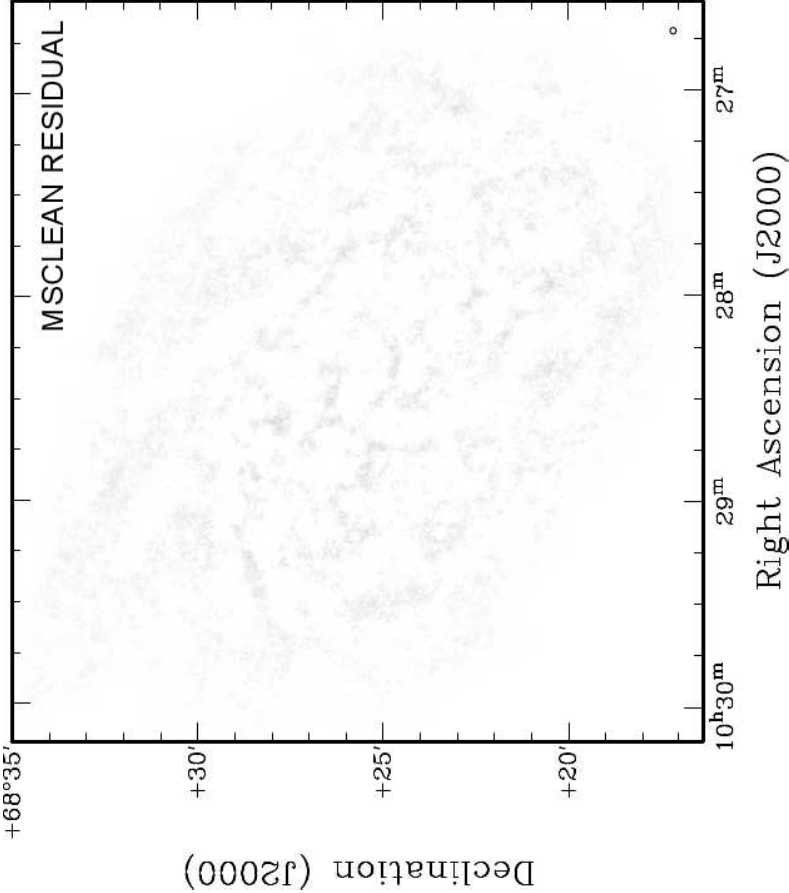}\\
  \end{tabular}
  \caption{\clean\ (left) and \msclean\ (right) integrated \HI\ maps
    (top row) for IC~2574. The maps were generated from the
    natural-weighted, masked, primary-beam corrected and for \clean,
    residual flux corrected cubes. The equivalent residual integrated \HI\
    maps are also shown (bottom row).  No primary-beam corrections and for
    \clean, no residual flux corrections have been applied. The gray-scale
    levels run from $0$ to $450$~\mjbeam~\kms. Beams are marked in bottom
    right corner of images.\label{fig:ic2574-msclean-mom0}} %min -21
\end{figure}

Integrated \HI\ maps, as well as integrated residual maps for each
galaxy are shown in Figures \ref{fig:ngc2403-msclean-mom0},
\ref{fig:hol2-msclean-mom0} and \ref{fig:ic2574-msclean-mom0} for the
\msclean\ and \clean\ natural-weighted, primary-beam corrected (and
for \clean, residual-scaled) data. The cubes were masked with the same
masks applied to the \clean\ data \citep[see][for a description of the
masking process]{THINGS}. The residual integrated maps were generated
from residual cubes, \ie\ cubes that were cleaned but did not have the
\clean\ components added to them. These integrated maps have not been
corrected for primary-beam attenuation and for the \clean\ data, no
residual scaling has been applied.

The masking applied to the integrated maps hides the signature of the
clean bowl seen in the channel maps of the \clean\ data in
Figures~\ref{fig:ngc2403-chanmaps} to \ref{fig:ic2574-chanmaps}. The
residual integrated \clean\ maps in
Figures~\ref{fig:ngc2403-msclean-mom0} to
\ref{fig:ic2574-msclean-mom0} do show a significant pedestal of
uncleaned flux, while the \msclean\ residual maps have no such
feature. There is trace source emission in the \msclean\ residual
integrated maps, but generally the residuals are much more
`noise-like'. Despite being on the same flux scale, there is a
definite visual difference between the \clean\ and \msclean\ data,
most clearly seen where there is significant source flux (the darker
regions) in Holmberg~II and IC~2574.  Conversely the low-level
extended structure is more clearly seen in the \msclean\ integrated
maps and extends out to the mask boundary.  The peak flux for compact
features is therefore higher in the \clean\ integrated images, while
the total flux of the underlying, extended structure is greater in
\msclean.

% \placefigure{fig:ngc2403-msclean-mom0}
% \placefigure{fig:hol2-msclean-mom0}
% \placefigure{fig:ic2574-msclean-mom0}

\begin{figure}[htbp]
  \centering
  \begin{tabular}{c}
    \includegraphics[angle=270,width=0.35\textwidth]{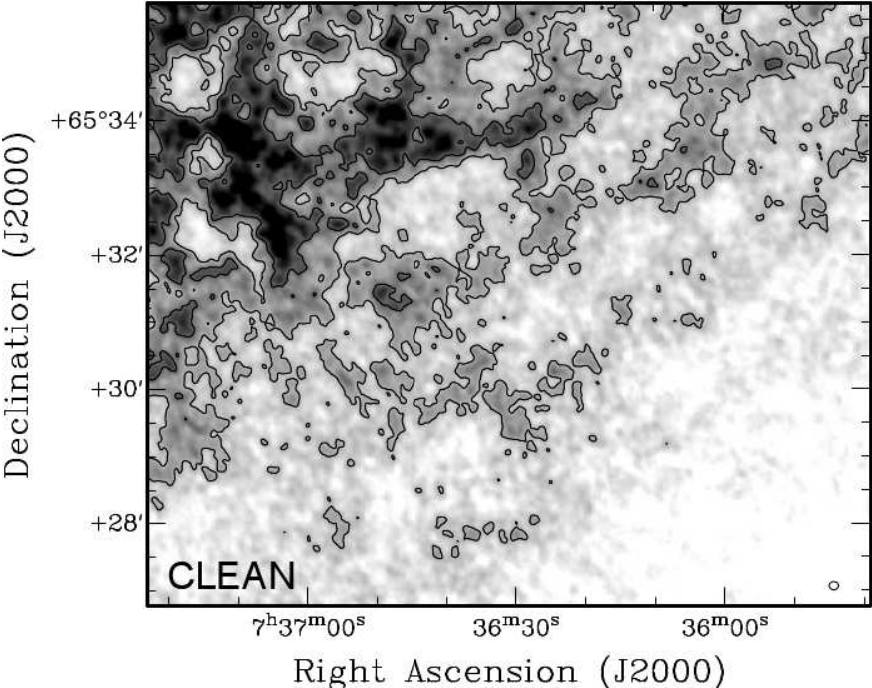} \\
    \includegraphics[angle=270,width=0.35\textwidth]{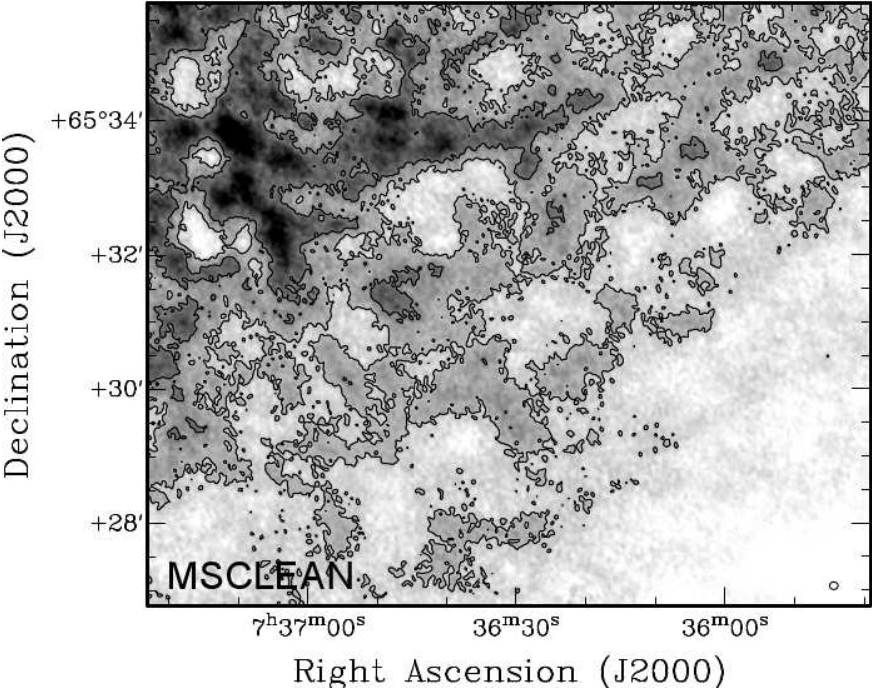} \\
  \end{tabular}
  \caption{Comparison of flux levels for NGC~2403.  On the top is the
    \clean\ integrated moment map and the bottom is the \msclean\ version.
    Images were generated from the natural-weighted, masked and
    primary-beam corrected data.  The \clean\ image has been residual flux
    corrected.  Contour levels of column density are plotted at
    $1\cdot10^{21}$ and $2\cdot10^{21}$~\coldensity.  The gray-scale
    levels run from $0$ to $200$~\mjbeam~\kms. Beams are marked in bottom
    right corner of images.\label{fig:ngc2403-fluxcomp}}
\end{figure}

\begin{figure}[htbp]
  \centering
  \begin{tabular}{c}
    \includegraphics[angle=270,width=0.3\textwidth]{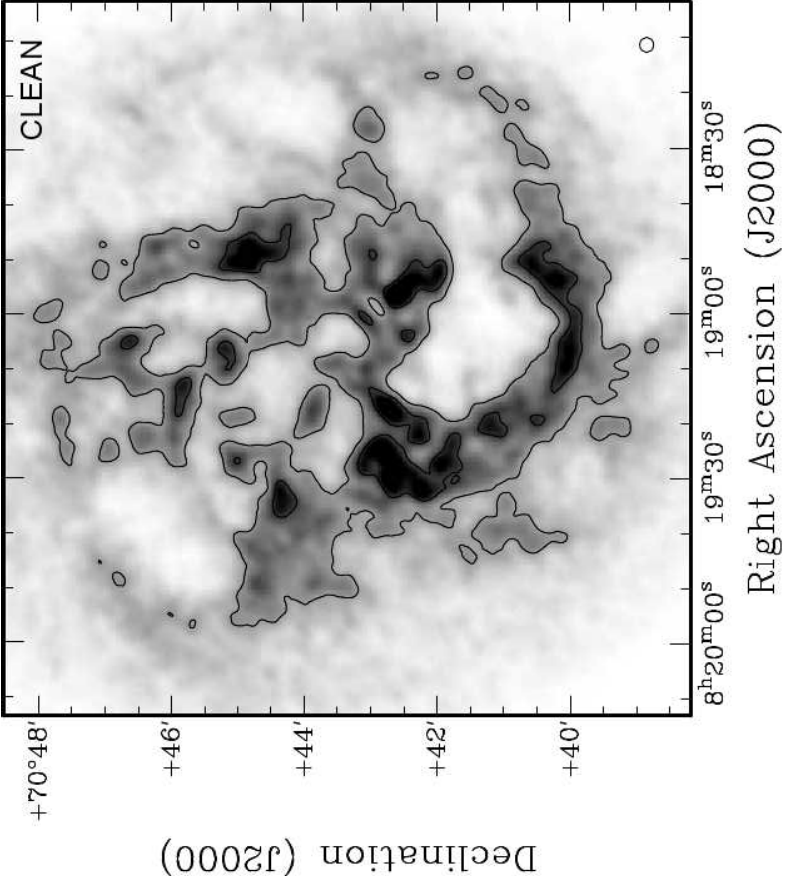}\\
    \includegraphics[angle=270,width=0.3\textwidth]{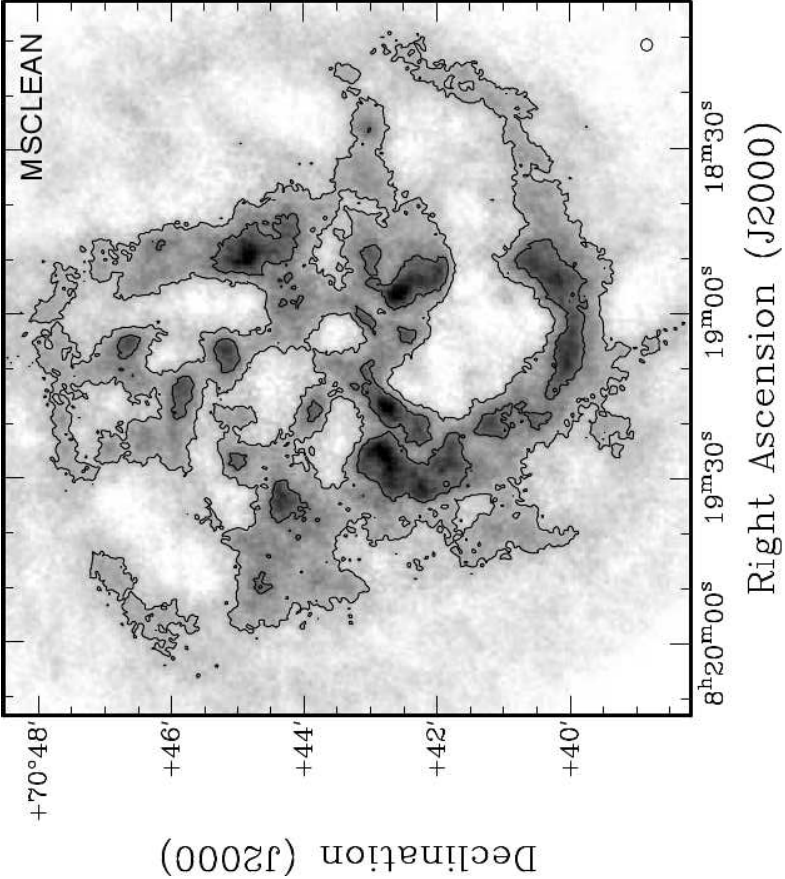}\\
  \end{tabular}
  \caption{Comparison of flux levels for Holmberg~II.  On the top is
    the \clean\ integrated moment map and the bottom is the \msclean\
    version.  Images were generated from the natural-weighted, masked and
    primary-beam corrected data.  The \clean\ image has been residual flux
    corrected.  Contour levels of column density are plotted at
    $1\cdot10^{21}$ and $2\cdot10^{21}$~\coldensity.  The gray-scale
    levels run from $0$ to $500$~\mjbeam~\kms. Beams are marked in bottom
    right corner of images.\label{fig:hol2-fluxcomp}}
\end{figure}

\begin{figure}[htbp]
  \centering
  \begin{tabular}{c}
    \includegraphics[angle=270,width=0.3\textwidth]{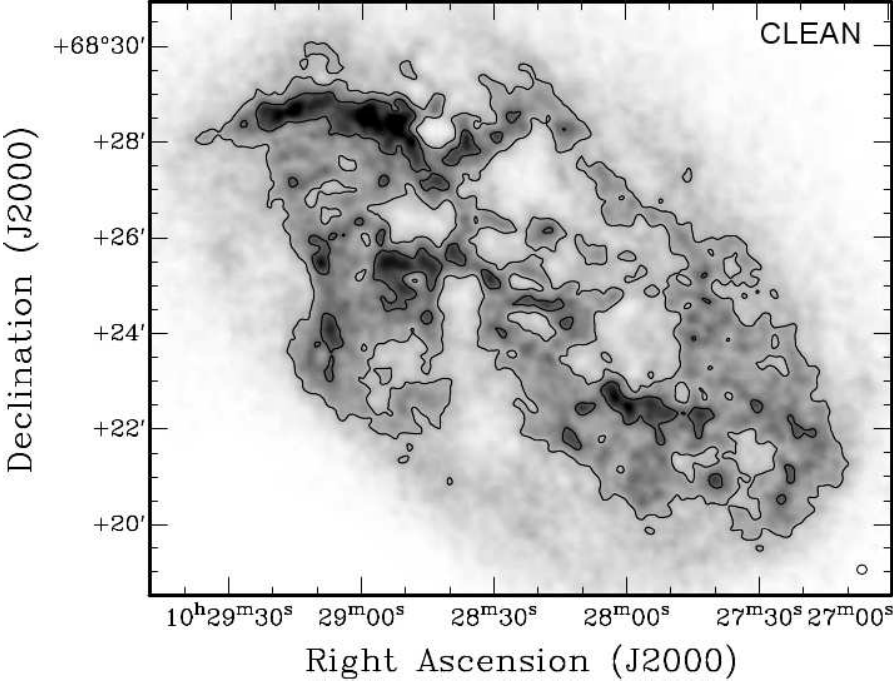}\\
    \includegraphics[angle=270,width=0.3\textwidth]{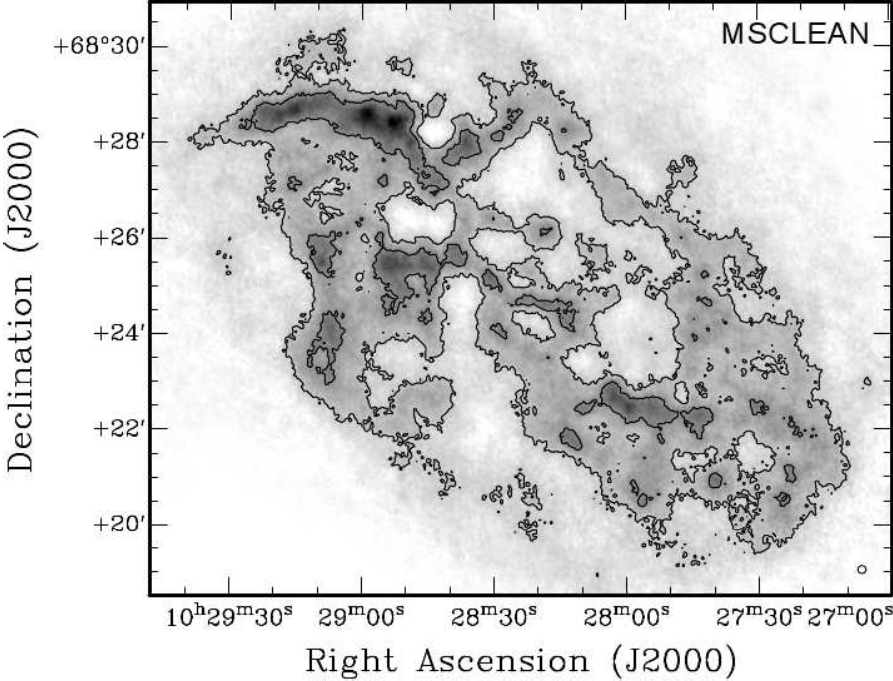}\\
  \end{tabular}
  \caption{Comparison of flux levels for IC~2574.  On the top is the
    \clean\ integrated moment map and the bottom is the \msclean\ version.
    Images were generated from the natural-weighted, masked and
    primary-beam corrected data.  The \clean\ image has been residual flux
    corrected.  Contour levels of column density are plotted at
    $1\cdot10^{21}$ and $2\cdot10^{21}$~\coldensity.  The gray-scale
    levels run from $0$ to $450$~\mjbeam~\kms. Beams are marked in bottom
    right corner of images.\label{fig:ic2574-fluxcomp}}
\end{figure}

To compare the flux scales between the data-sets, contour lines of
column density $1\cdot10^{21}$ and $2\cdot10^{21}$ cm$^{-2}$ have been
plotted on the (residual scaled) \clean\ and \msclean\ data for each
galaxy, shown in Figures \ref{fig:ngc2403-fluxcomp},
\ref{fig:hol2-fluxcomp} and \ref{fig:ic2574-fluxcomp}. Again, this
data has been masked and corrected for primary-beam attenuation.  The
location of the contours match closely across the \clean\ and
\msclean\ data, but they appear much smoother in the \clean\ data. The
contours in the \msclean\ images for each galaxy appears to trace a
much finer structure boundary. This is likely due to the pedestal of
leftover flux in classical \clean. The pedestal still has the dirty
beam as its PSF, the more extended wings of this beam will wash out
structure more severely than a Gaussian beam, and the low-level,
small-scale structure will be lost in the image. For low-level column
densities the \clean\ resolution is thus worse than one would expect
on the basis of the clean beam size, as we will show later. In
\msclean\ there is no pedestal, and all fine-scale structure is imaged
at the full resolution of the clean beam, enhancing the detailed
structures in the disk.

% \placefigure{fig:ic2574-fluxcomp}
% \placefigure{fig:hol2-fluxcomp}
% \placefigure{fig:ngc2403-fluxcomp}

\clearpage
\subsection{Flux and Noise Measurements}
\label{sec:comp-fluxnoise}

Table~\ref{tab:total-flux} shows the comparison of total flux for both
weightings for the \clean\ and \msclean\ data. The total flux for the
\msclean\ data was calculated from the masked cubes after primary-beam
correction, using the \textsc{stat} task of the
GIPSY\footnote{Groningen Image Processing System,
http://www.astro.rug.nl/$\sim$gipsy/} software package. The total flux
values for the \clean\ data are derived with the same method, using
the primary-beam corrected data with residual-scaling applied. The
uncertainties in the flux densities measured here is of the same order
as the flux calibration \citep[$\sim10$\%, see][]{THINGS}. From
Table~\ref{tab:total-flux}, in all three galaxies it can be seen that
\msclean\ recovers more flux than classical \clean\ for a given
weighting. This is despite apparently higher peak fluxes for compact
sources in the \clean\ integrated \HI\ maps of
Figures~\ref{fig:ngc2403-msclean-mom0} to
\ref{fig:ic2574-msclean-mom0}. The flux gains by \msclean\ are
therefore mostly in the recovery the extended source structure. The
relative differences in total fluxes as listed in
Table~\ref{tab:total-flux} are consistent with the results derived by
\cite{msclean} using artificial data.  Table~\ref{tab:rms-flux} shows
the \textit{rms} noise values for both weightings. The noise was
determined in line-free channels that were not used in determining the
continuum.

\begin{deluxetable}{lcccc}
  \tablecaption{Total flux comparison of \clean\ and \msclean\ for
    each galaxy.\label{tab:total-flux}}
  \tablewidth{0pt}
  \tablecolumns{5}
  \tablehead{
    \colhead{\multirow{2}{*}{Source}}&
    \colhead{\multirow{2}{*}{Weighting}}&
    \multicolumn{3}{c}{Total Flux (\jkms)}\\
    \colhead{} & \colhead{} &%
     \colhead{NGC~2403} & \colhead{Holmberg~II} & \colhead{IC~2574}}
  \startdata
  \multirow{2}{*}{\clean}%
            & Natural & $1055$ & $219$ & $387$\\
            & Robust  & $977$  & $210$ & $363$\\
  \multirow{2}{*}{\msclean}%
            & Natural & $1200$ & $261$ & $418$\\
            & Robust  & $1205$ & $271$ & $426$\\
  Single-dish & & $1172 \pm 553$ & $245 \pm 54$ & $466 \pm 69$ \\
  \enddata
  \tablecomments{The measurements for \clean\ were made on data which had
    been residual flux corrected.}
\end{deluxetable}

\begin{deluxetable}{lcccc}
  \tablecaption{Noise level Comparison for \clean\ (without
    residual-scaling) and \msclean\
    images.\label{tab:rms-flux}}
  \tablewidth{0pt}
  \tablecolumns{5}
  \tablehead{
    \colhead{\multirow{2}{*}{Source}}&
    \colhead{\multirow{2}{*}{Weighting}}&
    \multicolumn{3}{c}{Noise (\mjbeam)}\\
    \colhead{} & \colhead{} &%
     \colhead{NGC~2403} & \colhead{Holmberg~II} & \colhead{IC~2574}}
    \startdata
    \multirow{2}{*}{\clean}% 
             & Natural & $0.38$ & $0.92$ & $0.56$ \\
             & Robust  & $0.45$ & $1.06$ & $0.69$ \\
    \multirow{2}{*}{\msclean}%
             & Natural & $0.41$ & $0.93$ & $0.57$ \\
             & Robust  & $0.50$ & $1.11$ & $0.68$ \\
    \enddata
    \tablecomments{The measurements for \clean\ were made on data
      which had not been residual flux corrected and all measurements were
      made in line-free channels that were not used in determining the
      continuum.}
\end{deluxetable}
% \placetable{tab:total-flux}
% \placetable{tab:rms-flux}

\begin{figure}[htbp]
  \centering
  \begin{tabular}{cc}
    \includegraphics[width=0.4\textwidth]{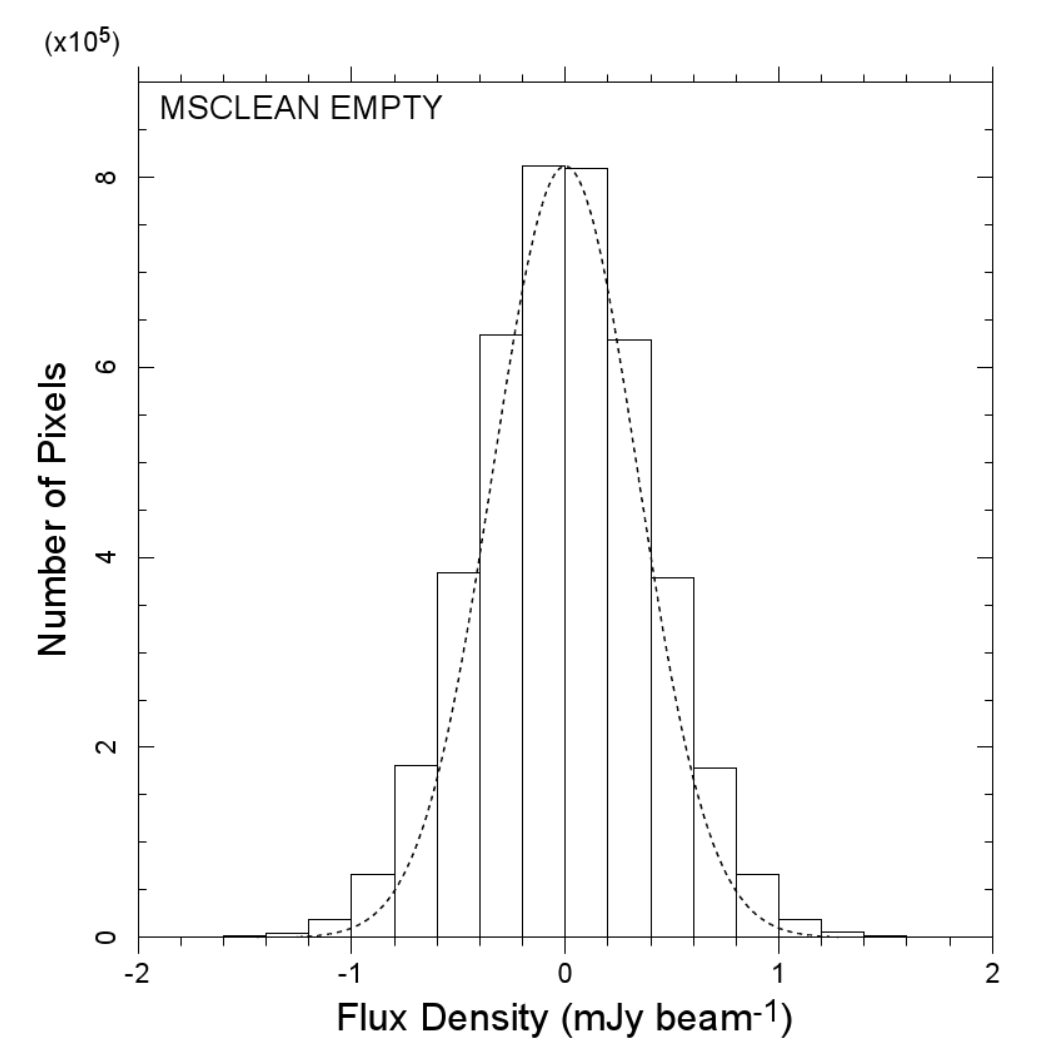}&
    \includegraphics[width=0.4\textwidth]{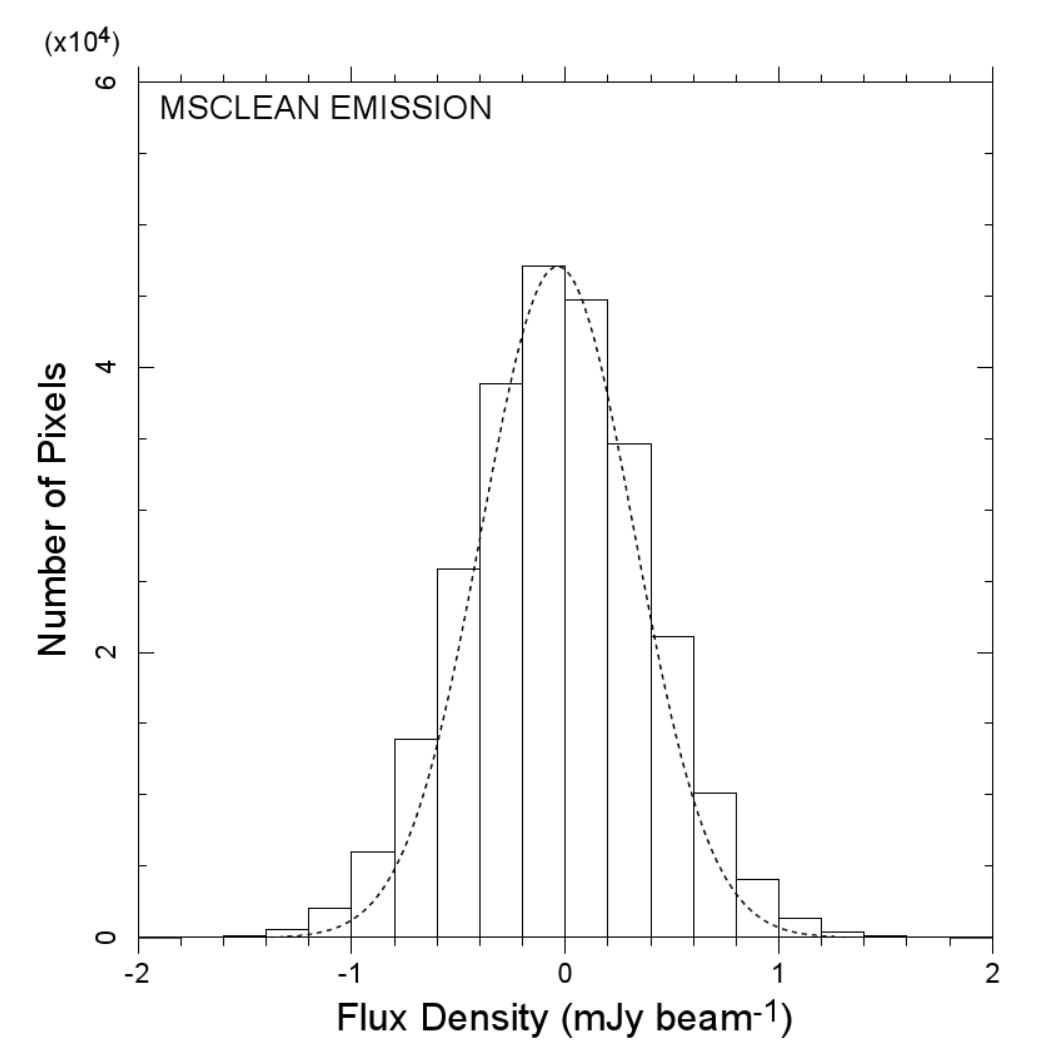}\\
    \includegraphics[width=0.4\textwidth]{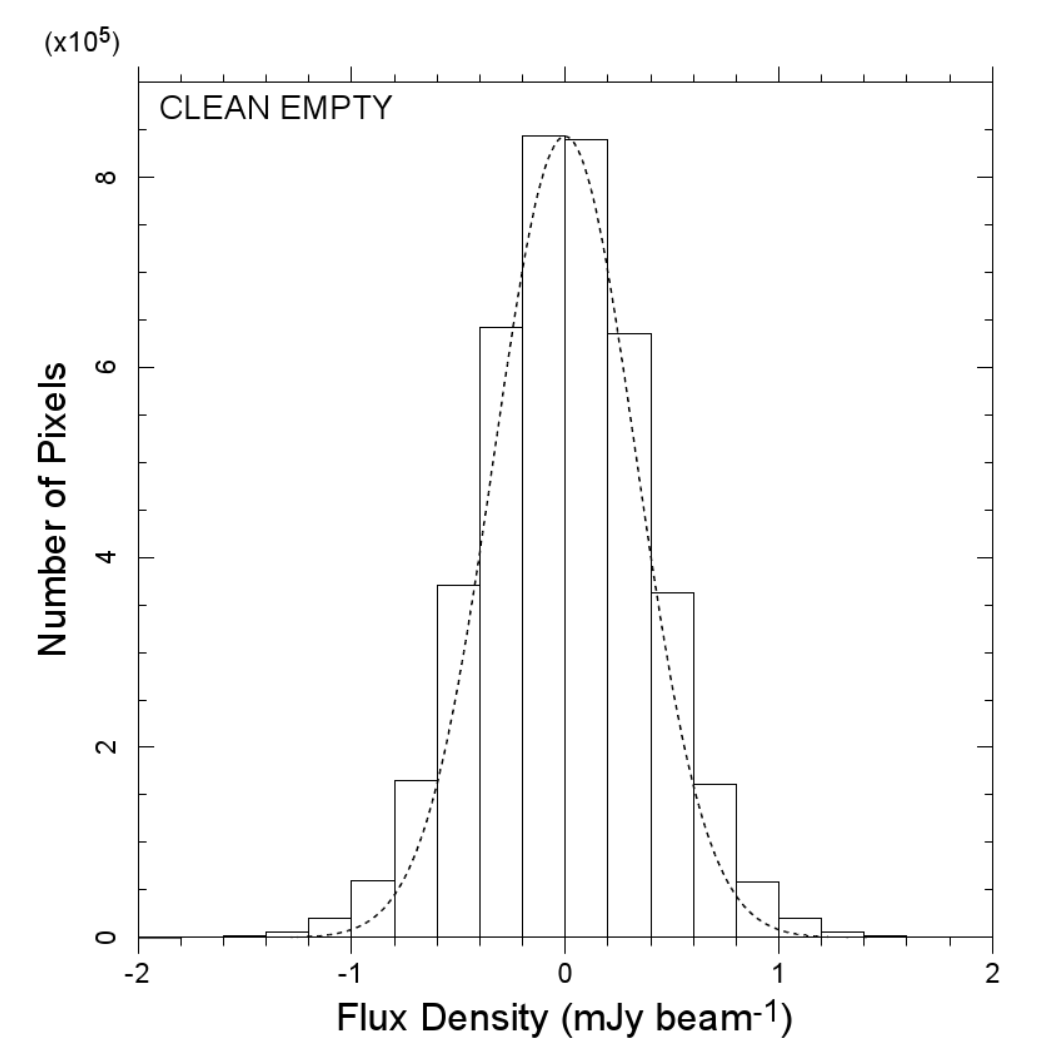}&
    \includegraphics[width=0.4\textwidth]{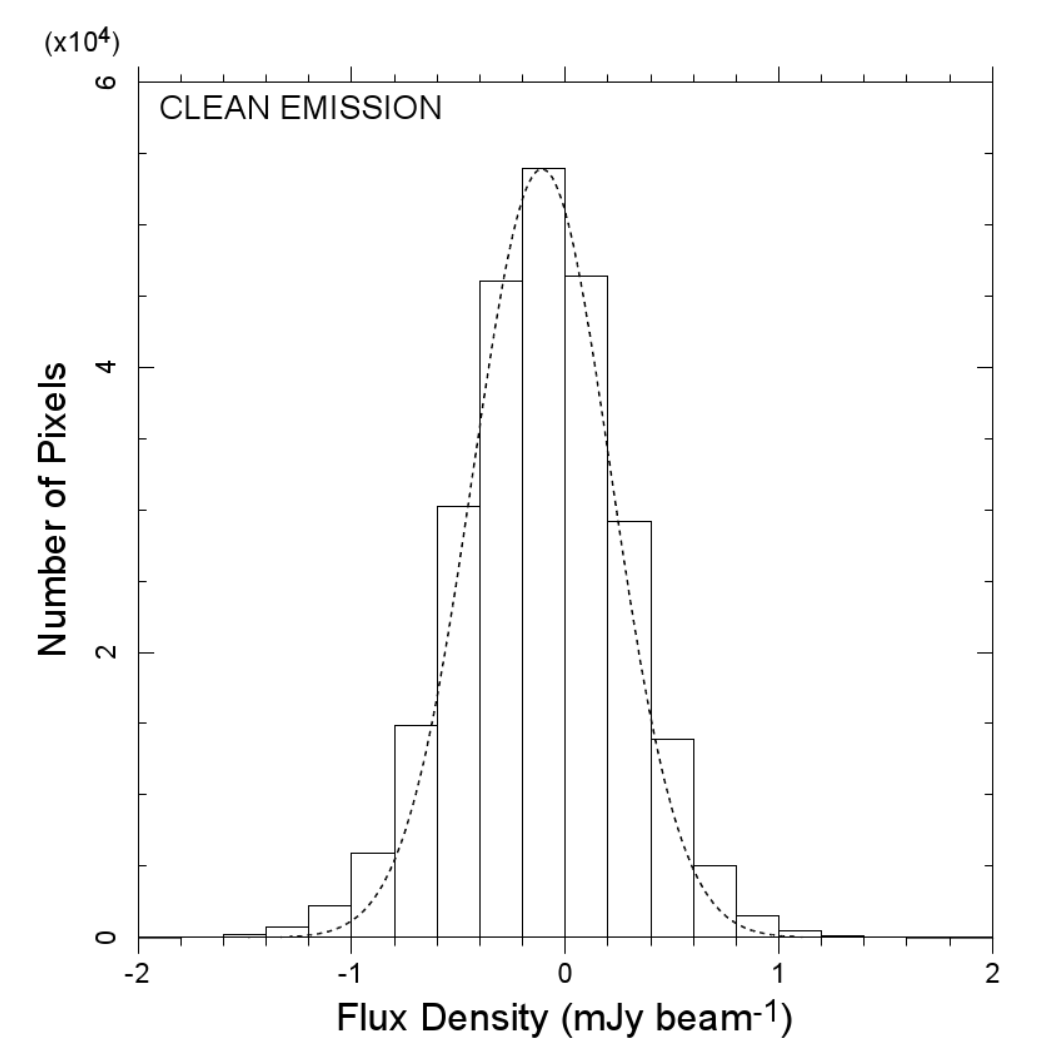}\\
  \end{tabular}
  \caption{Histograms of the flux density in the galaxy NGC~2403 for
    both \msclean\ (top) and \clean\ (bottom).  On the left are histograms
    calculated an empty channel in the unmasked, non primary-beam
    corrected (and non residual-scaled for \clean) data cube located at a
    heliocentric velocity of $278$~\kms.  On the right are histograms
    calculated in a $500\times500$ pixel box situated in an empty part of
    the channel containing galaxy emission at a heliocentric velocity of
    $201$~\kms. For each plot, $25$ bins of width $0.18$~\mjbeam\ have
    been used.  Also, for each histogram, the Gaussian fit to the
    histogram values has been plotted (dotted
    line).\label{fig:ngc2403-noise-histograms}}
\end{figure}

An important issue is whether the algorithms alter the amplitude or
distribution of noise in the data in a significant way. Figure
\ref{fig:ngc2403-noise-histograms} shows histograms of the flux values
in a channel that is free of galaxy emission and within an empty
region of a channel containing galaxy emission.  For \clean, the non
residual-scaled data has been used.  For the noise in both emission
and emission-free channels, and across both data-sets the noise has a
Gaussian distribution. The noise in the empty channels of the
\msclean\ and \clean\ data (left column) is identical.  For channels
containing emission there is a slight bias to negative values in the
\clean\ data, due to the clean bowl.  We also note that there is a
similar but smaller bias in the noise in the \msclean\ emission
channel. It is clear however, that neither algorithm alters the noise
significantly.

\subsubsection{Comparison to Single-dish Measurements}

Ideally one would like to compare the derived flux values with total
flux values derived using single-dish instruments.  Unfortunately for
the THINGS galaxies their large extent leads to single-dish
measurements which are more uncertain than one would like. We list in
Table~\ref{tab:total-flux} the single-dish fluxes derived by computing
the average values and standard deviations of the single dish fluxes
listed in the NASA/IPAC Extragalactic Database (NED).  In general we
see a good agreement, but a more precise comparison is precluded by
the uncertainty in the single dish values. We also note there is a
general satisfactory agreement between the fluxes derived here and
those listed in \cite{THINGS}.  Finally, we refer to Figure~4 in
\cite{wb_ic2574} which plots the VLA \HI\ (residual-corrected)
spectrum for IC~2574 on top of a single-dish spectrum by
\cite{rots_singledish} which shows excellent agreement. 

\subsection{\texorpdfstring{\HIfat}{HI} Spectra}
\label{sec:comp-hispectra}

\begin{figure}[htbp]
  \centering
  \includegraphics[width=0.5\textwidth]{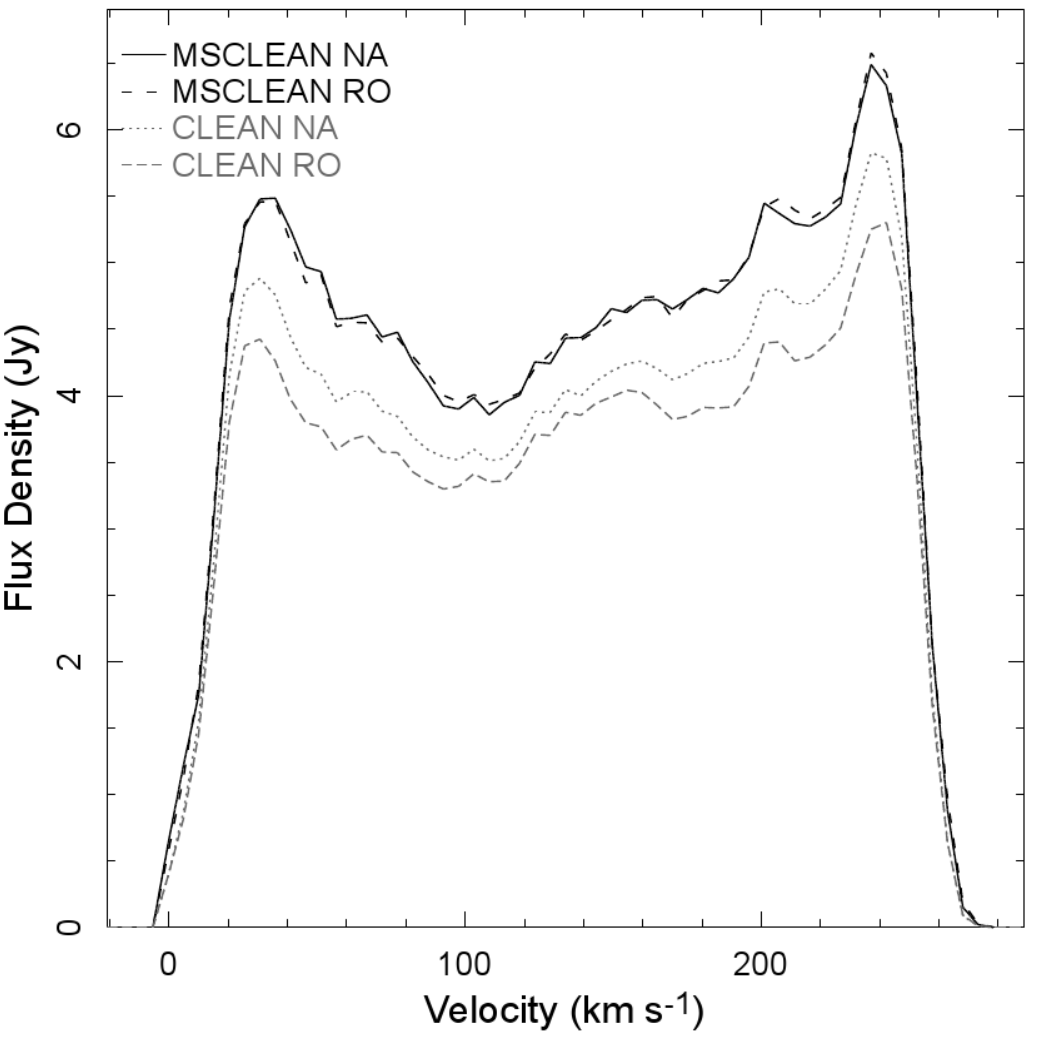}%
  \caption{Global \HI\ Profile of NGC~2403 for both natural and
    robust \clean\ and \msclean\ cubes.  The profiles are derived
    from masked, primary-beam corrected (and for \clean, residual
    scaled) data.\label{fig:ngc2403-globalflux}}
\end{figure}

\begin{figure}[htbp]
  \centering
  \includegraphics[width=0.5\textwidth]{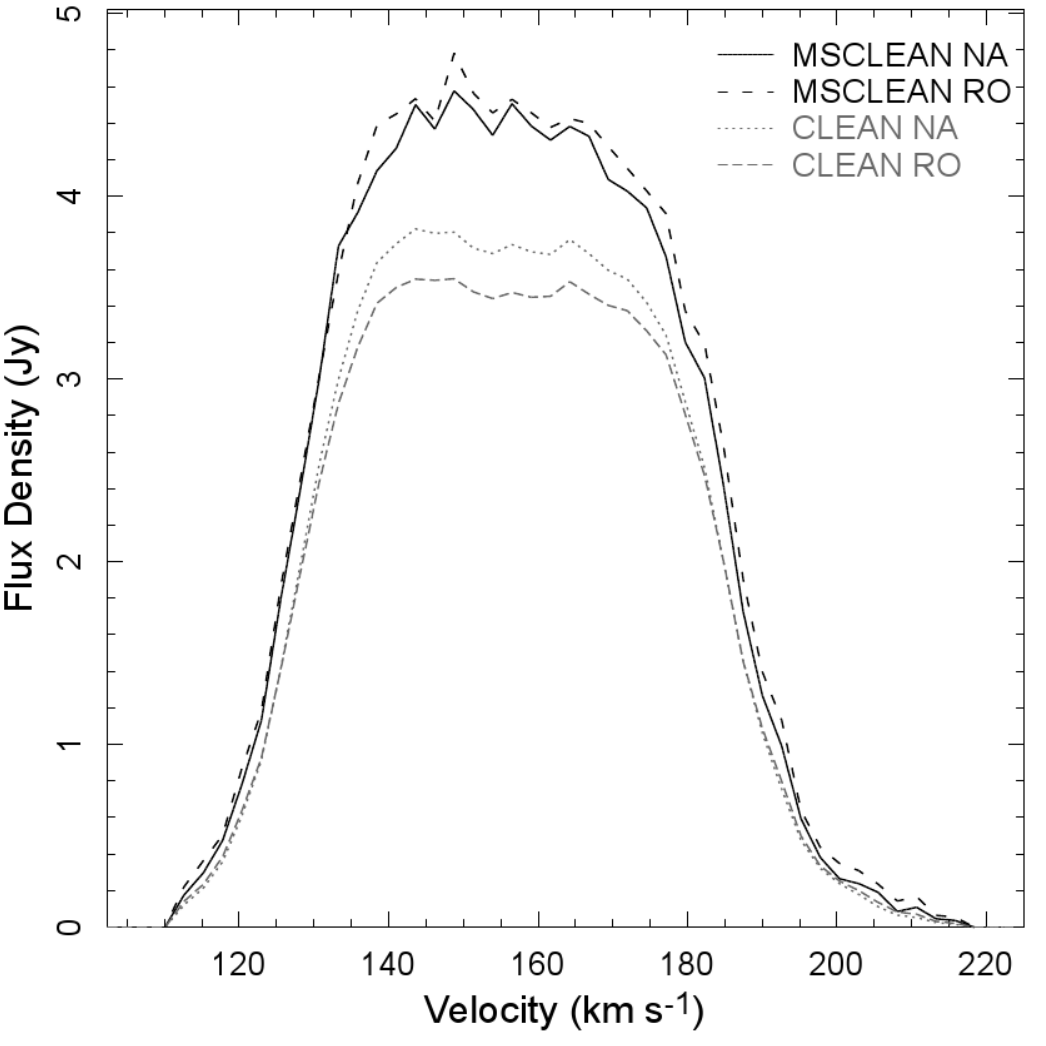}%
  \caption{Global \HI\ Profile of Holmberg~II for both natural and
    robust \clean\ and \msclean\ cubes.   The profiles are derived
    from masked, primary-beam corrected (and for \clean, residual
    scaled) data.\label{fig:hol2-globalflux}}
\end{figure}

\begin{figure}[htbp]
  \centering
  \includegraphics[width=0.5\textwidth]{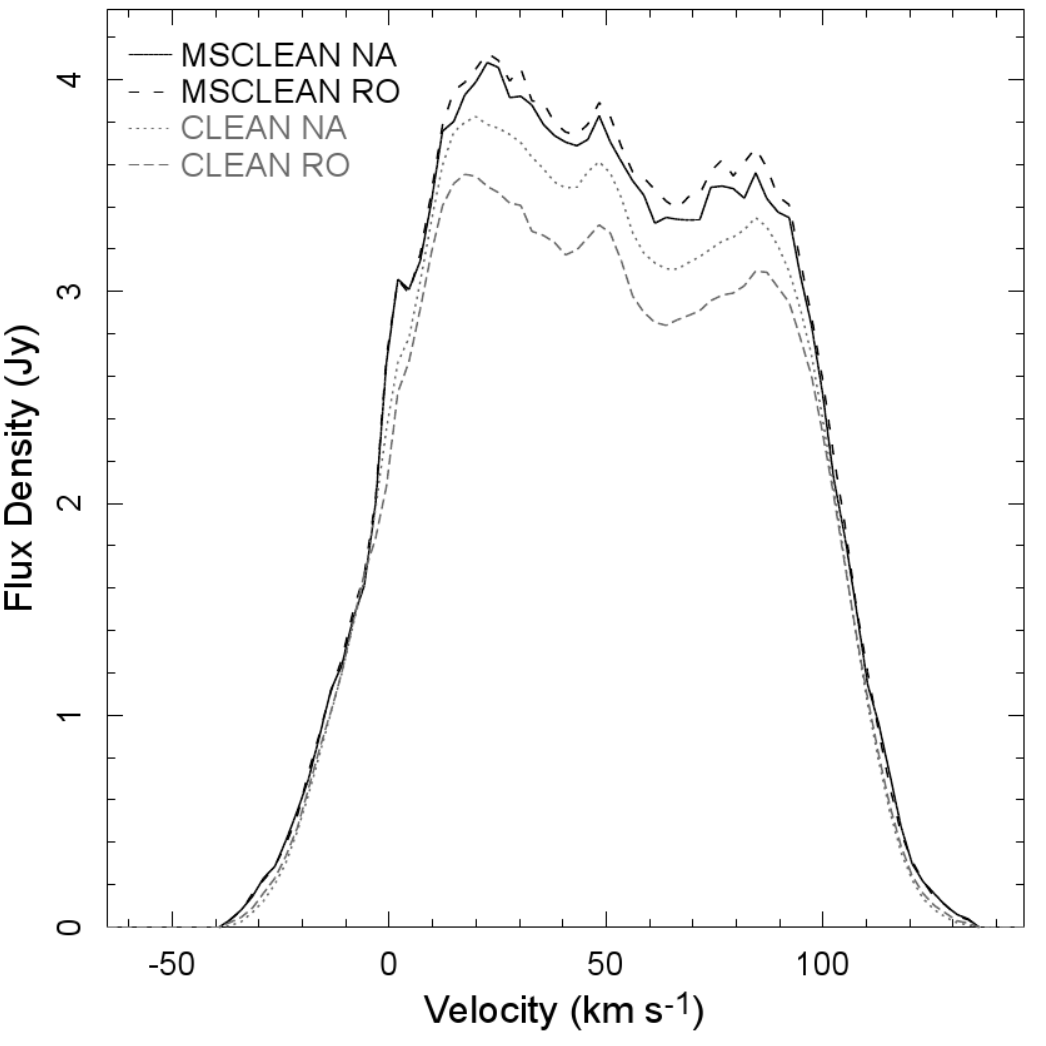}%
  \caption{Global \HI\ Profile of IC~2574 for both natural and
    robust \clean\ and \msclean\ cubes.   The profiles are derived
    from the masked, primary-beam corrected (and for \clean, residual
    scaled) data.\label{fig:ic2574-globalflux}}
\end{figure}

\begin{figure}[htbp]
  \centering
  \includegraphics[width=0.5\textwidth]{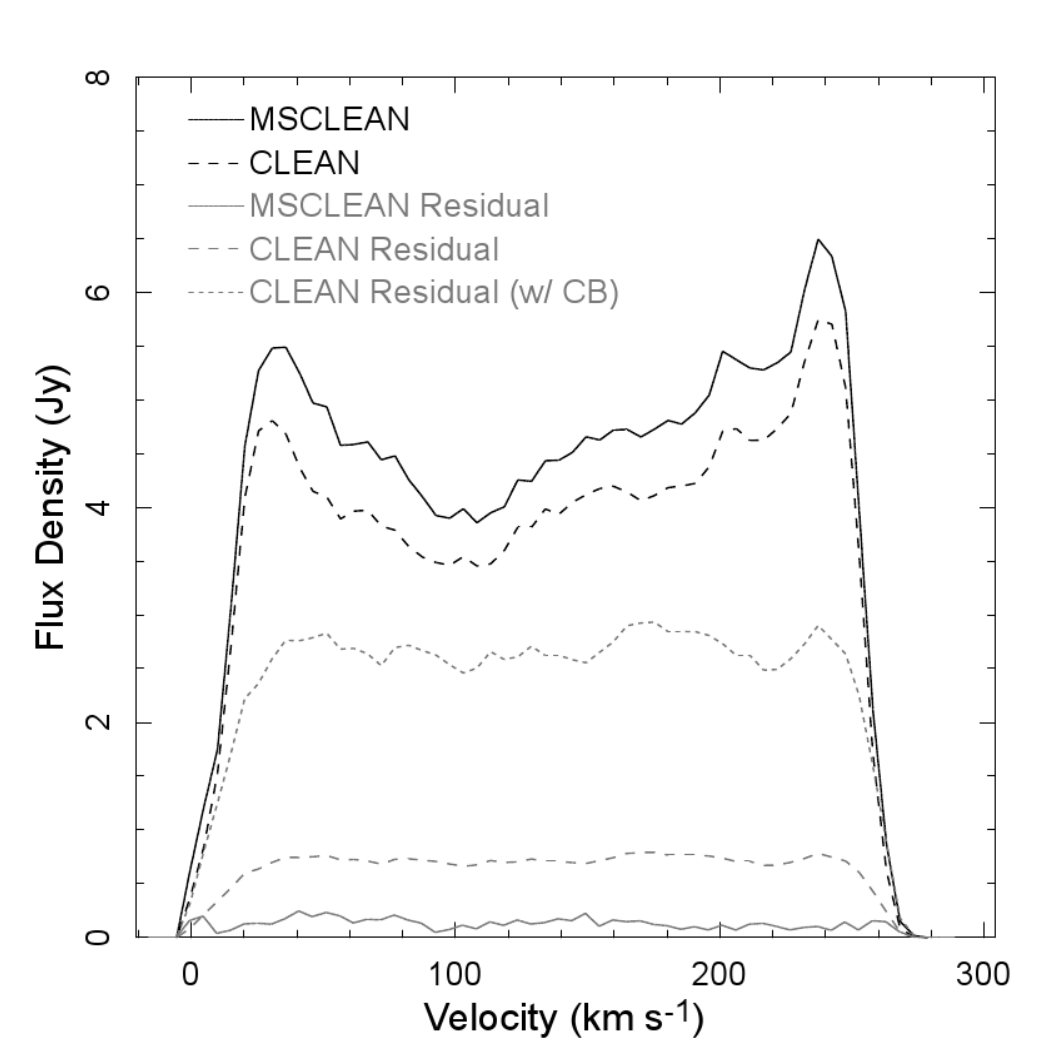}%
  \caption{Total (black) and residual (gray) \HI\ flux profiles for
    the natural-weighted \msclean\ (solid) and \clean\ (long dashed) data
    cubes of NGC~2403.  Also shown is the \clean\ residual flux
    profile calculated using the \clean\ beam (\ie\ without residual
    flux scaling) as the short-dashed gray line. Total flux profiles are
    derived from the masked, primary-beam corrected (residual scaled
    for \clean) data.  Residual flux profiles are derived from masked, non
     primary-beam corrected data.\label{fig:globalhiprof-comp}}
\end{figure}

Figures \ref{fig:ngc2403-globalflux}, \ref{fig:hol2-globalflux} and
\ref{fig:ic2574-globalflux} show the flux density plotted as a
function of velocity for NGC~2403, Holmberg~II and IC~2574
respectively. The \msclean\ natural (solid black line) and robust
(dashed black line) result in higher flux densities than the standard
\clean\ natural (dotted gray line) and robust (dashed gray line)
cubes. Additionally, \msclean\ natural and robust weighting are
equally good at recovering flux, whereas robust weighting is worse
than natural weighting for \clean.  We also compare the residual flux
density profiles of the two algorithms in
Figure~\ref{fig:globalhiprof-comp} for NGC~2403. For the \clean\
algorithm, the residual profile has been corrected using both the
dirty beam (\ie\ residual flux scaling) and with the \clean\ beam
(without residual flux scaling).  Because \msclean\ cleans close to
the noise level, there is essentially little source emission remaining
in its residual profile in Figure~\ref{fig:globalhiprof-comp} (gray
solid line).  On the other hand, the classical \clean\ residual
profile (long-dashed gray line) is a constant, flat line at a flux
density of approximately $0.5$~Jy, showing that the algorithm has left
source emission uncleaned (due to the flux cut-off used in the \clean\
process).  Correcting this residual profile using the \clean\ beam
leads to a significantly higher residual profile (gray short dashed
line) that will result in erroneous flux measurements, as opposed
to measurements using the same residual profile but corrected using
the dirty beam (gray long dashed line).

% \placefigure{fig:ngc2403-globalflux}
% \placefigure{fig:hol2-globalflux}
% \placefigure{fig:ic2574-globalflux}
% \placefigure{fig:globalhiprof-comp}

\begin{figure}[htbp]
  \centering
  \includegraphics[width=0.5\textwidth]{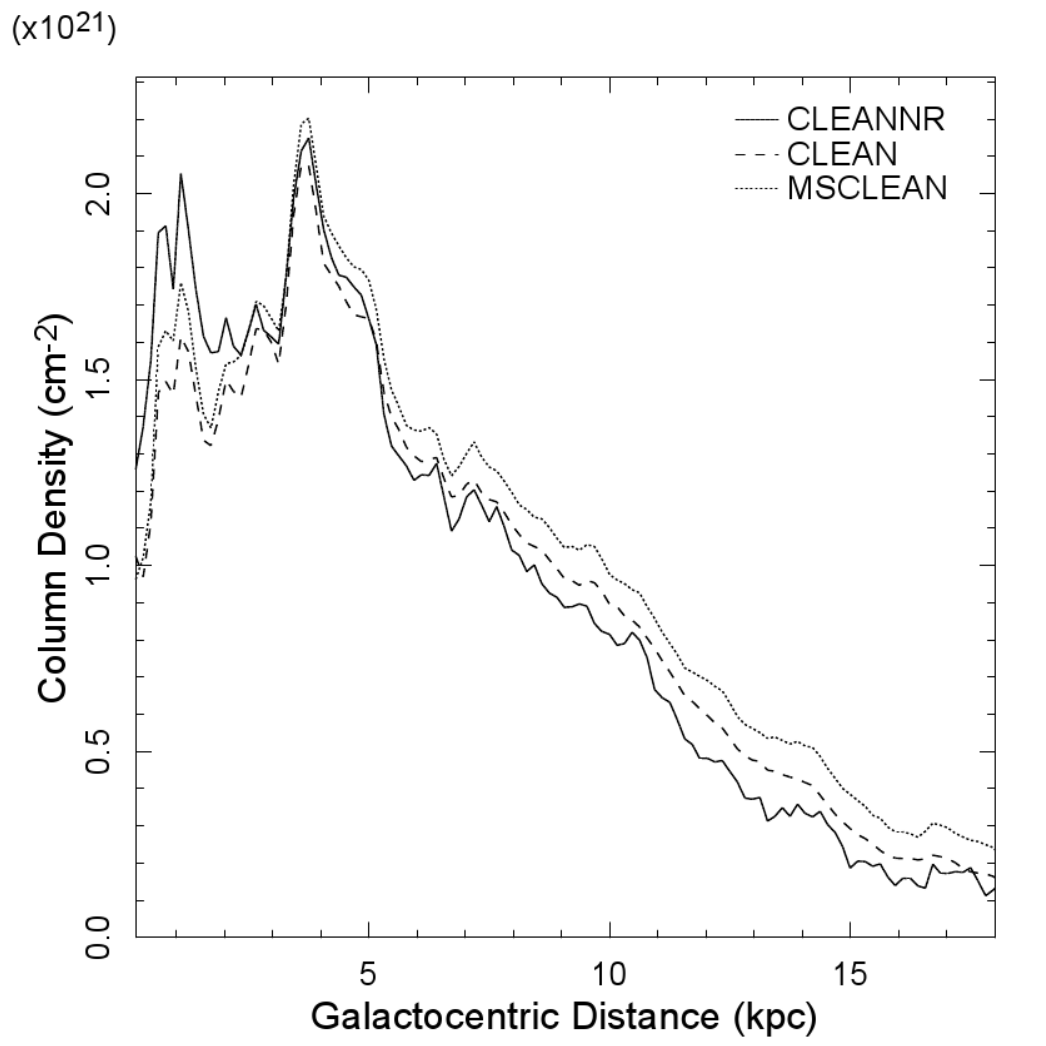}%
  \caption{The radial \HI\ column density profile of the galaxy
    NGC~2403 for \clean\ without residual scaling (solid line), \clean\
    with residual scaling (dashed line) and \msclean\ (dotted line). The
    masked, primary-beam corrected data was used for the generation of the
    profiles.\label{fig:ngc2403-radhiprof}}
\end{figure}

We also present a radial \HI\ column density profile in
Figure~\ref{fig:ngc2403-radhiprof} for NGC~2403.  The \HI\ column
density profile has been plotted for the \clean\ data with and without
residual flux scaling applied and for the \msclean\ data.  All
profiles were generated using the \textsc{ellint} task in the MIRIAD
software package, in $10$\arcsec\ increments, correcting for the
position angle and inclination of the \HI\ disk. As can be seen, not
applying residual flux scaling can lead to an overestimate of the \HI\
flux (and thus column density) in the inner parts of the disk.  As the
disk radius increases, the non residual-scaled \clean\ profile begins
to equal the \msclean\ and \clean\ residual scaled profiles, then
drops below the latter two profiles out to the extremes of the disk.
There is therefore significant variations in the fluxes measured
without residual-scaling the data.  \clean\ with residual-scaling
applied agrees much more closely to \msclean\ than when not applying
residual scaling.  The figure also demonstrates how \msclean\ does not
suffer as much from the \clean\ bowl effect as classical \clean\ and
thus can probe the outer regions of the disk more effectively than
classical \clean.  In the inner parts of the disk, the \clean\ bowl is
less significant because overlapping emission located at the same
spatial position but different frequencies partly cancels its
effect. In the outer parts of the disk, there are fewer overlapping
flux regions, and so the bowl has a larger effect, depressing the flux
in the \clean\ profile.  \msclean\ on the other hand shows
consistently higher \HI\ column densities out to the extremes of the
disk.

% \placefigure{fig:ngc2403-radhiprof}

\subsection{Power Spectra}
\label{sec:comp-powerspec}

\begin{figure}[htbp]
  \centering
  \begin{tabular}{cc}
    \includegraphics[width=0.45\textwidth]{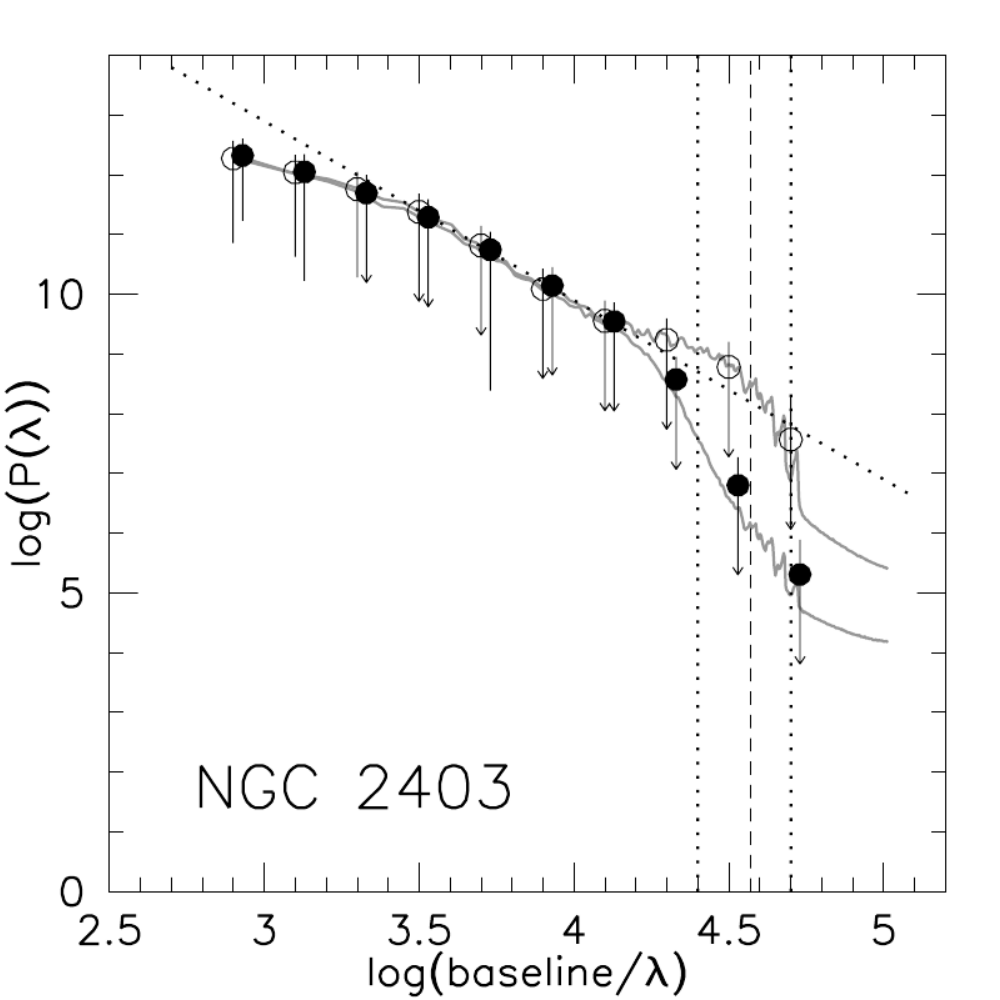}&
    \includegraphics[width=0.45\textwidth]{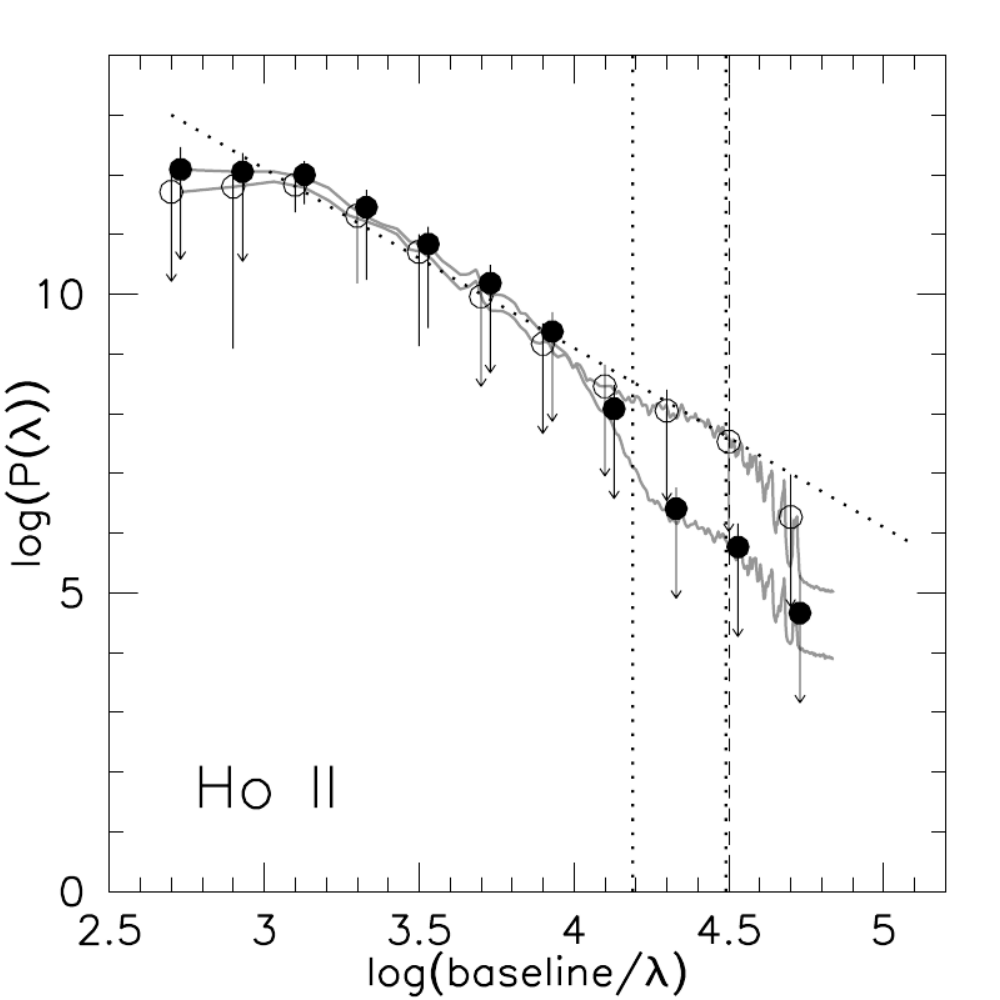}\\
    \multicolumn{2}{c}{
      \includegraphics[width=0.45\textwidth]{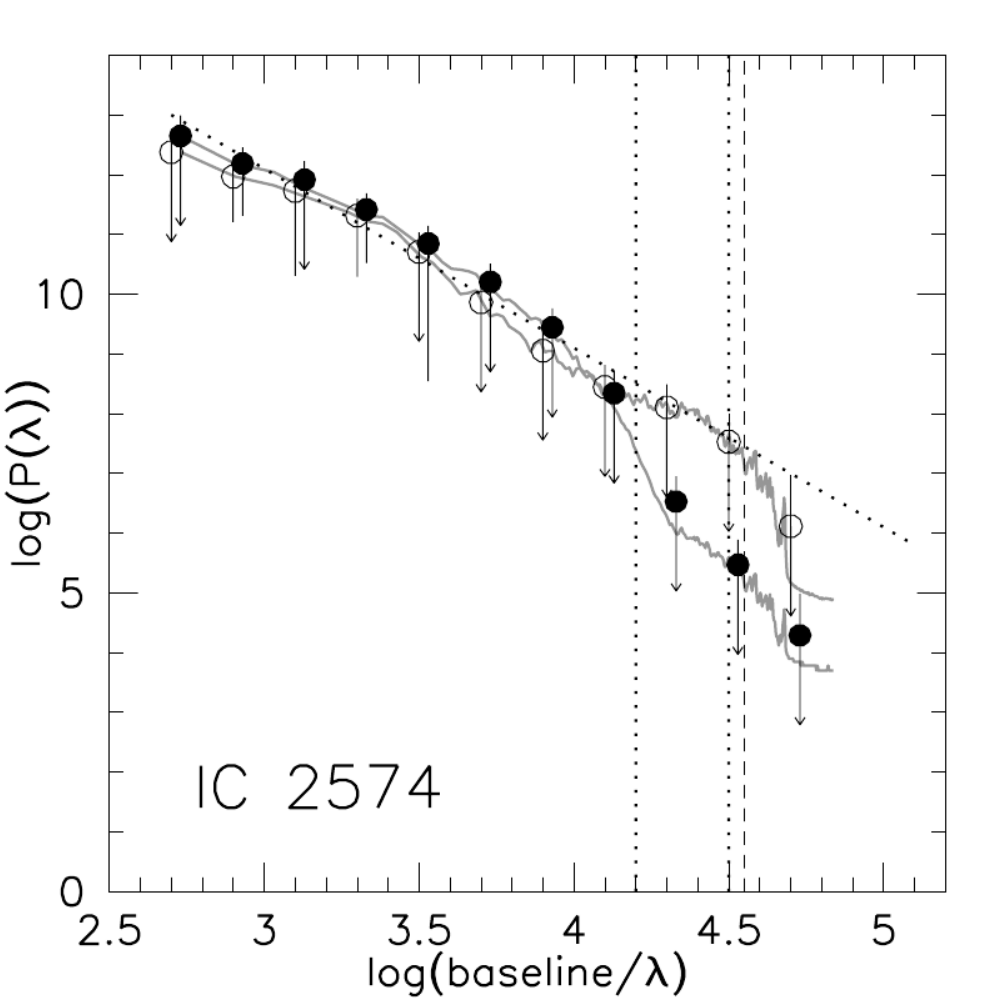}}\\
  \end{tabular}
  \caption{Power spectra of the natural-weighted integrated \HI\ 
    maps. Filled circles show the standard \clean\ data, open circles the
    \msclean\ data. The points show logarithmically binned data, the thin
    gray lines show the actual behaviour of the power in linearly sampled
    intervals. The diagonal dotted line has a power law slope of
    $-3$. Note that it is not a fit to the data. The left-most dotted
    vertical line shows the natural-weighted beam size, the right-most
    dotted line half the natural-weighted beam size. The vertical dashed
    line shows the robust-weighted beam size. Arrows indicate that the
    lower error-bars extend beyond the plotted
    range.\label{fig:powerspec}}
\end{figure}

The integrated moment maps in figures \ref{fig:ngc2403-msclean-mom0},
\ref{fig:hol2-msclean-mom0} and \ref{fig:ic2574-msclean-mom0} seem to
indicate that the \msclean\ maps contain more small-scale structure
than the \clean\ maps. The difference between the \msclean\ and
\clean\ maps can be quantified using the power spectrum of the
resulting \HI\ distributions in both types of maps. We derive the
power (defined as the square of the modulus of the Fourier transform
of the image) in azimuthally averaged rings with logarithmically
increasing baseline length. Figure~\ref{fig:powerspec} shows the power
spectra for our three sample galaxies for both sets of integrated \HI\
maps. Both the \clean\ and \msclean\ spectra flatten sightly at large
scales (\ie\ shortest baselines), possibly indicating that some of
these largest scales are not completely probed by the VLA, even at its
shortest baselines.  Alternatively it may indicate that these largest
scales are simply not present in the galaxies under consideration. It
is clear though that at large scales there is good agreement.

As the spatial scale probed approaches the size of the
natural-weighted beam however, the power in the \clean\ integrated
moment maps begins to decrease compared to the \msclean\ maps. At
about $\sim1.5$ times the natural-weighted beam size we start to see a
significant deviation. Also shown in Figure~\ref{fig:powerspec} is a
power law with slope $-3$ which is a reasonable description for the
power spectrum at intermediate scales. It is consistent with values
found in other galaxies \citep{msrs_stat}.

We will not attempt here to relate the power law slope to \eg\ the
turbulence or the energy input of the ISM, but we draw attention to
the fact that as the small-scale power in the \clean\ maps starts to
fall away, the \msclean\ power spectrum continues to follow this
power-law behavior. This indicates that the \msclean\ maps probe real
small-scale structure more efficiently than the classical \clean\
maps.  Note that the power in the \msclean\ maps only starts to drop
away at scale sizes of half a natural-weighted beam. Scales probed in
classical \clean\ are only completely sampled down to $\sim1.5$ times
the natural-weighted beam size.  Looking at
Table~\ref{tab:beam-size-comp} it can be seen that the robust beam
sizes for IC~2574 and Holmberg~II are about half the size of the
natural beams. This means that \msclean\ allows the probing of scales
at close to robust-weighted resolution but with natural-weighted
sensitivity, and at correct flux scales. This does not mean that
\msclean\ can super-resolve data, just that it is simply more
efficient at extracting the small-scale information that is present in
the data.

%%%%####################################################################

\section{Multi-Scale, Multi-Resolution and Windowed
  \texorpdfstring{\bclean}{CLEAN} Comparisons}
\label{sec:ms-mr-w-c}

So far we have only explicitly compared \msclean\ with classical
\clean. We now extend our comparisons to also include \clean\ windows
in combination with classical \clean, and include the version of
\msclean\ implemented in AIPS.  We refer to this algorithm as
\mrclean\ in keeping with the terminology used within the AIPS
documentation.  We note however, that it is not an
implementation of the original multi-resolution \clean\
\citep{ws_mrclean} and so is similar in name only.  The algorithm
behind \mrclean\ is the similar to the algorithm used by \msclean.
\msclean\ and \mrclean\ can therefore be thought of as differing
implementations of the same scale-sensitive deconvolution algorithm.

A small variation on standard \clean, which does not alter the
algorithm in any way, is to define a \clean\ `window', restricting the
\clean\ processing to a defined area of the dirty map.  The advantage
of this windowed-\clean\ is a greatly increased processing speed,
particularly for extended sources.  By defining a \clean\ window,
\clean\ is restricted to where it can find components.  However, care
must be taken not to define a window that is too small to encompass
low-level extended structure.  Obviously, for the case of extended
emission throughout a large fraction of the primary beam, the \clean\
window has to become so large that a windowed-\clean\ reverts to a
simple un-windowed \clean. 

For these comparisons between the algorithms, we have used a channel
map from Holmberg~II located at heliocentric velocity $165$~\kms. For
\msclean\ and \mrclean, six scales/resolutions were chosen, using the
same method as described in Section~\ref{subsec:mscleanscale}. The
diameters from smallest to largest were $0$\arcsec, $4$\arcsec,
$13$\arcsec, $40$\arcsec, $133$\arcsec\ and $400$\arcsec. A gain of
$0.7$ was used for \msclean\ and \mrclean\ (the same gain as used for
\msclean\ before) while a gain of $0.1$ was used
for the window \clean\ (in line with the gain used in classical
\clean).  For all algorithms, a flux threshold of $2.3$~\mjbeam\
($2.5\sigma$) was used in all algorithms. Maximum iteration limits of $5000$
were set for \msclean\ and \mrclean\ and $100000$ for the window
\clean.  We note that for all algorithms the flux threshold was
reached before hitting the iteration limit.

\subsection{Beams, Total Flux and Noise Measurements}
\label{sec:ms-mr-w-c-flux}

Values for the \clean\ beams, total flux recovered and the
\textit{rms} noise can be found in Table~\ref{tab:ms-mr-w-c-comp}.  To
measure the flux, a mask was created from the \msclean\ restored image
by blanking out all emission below a $2\sigma$ level and applying this
mask to all the images. The resulting masked image was corrected for
the primary beam and used to measure the flux. For \clean\ and
\wclean, residual-corrected images were used for measuring the total
flux. For measuring the \textit{rms} noise, unmasked, non primary beam
corrected (and for \clean\ with/without windows, non residual-scaled)
images were used. The \textit{rms} noise was measured in a
$50\times50$ pixel box within the cleaned region of each image, and
in a location with no source emission.

\begin{deluxetable}{lcccc}
  \tablecaption{Beam, Total Flux and \textit{rms} Noise comparison for
    \msclean, \mrclean, \wclean\ and \clean\ in
    Holmberg~II.\label{tab:ms-mr-w-c-comp}}
  \tablewidth{0pt}
  \tablecolumns{4}
  \tablehead{\colhead{\multirow{2}{*}{Algorithm}} & \colhead{Beam Size} &%
    \colhead{Flux Recovered} & \colhead{\textit{rms} Noise} \\ 
    \colhead{} & \colhead{($\arcsec$)}  & \colhead{(Jy)} &%
    \colhead{($\mjbeam$)}}
    \startdata
    \msclean & $11.3\times10.8$ & $3.11$ & $0.83$ \\
    \mrclean & $13.7\times12.6$ & $3.02$ & $0.87$ \\
    \clean   & $13.7\times12.6$ & $2.77$ & $0.88$ \\
    \cwclean  & $13.7\times12.6$ & $2.88$ & $0.91$ \\
    \cwclean\ ($1\sigma$) & $13.7\times12.6$ & $2.93$ & $1.3$ \\
    \cwclean\ ($0.5\sigma$) & $13.7\times12.6$ & $2.95$ & $1.6$ \\
    \enddata 
    \tablecomments{Total flux measurements for \clean\ and \wclean\
      were made with data that was residual flux corrected, while
      \textit{rms} noise measurements were made with data which was
      not residual flux corrected.}
\end{deluxetable}
% \placetable{tab:ms-mr-w-c-comp}

\cwclean\ recovers $\sim5$\% more flux than \clean, but does not
recover as much flux as the two scale-sensitive algorithms.  The
number of iterations taken for each to reach the flux threshold were
$3500$ for \msclean, $4050$ for \mrclean, $5100$ for \wclean\ and
$9720$ for \clean.  The flux gains made here on real data by the
scale-sensitive algorithms are in line with the gains made by these
algorithms on simulated data \citep{msclean}. The \textit{rms} noise
measurements are in good agreement, which indicates that none of the
algorithms significantly changes the noise properties of the data. We
note that the small differences in noise levels between those listed
in Table~\ref{tab:ms-mr-w-c-comp} (as derived for Holmberg~II using
\msclean, \mrclean\ and \clean) are slightly different from those
listed in Table~\ref{tab:rms-flux} (as determined for a line-free
channel in the Holmberg~II data cube). This difference is mostly
attributable to small channel-to-channel changes in the average noise
level, as well as small statistical spatial noise fluctuations in each
channel. We find these noise measurements to have a dispersion of
$\sim0.03$~mJy for this particular case.

When using \clean\ windows, the search for \clean\ components is
restricted to the area of known flux (\ie\ the window) and it should
therefore in principle be possible to \clean\ to a deeper level, thus
increasingly negating the need for residual scaling. We have tested
this by running \wclean\ down to additional flux thresholds of
$1\sigma$ and $0.5\sigma$ in Holmberg~II. The total flux recovered and
the \textit{rms} noise are listed in Table~\ref{tab:ms-mr-w-c-comp}.
We note there is a significant increase in the number of iterations
required to \clean\ down to these flux thresholds, with a marginal
gain in recovered flux.  For a $1\sigma$ threshold, $38000$ iterations
are required and for a $0.5\sigma$ level, $198000$ iterations are
needed.  For the original $2.5\sigma$ threshold, $5000$ iterations were
required.  In comparison, \msclean\ and \mrclean\ recover more flux without
an increase in \textit{rms} noise, and with fewer iterations. 

\begin{figure}[htbp]
  \centering
  \begin{tabular}{ccc}
    \includegraphics[angle=270,width=0.3\textwidth]{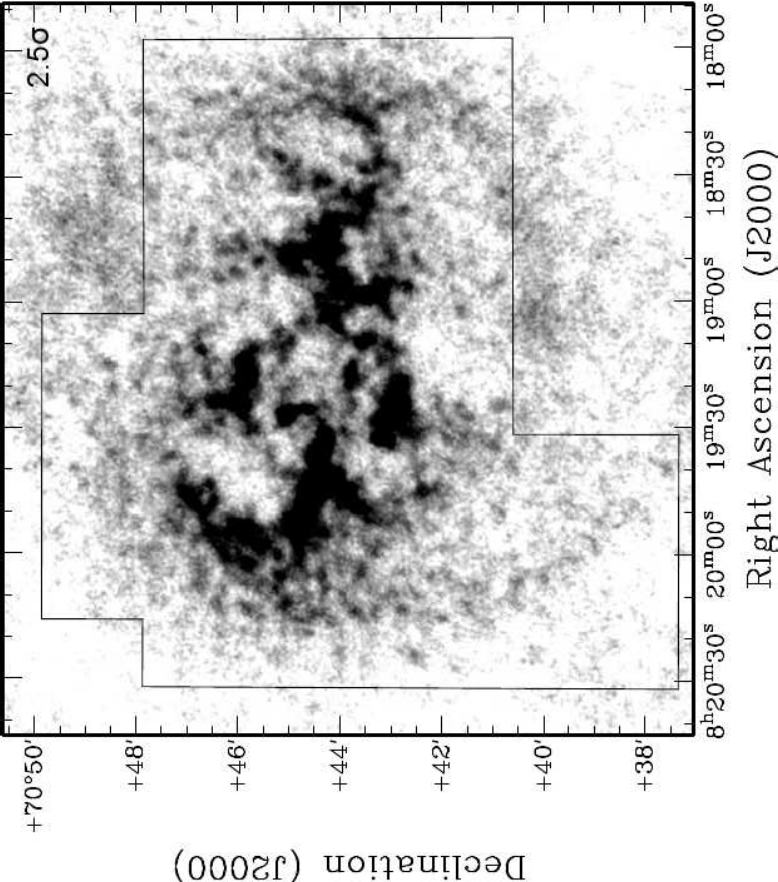}&
    \includegraphics[angle=270,width=0.3\textwidth]{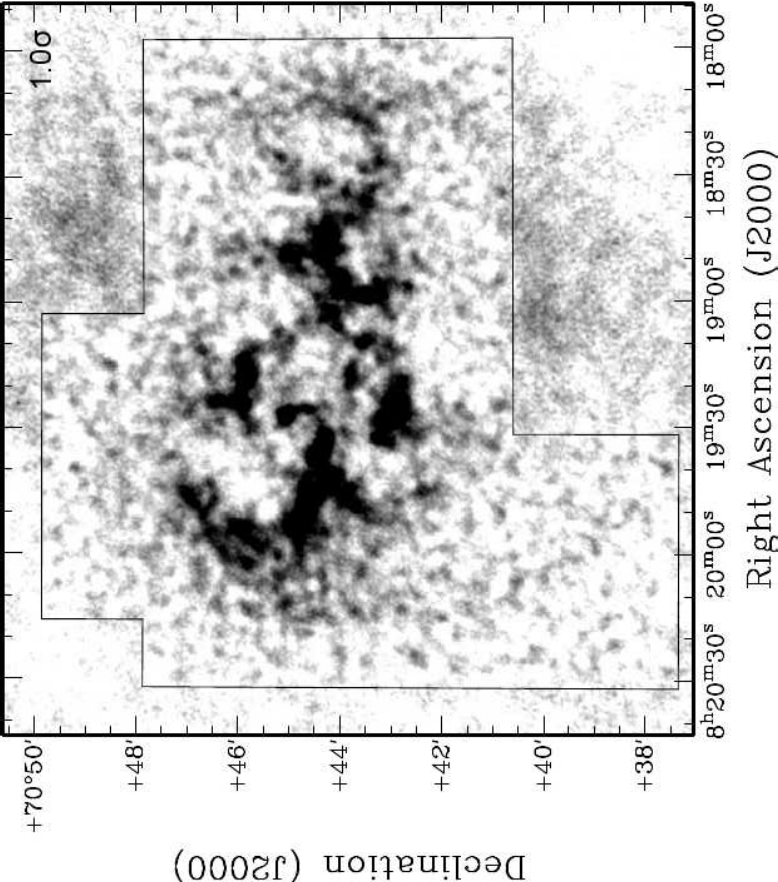}&
    \includegraphics[angle=270,width=0.3\textwidth]{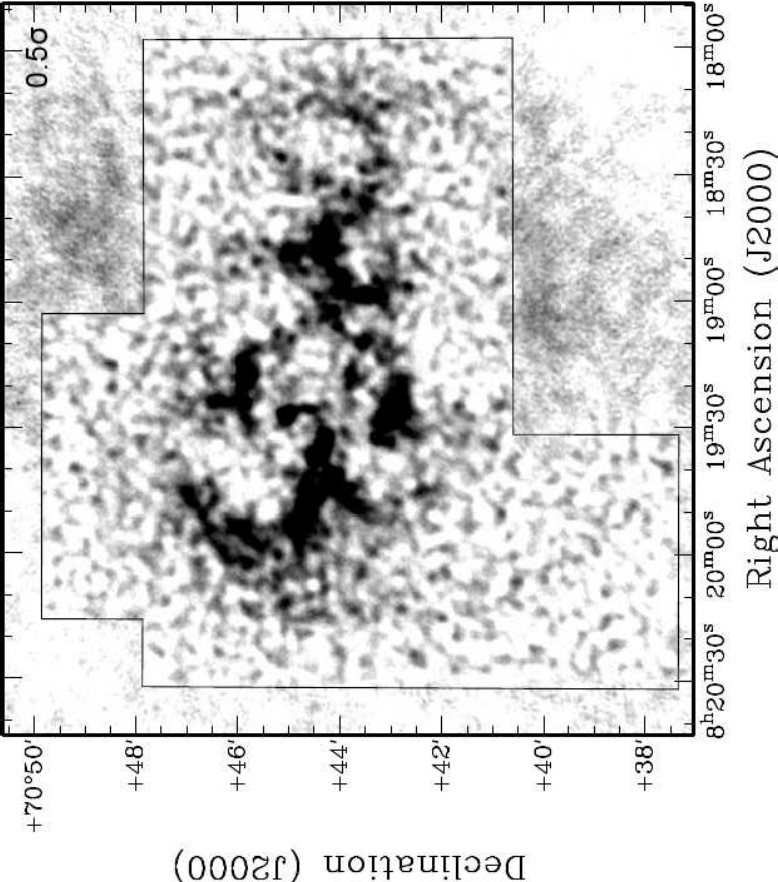}\\
    \includegraphics[width=0.3\textwidth]{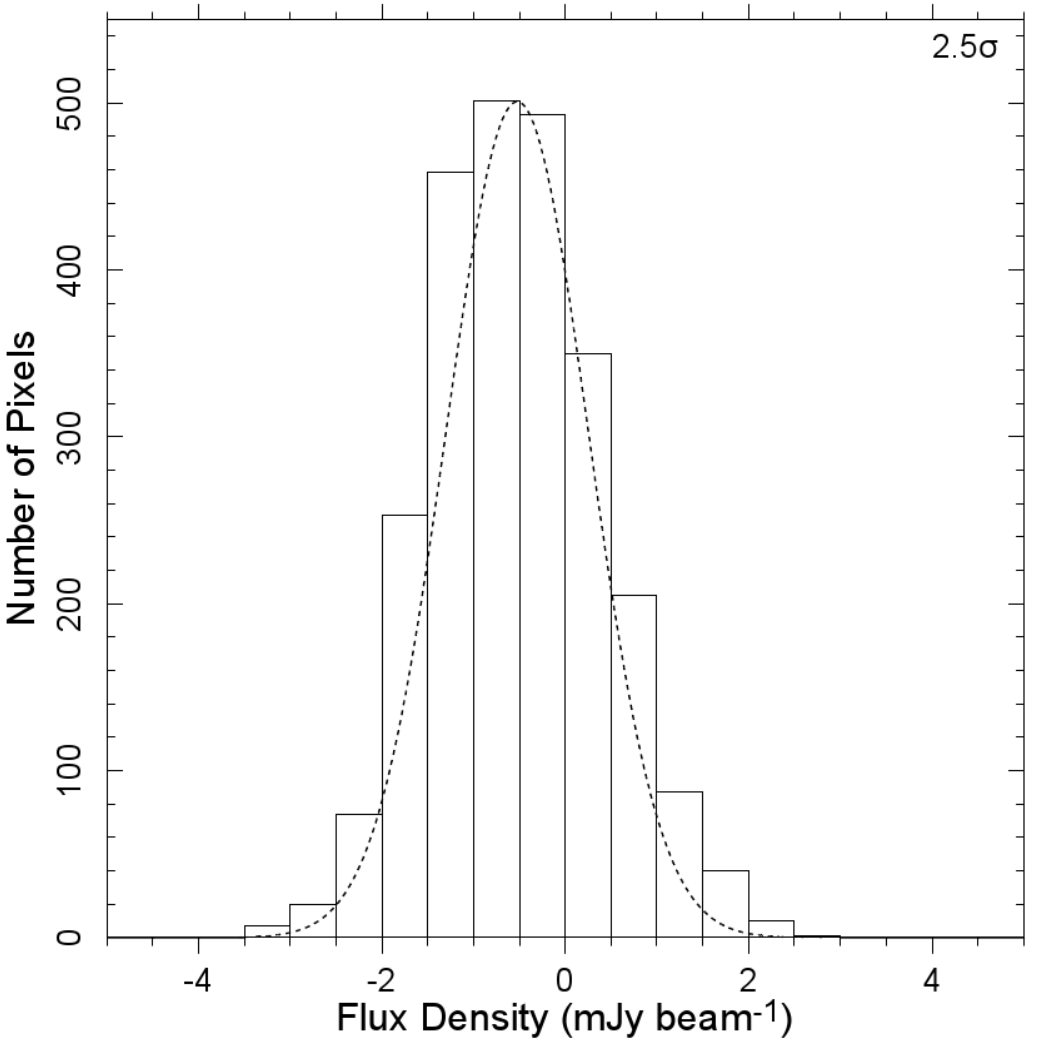}&
    \includegraphics[width=0.3\textwidth]{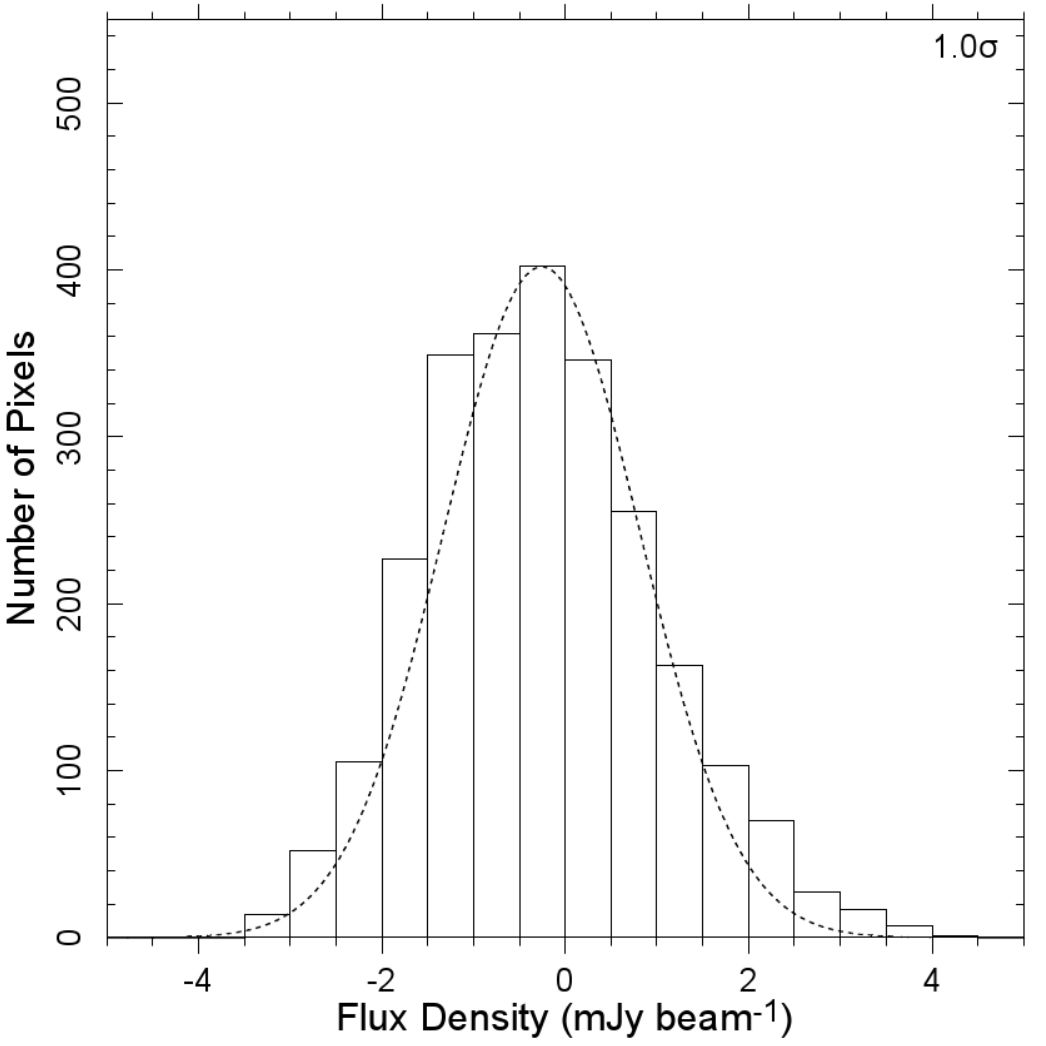}&
    \includegraphics[width=0.3\textwidth]{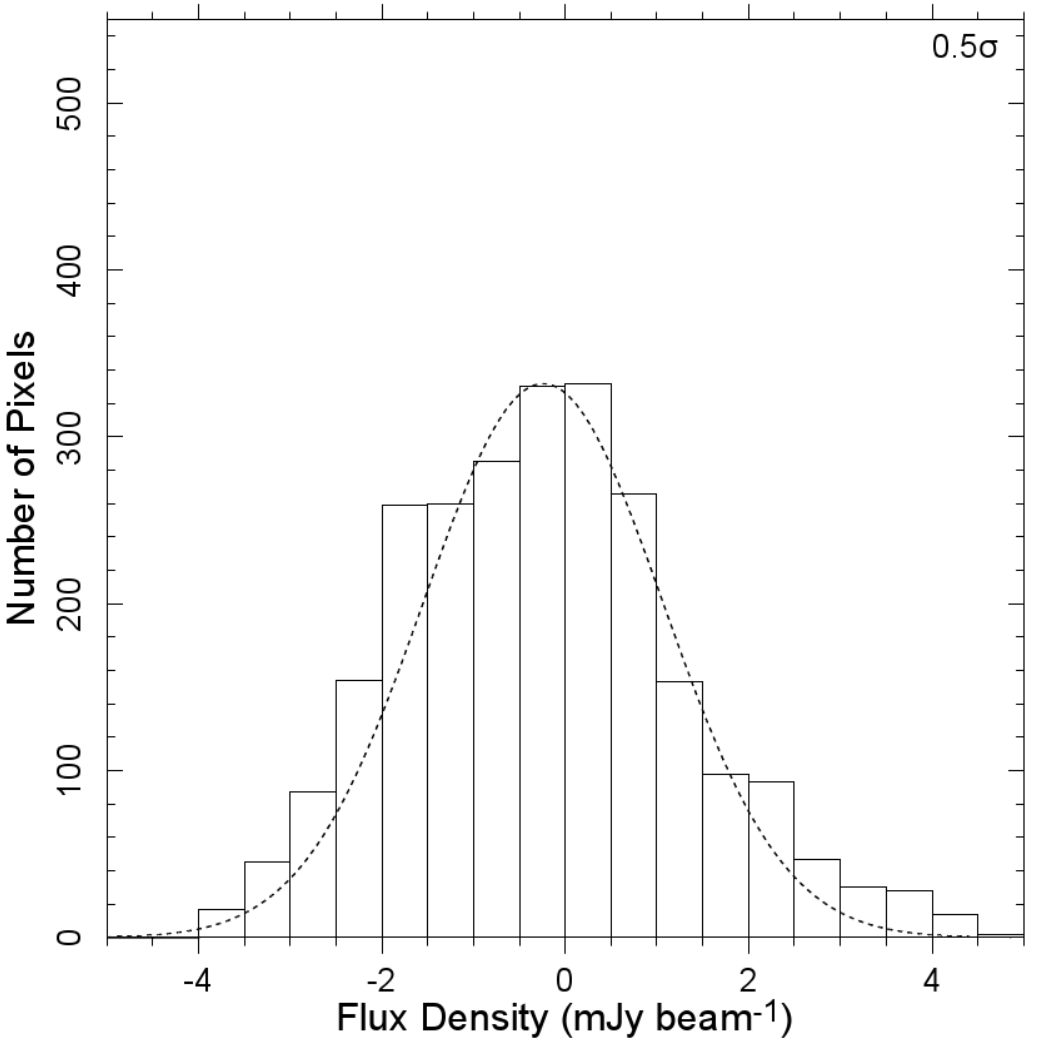}\\
  \end{tabular}
  \caption{Images (top row) of \wclean\ to $2.5\sigma$ (left),
    $1\sigma$ (middle) and $0.5\sigma$ (right) for a channel for the
    galaxy Holmberg~II data. The outline of the window used is shown in
    black. Images are residual flux corrected, unmasked and without
    primary-beam correction applied. The gray-scale levels run from $0$ to
    $8$~\mjbeam.  Also shown are histograms (and the Gaussian fit) of the
    flux values (bottom row). The histogram values are sampled in a
    $50\times50$ pixel box within the cleaned region and where there is no
    source emission. Images without residual scaling applied were used for
    the histogram generation.\label{fig:deep-wclean-hol2}}
\end{figure}
% \placefigure{fig:deep-wclean-hol2}

The resulting channel maps along with the channel map for the original
\wclean\ to $2.5\sigma$ flux threshold are shown in
Figure~\ref{fig:deep-wclean-hol2}. The window applied is marked by the
black box around the emission in the images. The box was chosen based
on emission visible in the dirty image. After deconvolution,
additional emission became visible outside our \clean\ window (as
shown in Figure~\ref{fig:deep-wclean-hol2}). Normally one would
adjust the \clean\ window and perform further \clean\ runs. Here,
however, we only show the result from our first run, as this clearly
illustrates the differences in noise properties inside and outside the
\clean\ box. Immediately noticeable in these images is the increased
`spottiness' that comes with cleaning to a deeper flux level.  This is
most likely due to \clean\ operating into the noise and thus cleaning
noise spikes as opposed to real emission with a distortion of the
noise.  This can be seen in the histograms of
Figure~\ref{fig:deep-wclean-hol2} which show the increased
\textit{rms} noise for the deepest {\wclean}s and is also clear from
the measured \textit{rms} listed in Table~\ref{tab:ms-mr-w-c-comp}.
The same negative bias as observed in
Figure~\ref{fig:ngc2403-noise-histograms} is seen in the histogram of
the \wclean\ down to a flux threshold of $2.5\sigma$ (left histogram),
but this bias disappears for deep {\wclean}s (middle and right
histograms). A deep \wclean\ can therefore eliminate the negative
bowl, but does so at the cost of distorted (and increased) noise.
These results are in line with those on simulated data by
\cite{msclean}.

We note that the definition of a \clean\ window is trivial for the
case of a single channel map discussed here.  For a large data-cube of
a complicated extended source, an individual \clean\ window would need
to be defined for each channel of the cube, as the structure of the
source in each channel changes in both shape and position.  It is
therefore impractical to use \wclean\ for the THINGS sample, where the
number of spectral channels for a single galaxy can be as high as
$100$.

\subsection{Channel Map Comparison}
\label{sec:ms-mr-w-c-chanmaps}

\begin{figure}[htbp]
  \centering
    \begin{tabular}{ccc}
      \includegraphics[angle=270,width=0.25\textwidth]{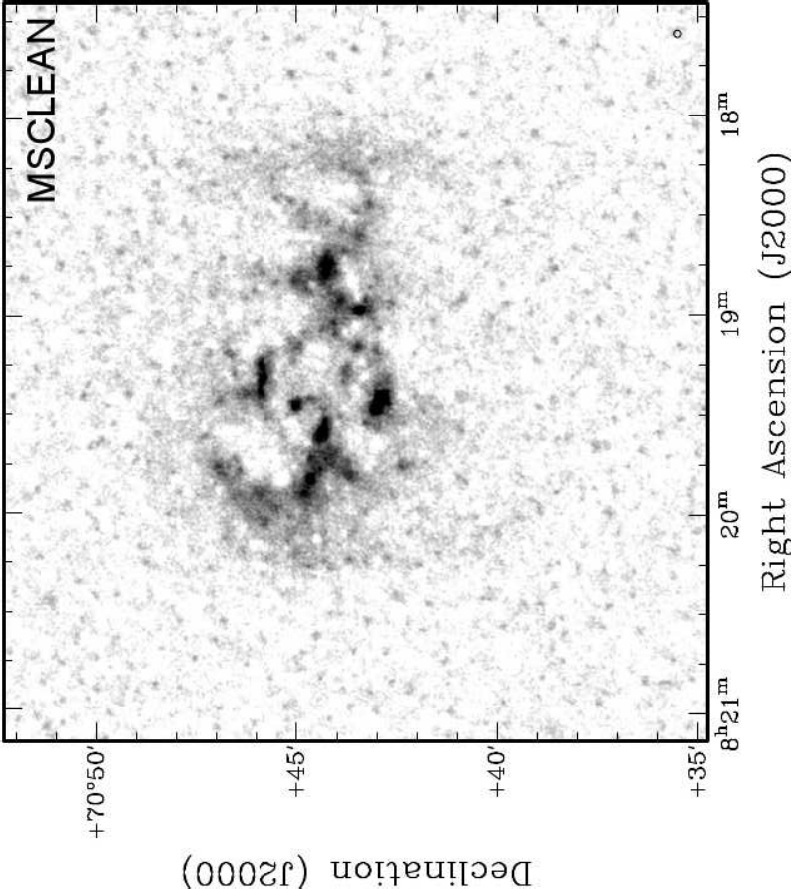}&
      \includegraphics[angle=270,width=0.25\textwidth]{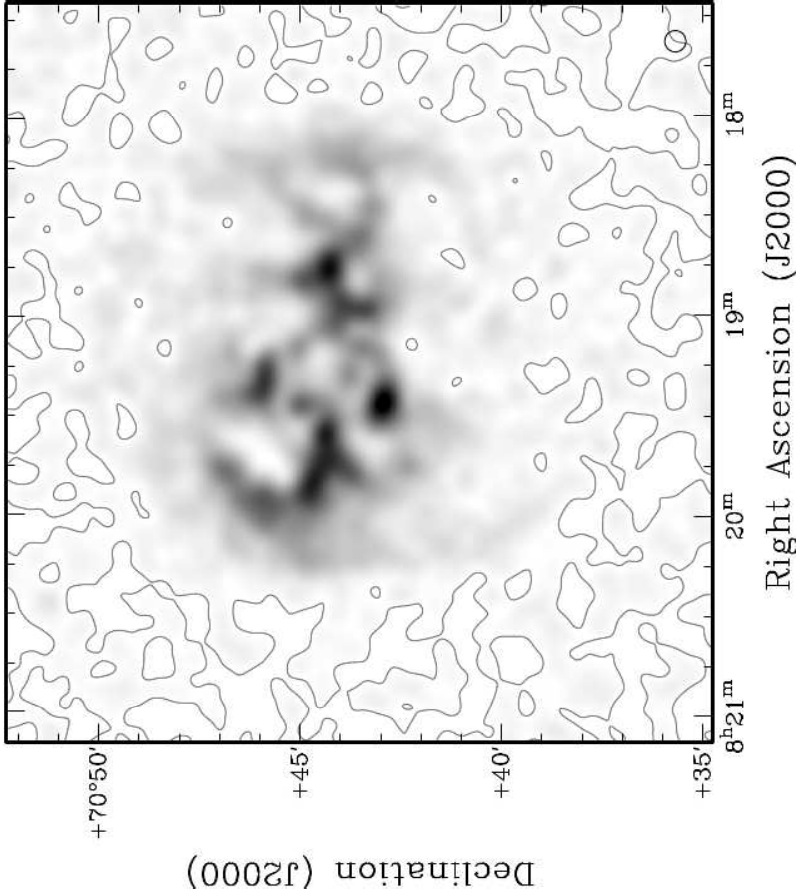}&
      \includegraphics[angle=270,width=0.25\textwidth]{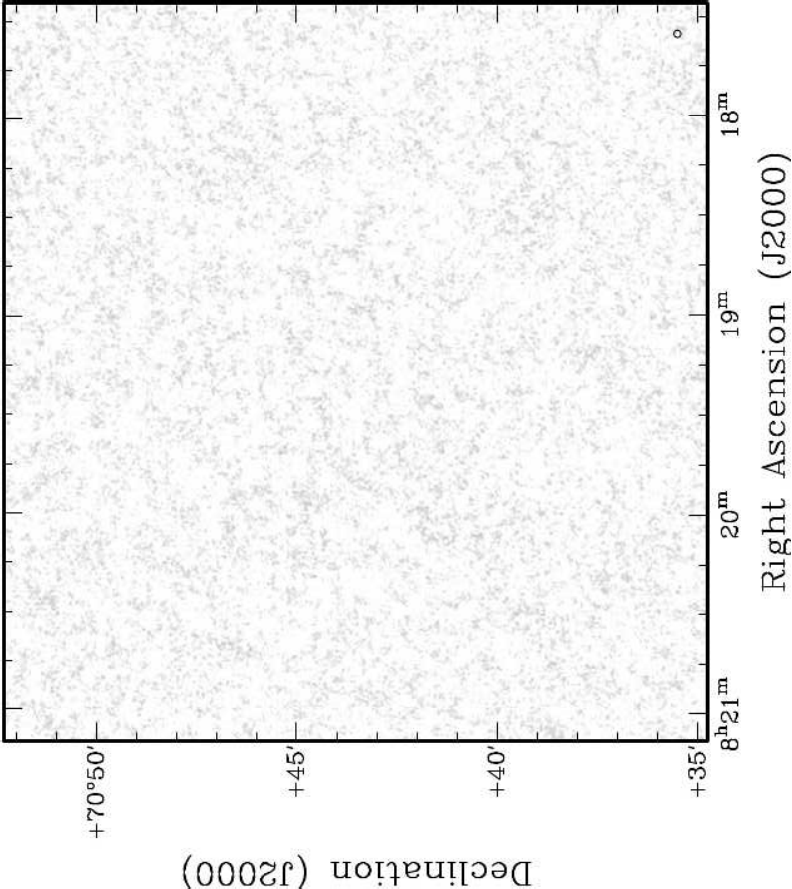}\\
      \includegraphics[angle=270,width=0.25\textwidth]{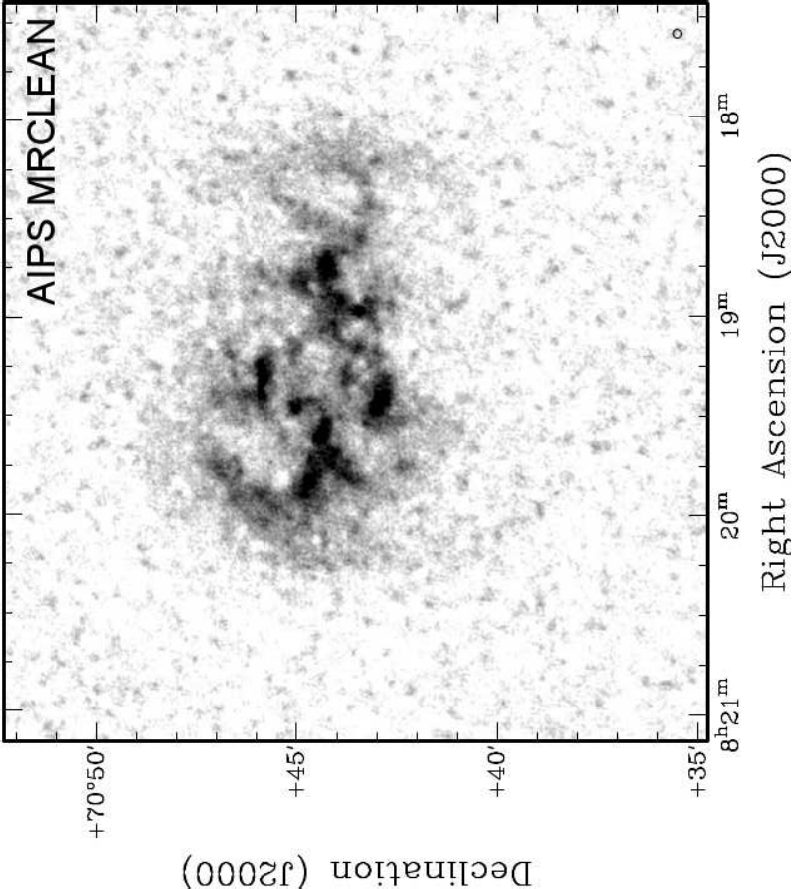}&
      \includegraphics[angle=270,width=0.25\textwidth]{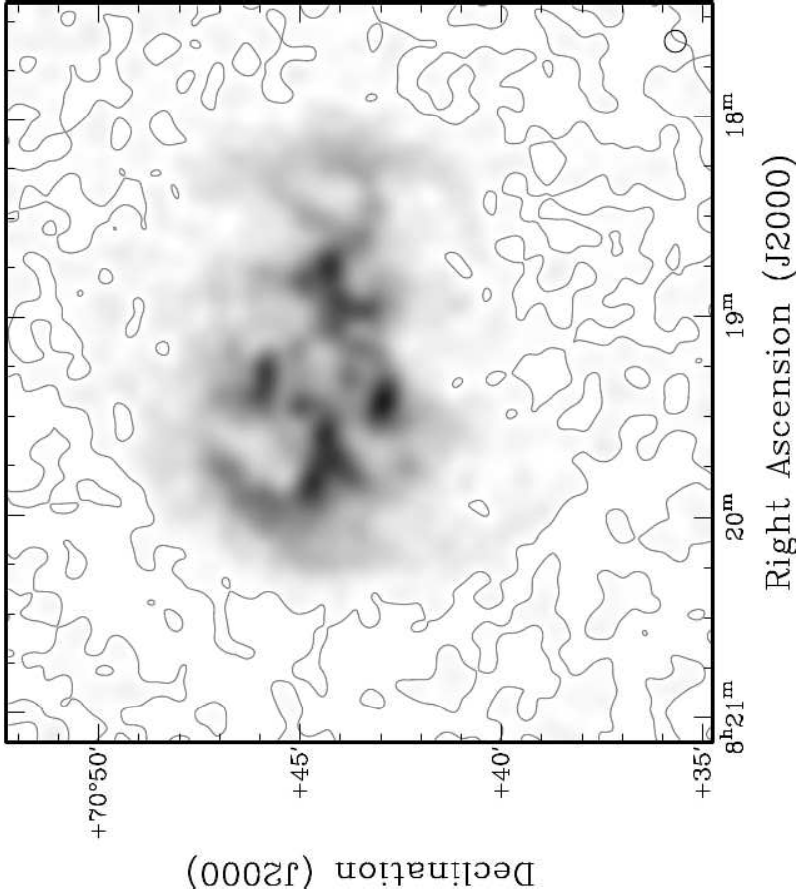}&
      \includegraphics[angle=270,width=0.25\textwidth]{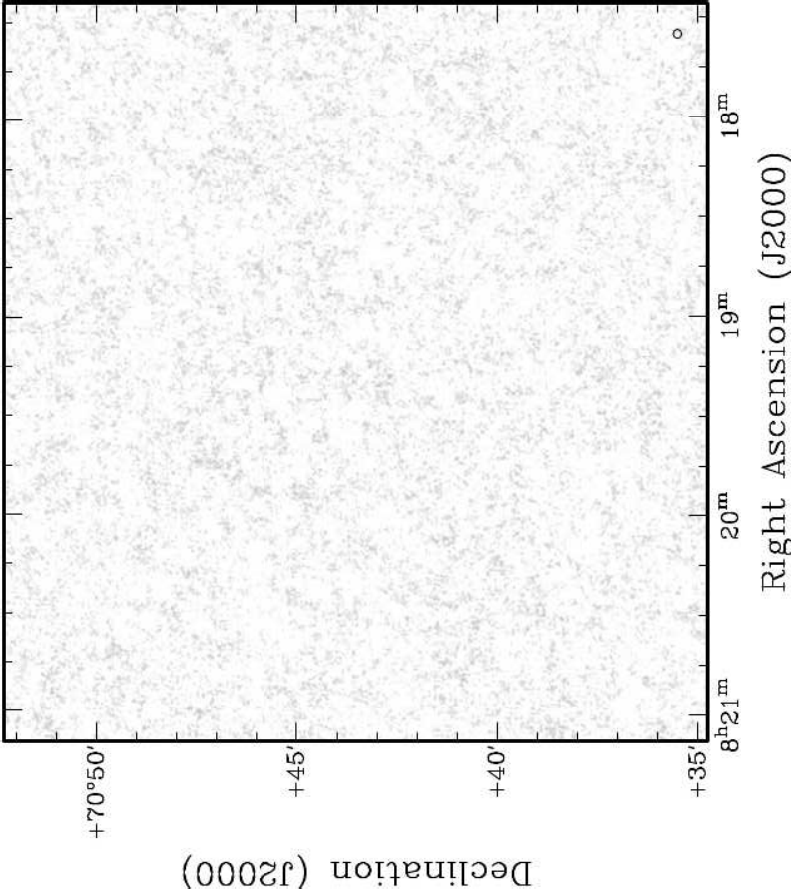}\\
      \includegraphics[angle=270,width=0.25\textwidth]{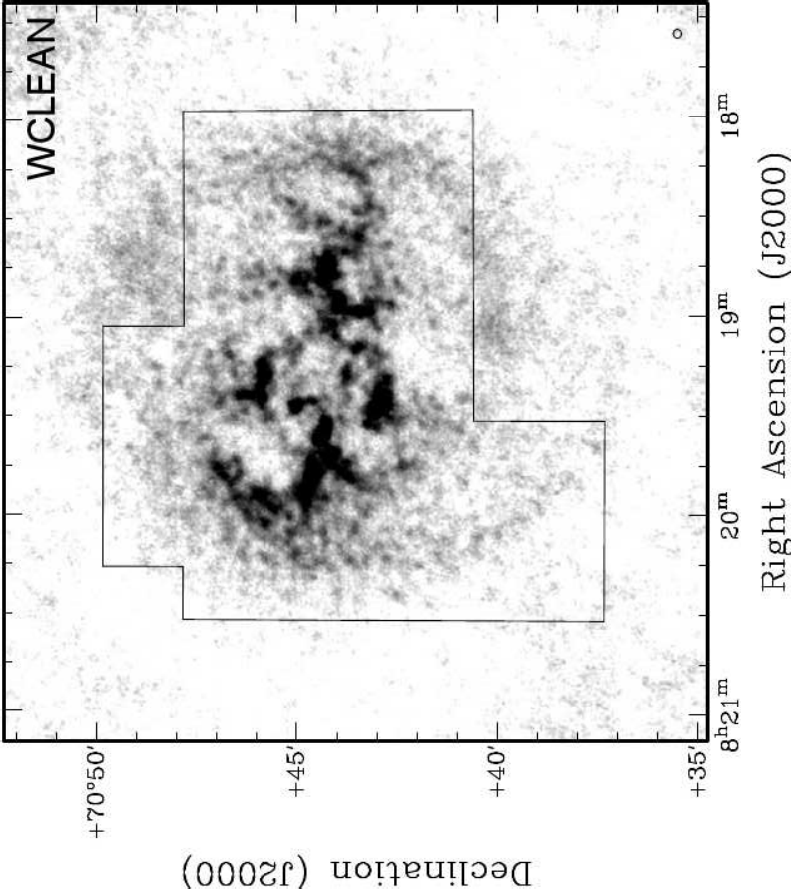}&
      \includegraphics[angle=270,width=0.25\textwidth]{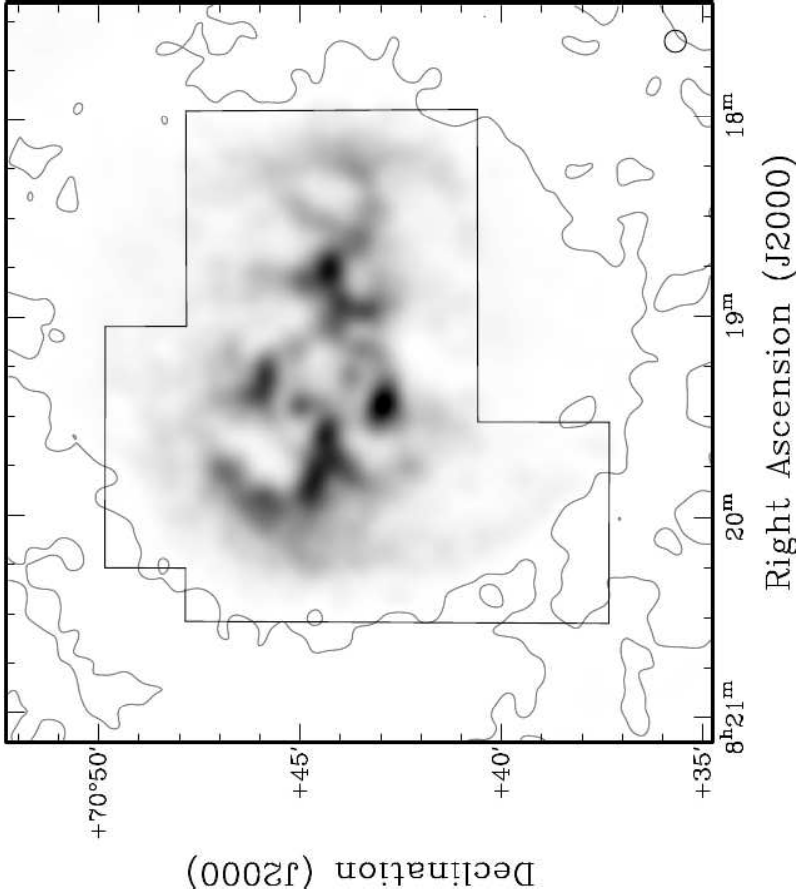}&
      \includegraphics[angle=270,width=0.25\textwidth]{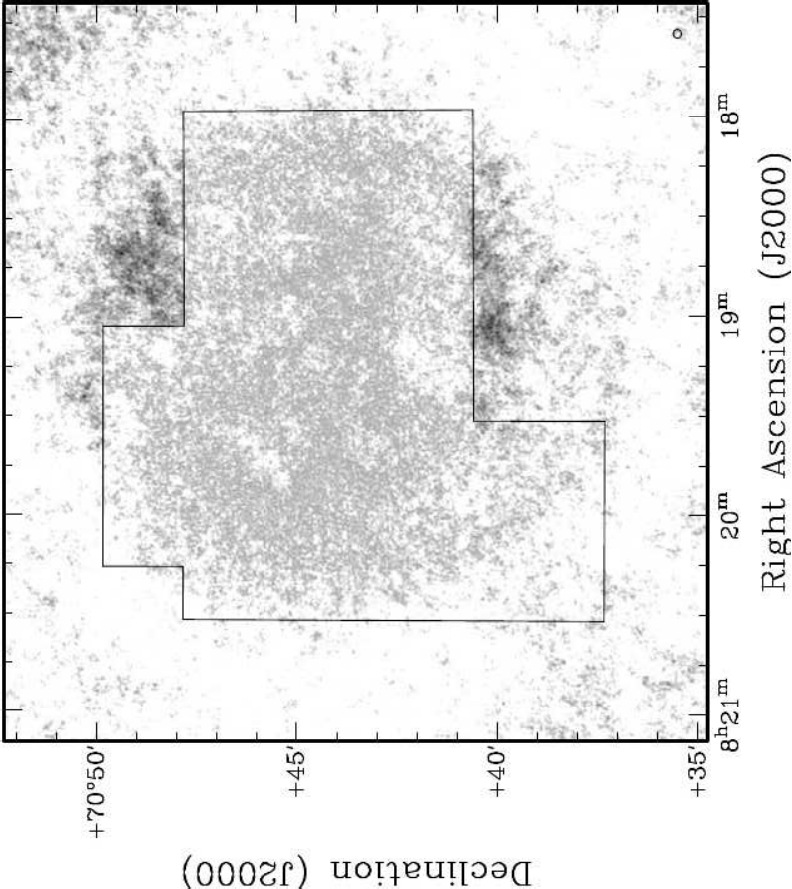}\\
      \includegraphics[angle=270,width=0.25\textwidth]{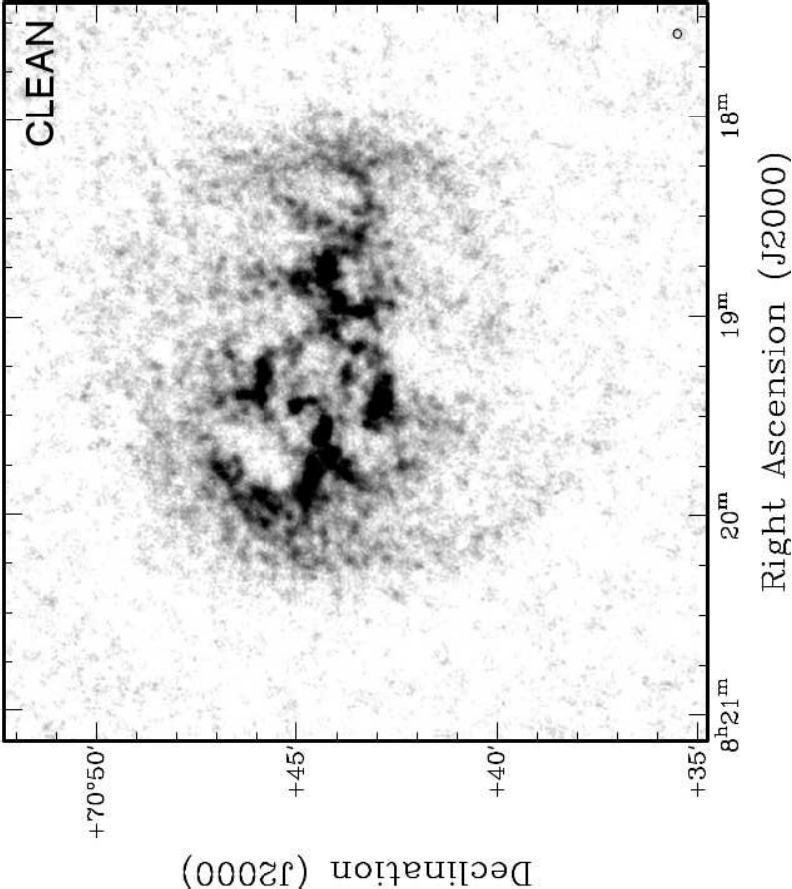}&
      \includegraphics[angle=270,width=0.25\textwidth]{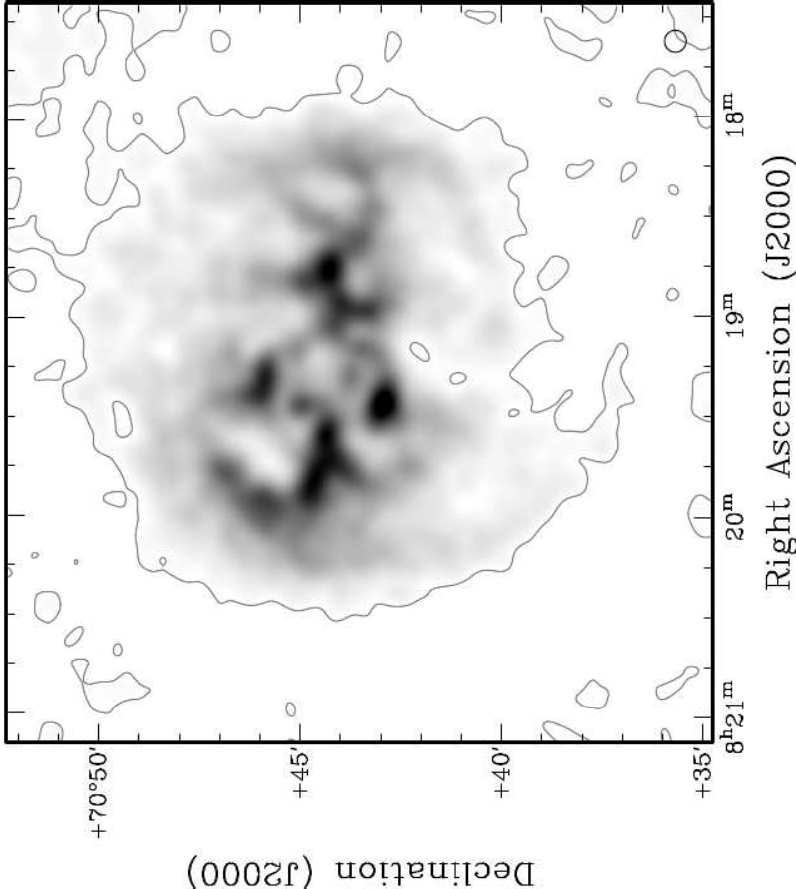}&
      \includegraphics[angle=270,width=0.25\textwidth]{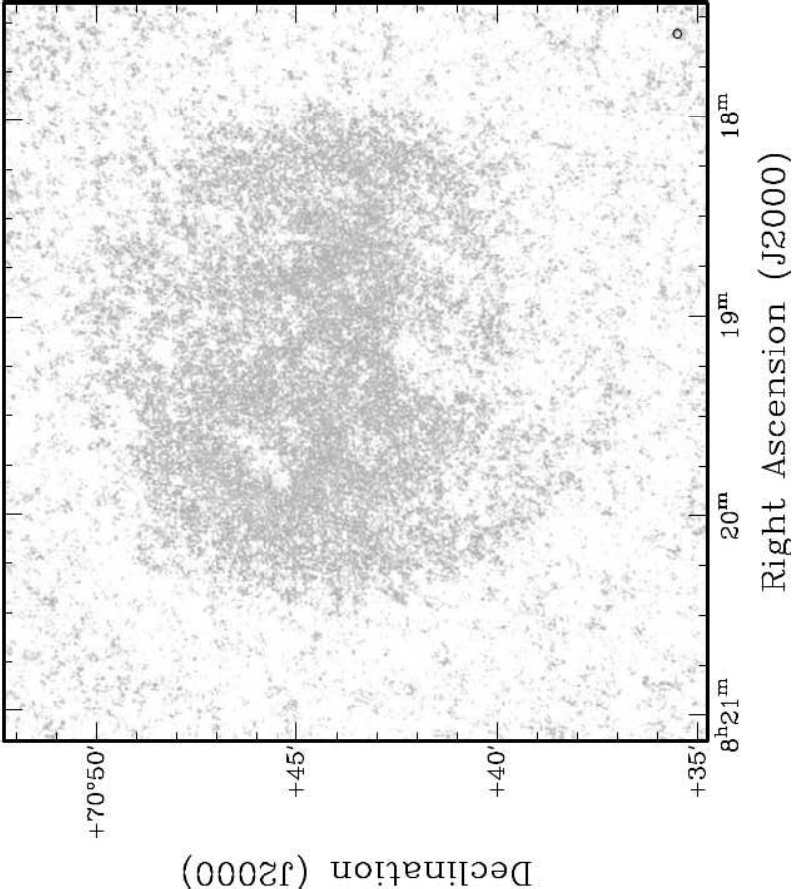}\\
    \end{tabular}
    \caption{The cleaned image (left), smoothed to $30$\arcsec\
      (middle) and residual (right) maps for, from top row to bottom row,
      \msclean, \mrclean, \wclean\ and \clean, for the galaxy Holmberg~II.
      Images are un-masked and no primary-beam correction has been
      applied. No residual-scaling has been applied to the \clean\ and
      \wclean\ images. Restoring beams are as marked in the bottom-right
      corner of each image.  For each image in the center column, a
      (smoothed) contour has been plotted at a flux density level of
      $-0.45$~\mjbeam\ ($-1.5\sigma$ of the \clean\ noise).  For the
      \wclean\ images and residual, the outline of the window used is also
      shown in black. The gray-scale levels for the left and right columns
      run from $0$ to $15$~\mjbeam and for the middle column, $0$ to
      $70$~\mjbeam.\label{fig:ms-mr-w-c-hol2}}
  \end{figure}

In Figure \ref{fig:ms-mr-w-c-hol2}, we compare the images produced by
each algorithm.  From top to bottom in each figure are the images for
\msclean, \mrclean, \wclean\ and \clean.  The left column is the
restored image. The center column is the restored image smoothed to
$30$\arcsec\ with a contour plotted at level equal to $-1.5\sigma$ of
the (smoothed) noise to show the effect of the negative bowl.  The
right column is the residual image. For each column, all of the
images are plotted to the same gray-scale levels.  Data without masks
and not corrected for primary-beam attenuation was used.  For \clean\
and \wclean, no residual scaling was applied. The contours in the
smoothed images (middle column) show that the \clean\ bowl is notably
less present for the \msclean\ and \mrclean\ images.  There is also no
trace of emission in the residual images for these two algorithms
(right column), while both windowed and un-windowed \clean\ leave an
obvious pedestal.

% \placefigure{fig:ms-mr-w-c-hol2}

\subsection{Difference Maps}
\label{sec:ms-mr-w-c-diffmaps}

\begin{sidewaysfigure}[p]
  \centering
    \begin{tabular}{cccc}
      \includegraphics[angle=270,width=0.225\textwidth]{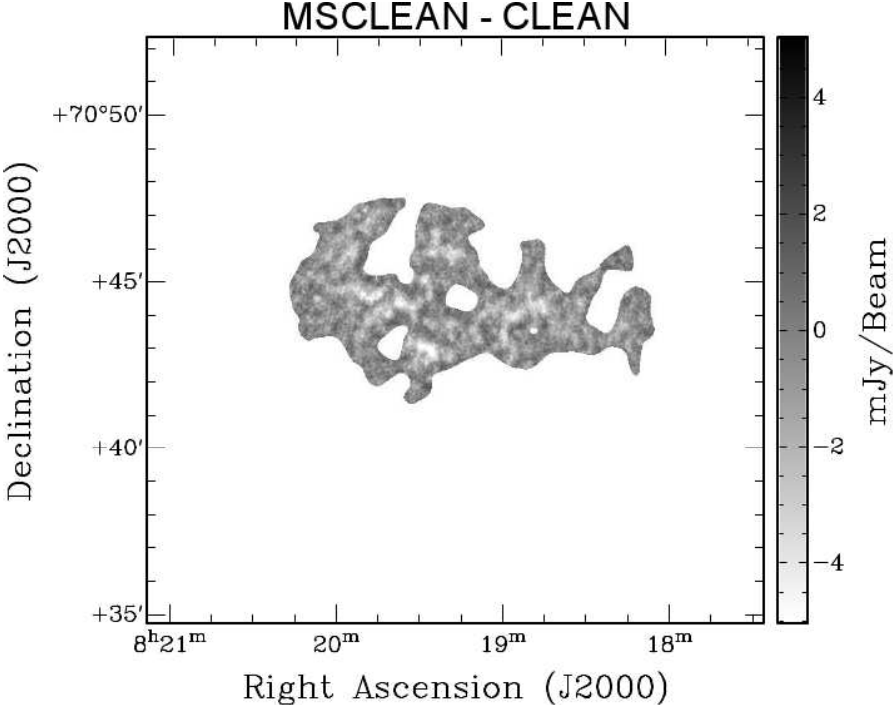}&
      \includegraphics[angle=270,width=0.225\textwidth]{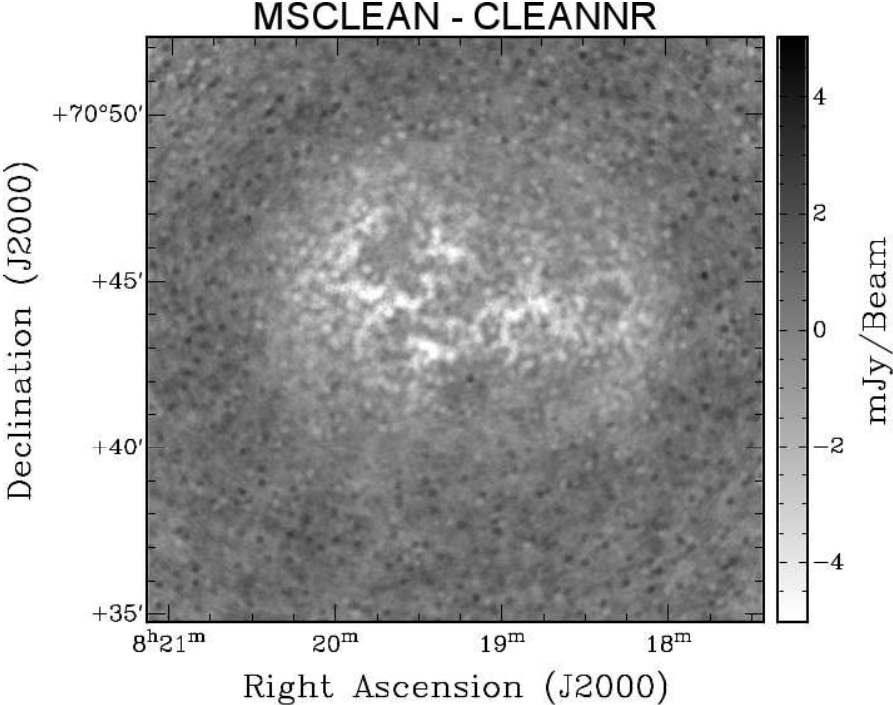}&
      \includegraphics[angle=270,width=0.225\textwidth]{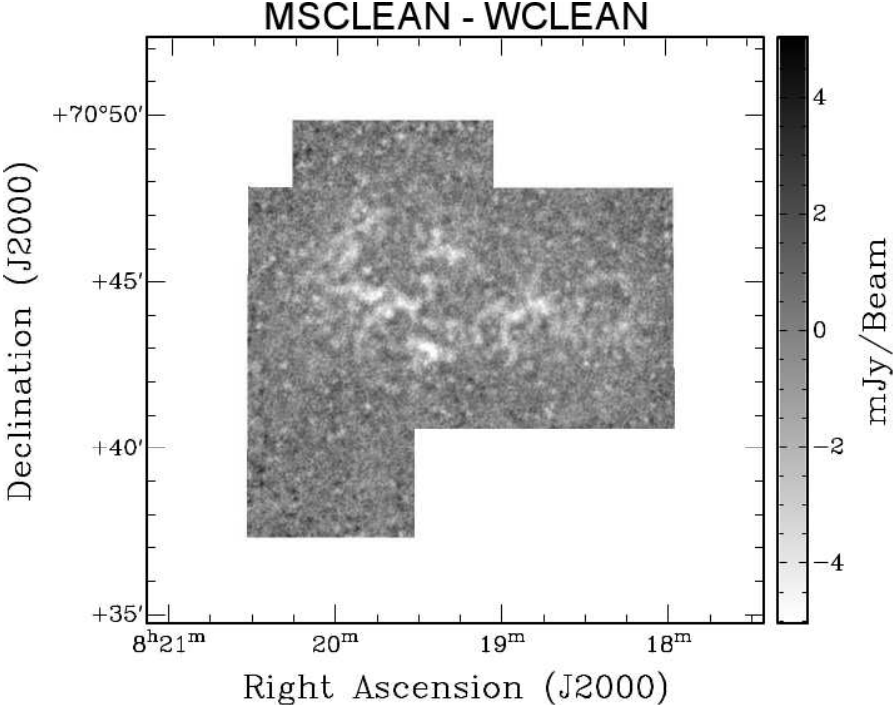}&
      \includegraphics[angle=270,width=0.225\textwidth]{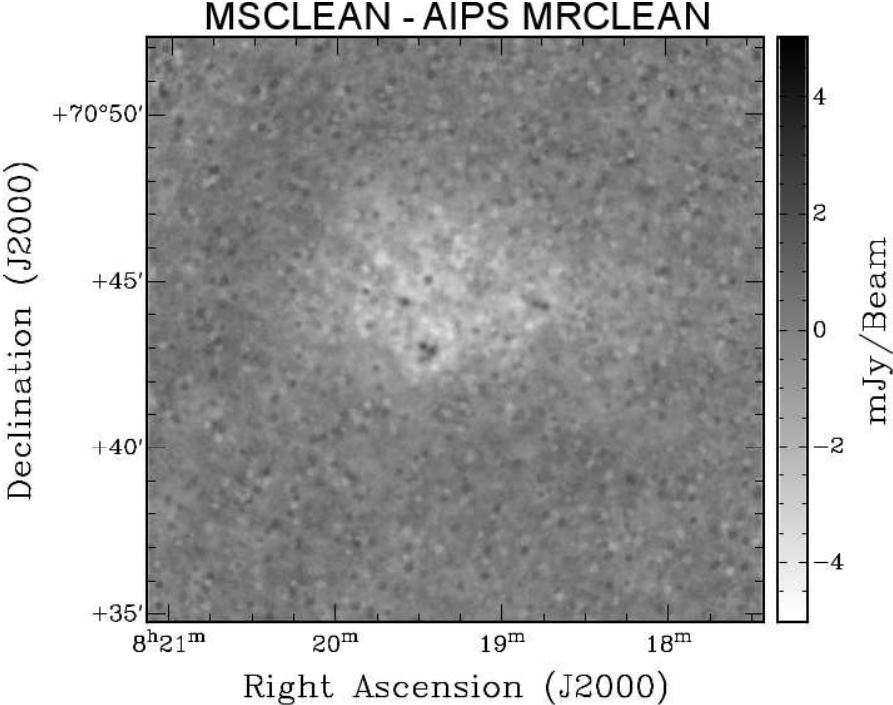}\\
      \includegraphics[width=0.2\textwidth]{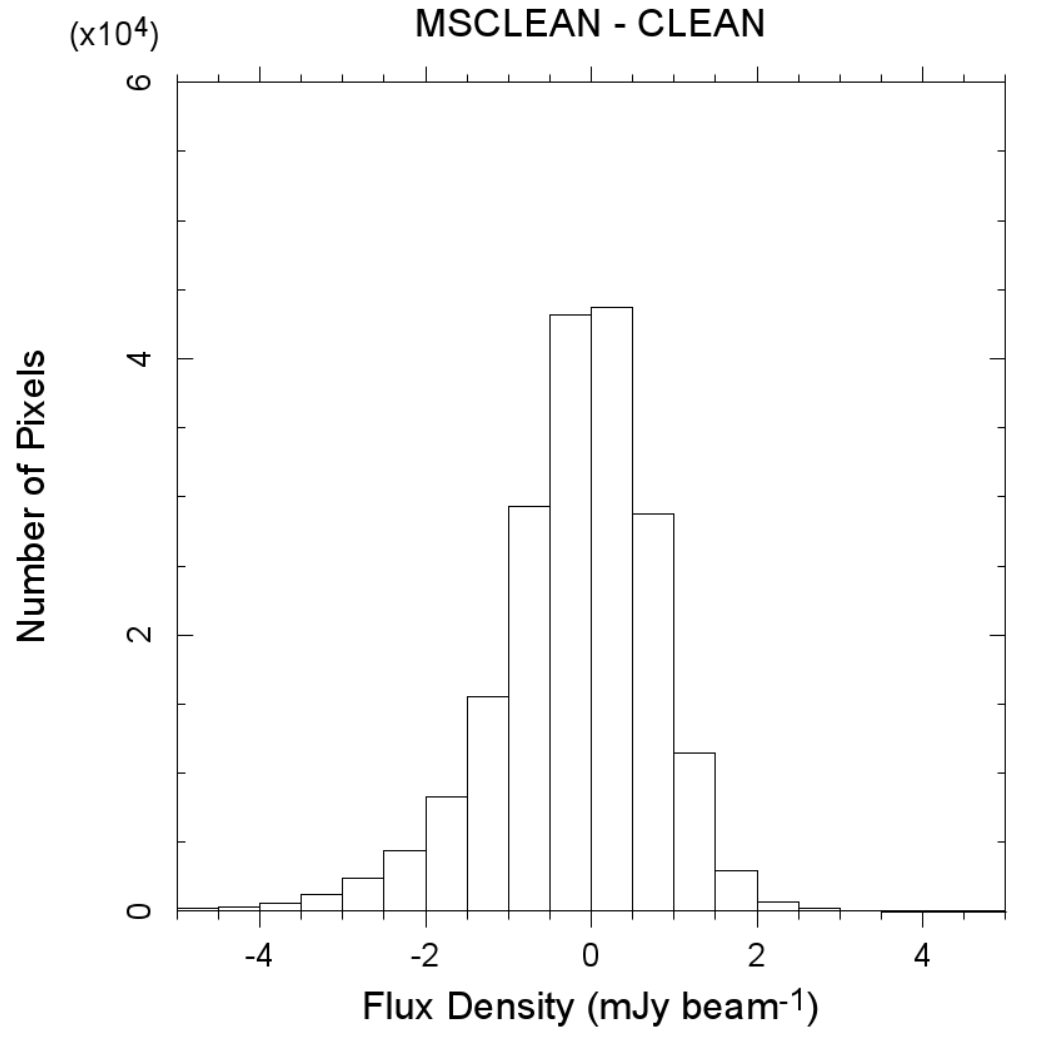}&
      \includegraphics[width=0.2\textwidth]{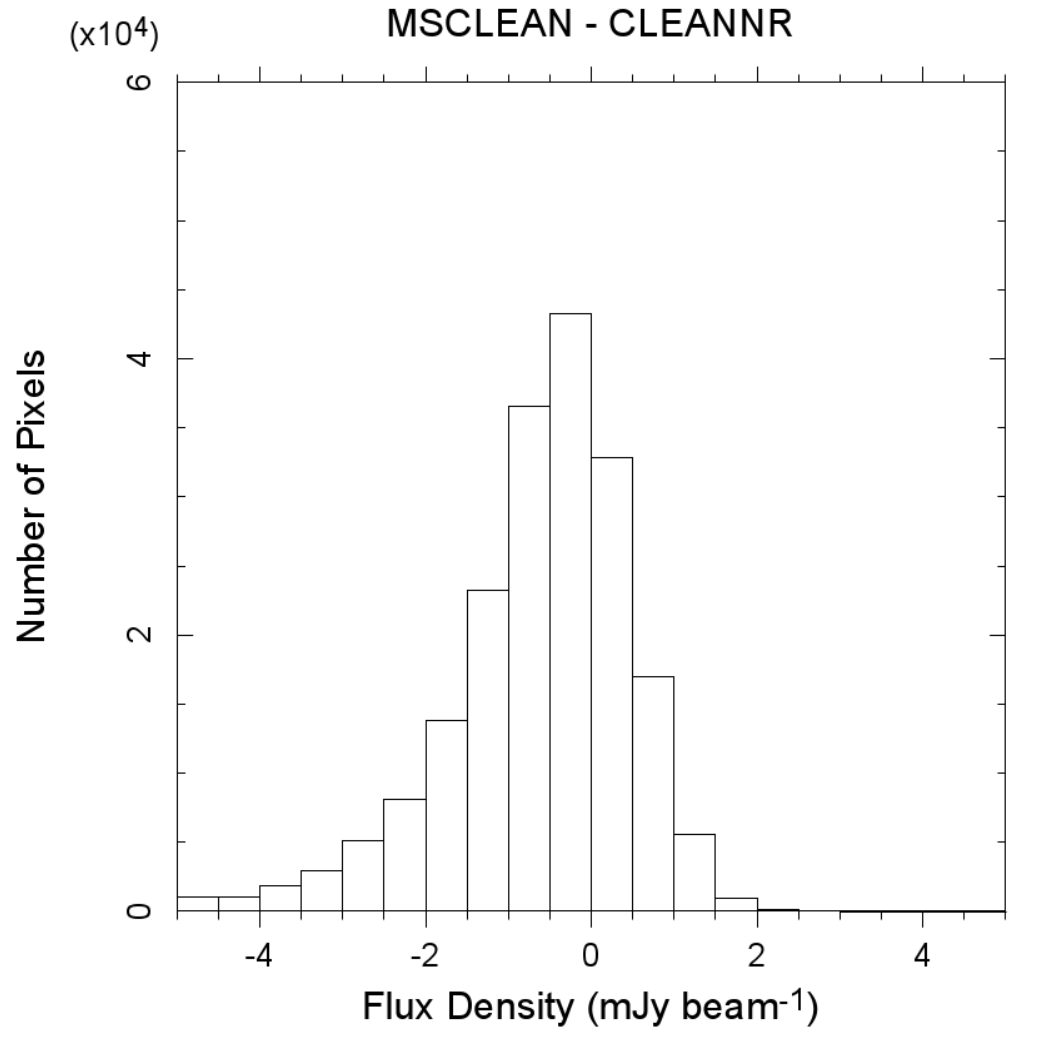}&
      \includegraphics[width=0.2\textwidth]{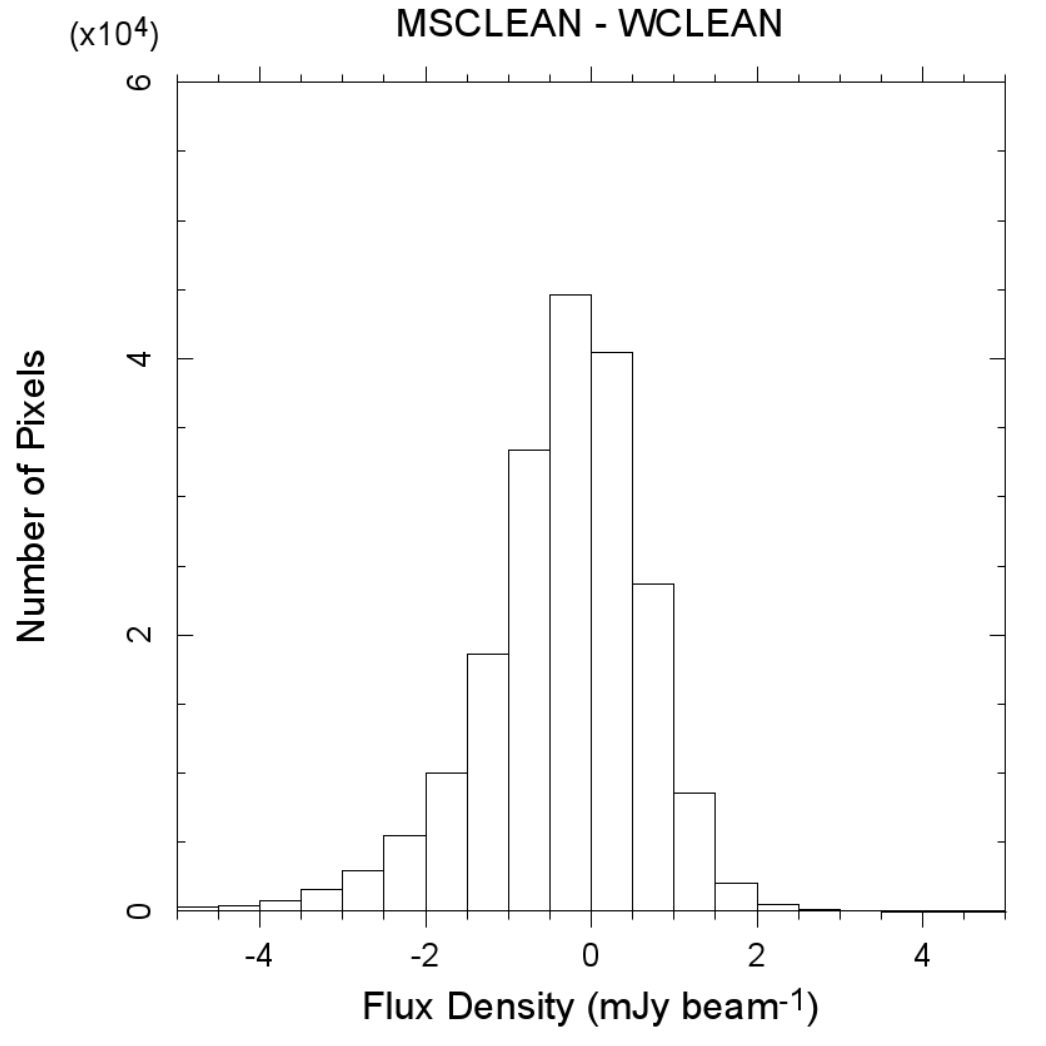}&
      \includegraphics[width=0.2\textwidth]{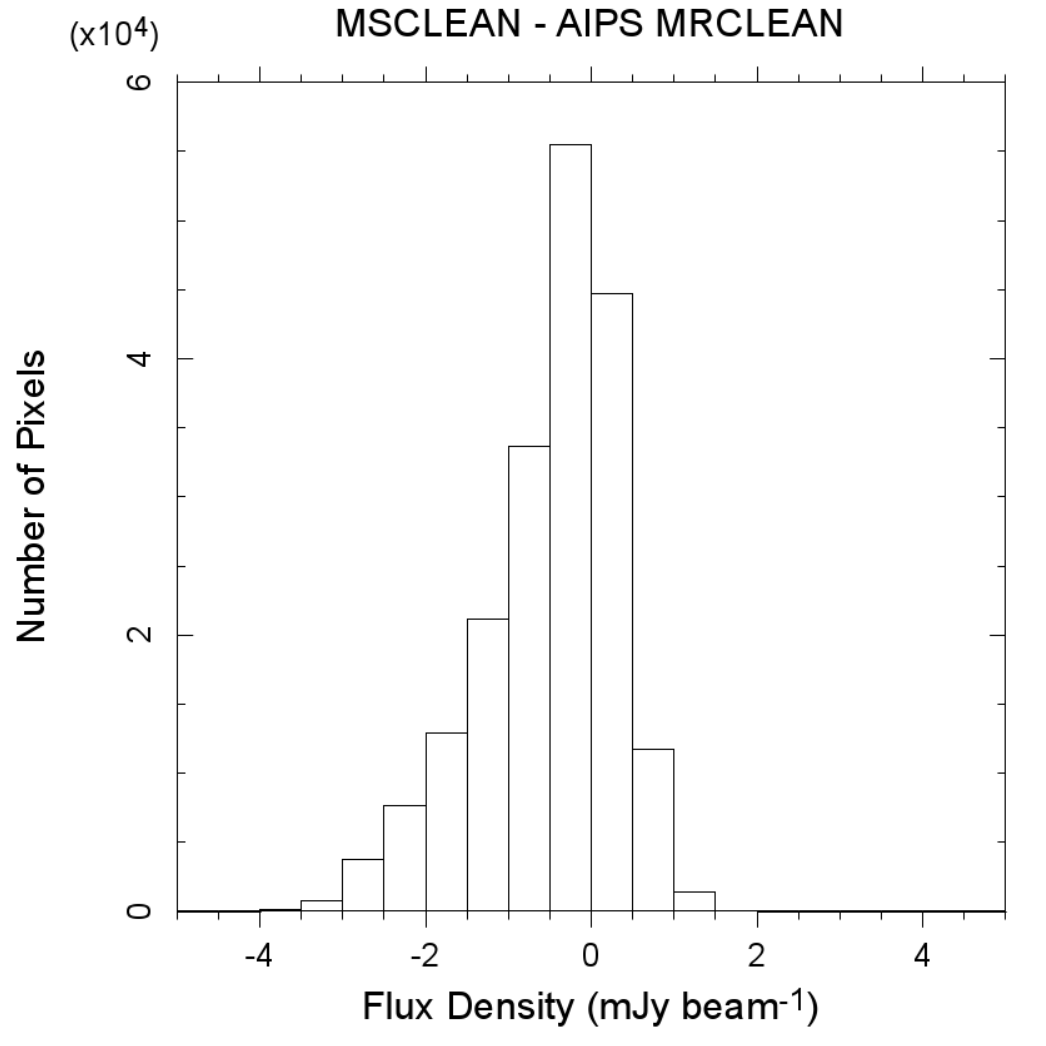}\\
    \end{tabular}
    \caption{Difference images (top row) of \msclean\ minus
      residual-scaled \clean, non residual-scaled \clean, residual-scaled
      \wclean\ and \mrclean\ for Holmberg~II.  The original images were
      corrected for the primary-beam. For residual-scaled \clean\ all flux
      below $2\sigma$ of the \text{rms} noise in the \msclean\ restored
      image has been masked. For \wclean, only the difference within the
      \clean\ window used is shown.  Histograms (bottom row) of the flux
      values calculated within the same area of each image and equal to the
      area inside the window used for the \wclean\ are also shown.
      Gray-scale levels in the images run from $-5$ to
      $5$~\mjbeam.\label{fig:ms-minus-hol2}}
\end{sidewaysfigure}

\begin{sidewaysfigure}[p]
  \centering
    \begin{tabular}{ccc}
      \includegraphics[angle=270,width=0.225\textwidth]{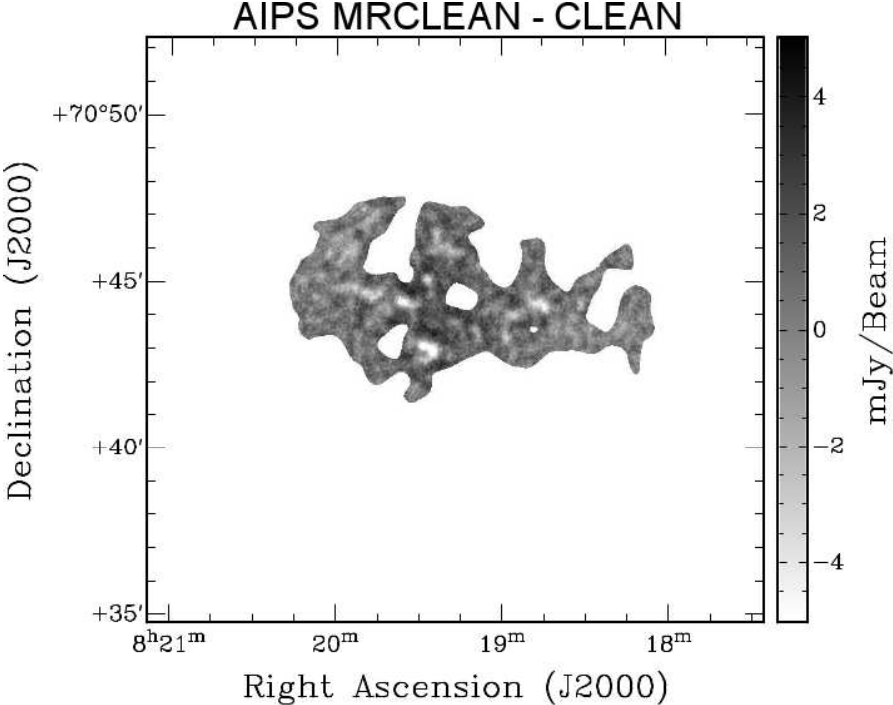}&
      \includegraphics[angle=270,width=0.225\textwidth]{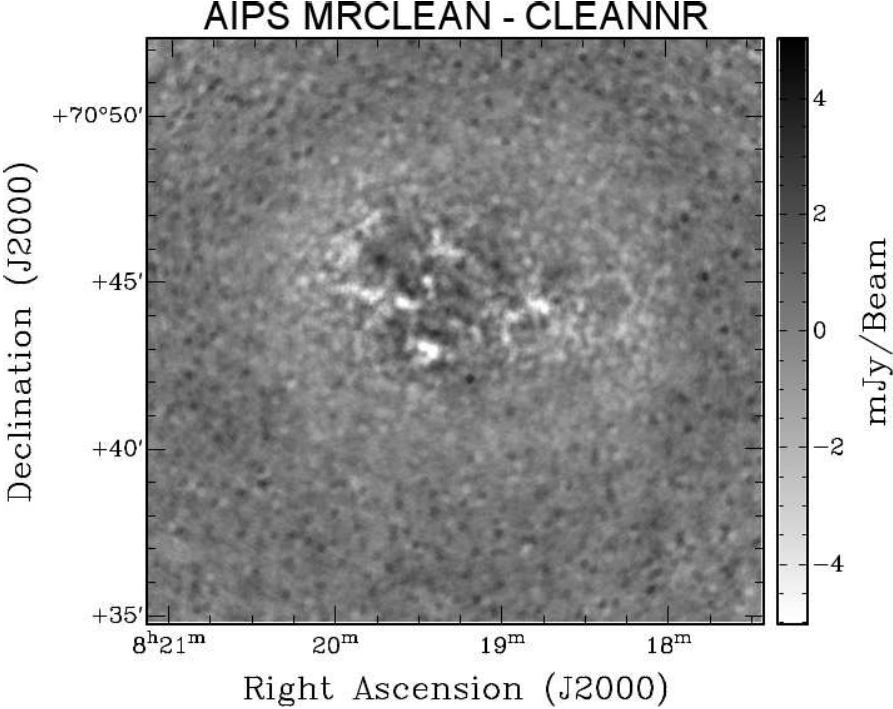}&
      \includegraphics[angle=270,width=0.225\textwidth]{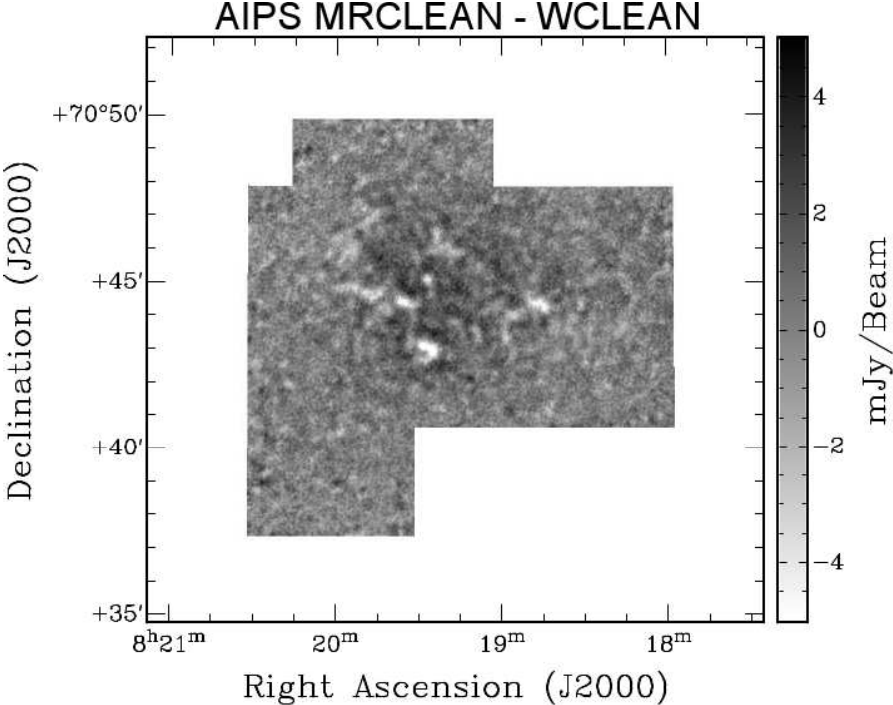}\\
      \includegraphics[width=0.2\textwidth]{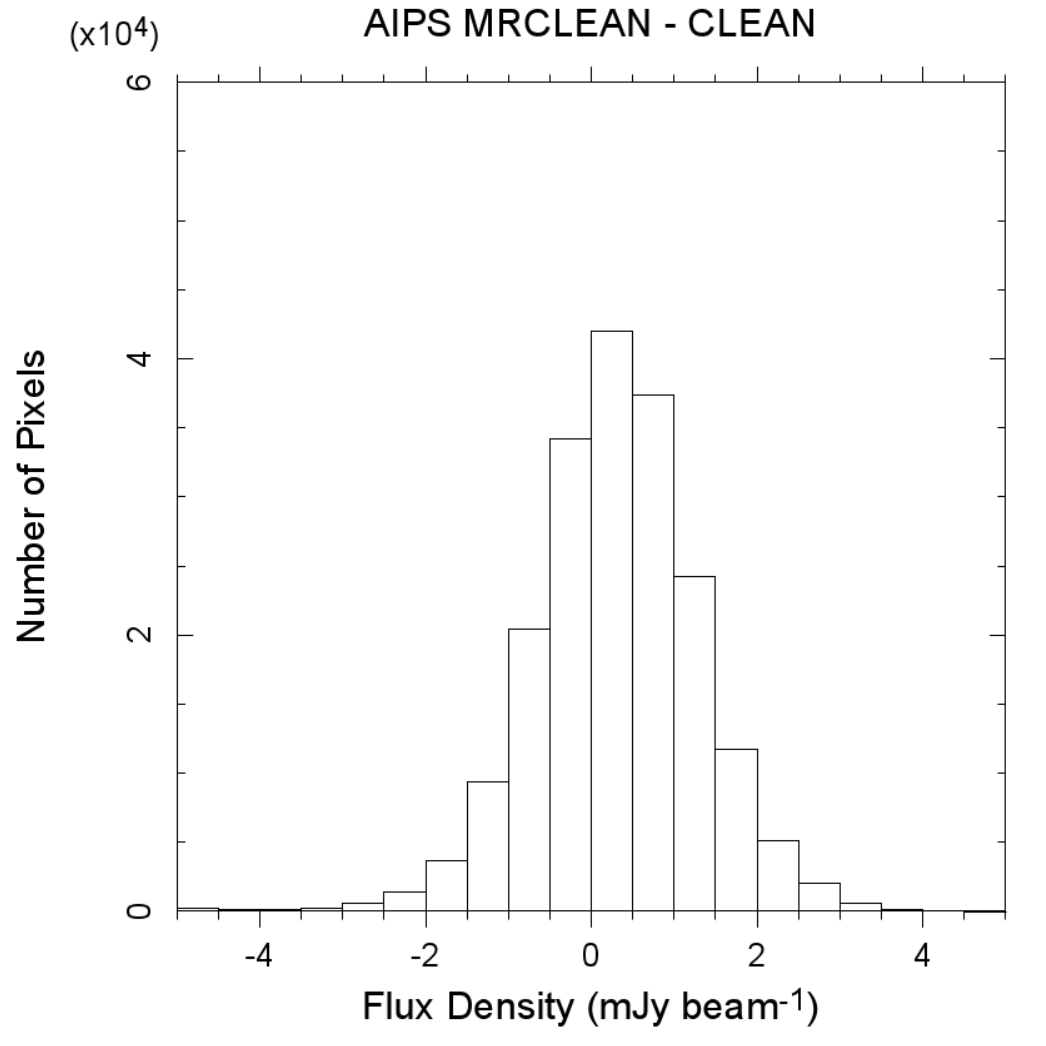}&
      \includegraphics[width=0.2\textwidth]{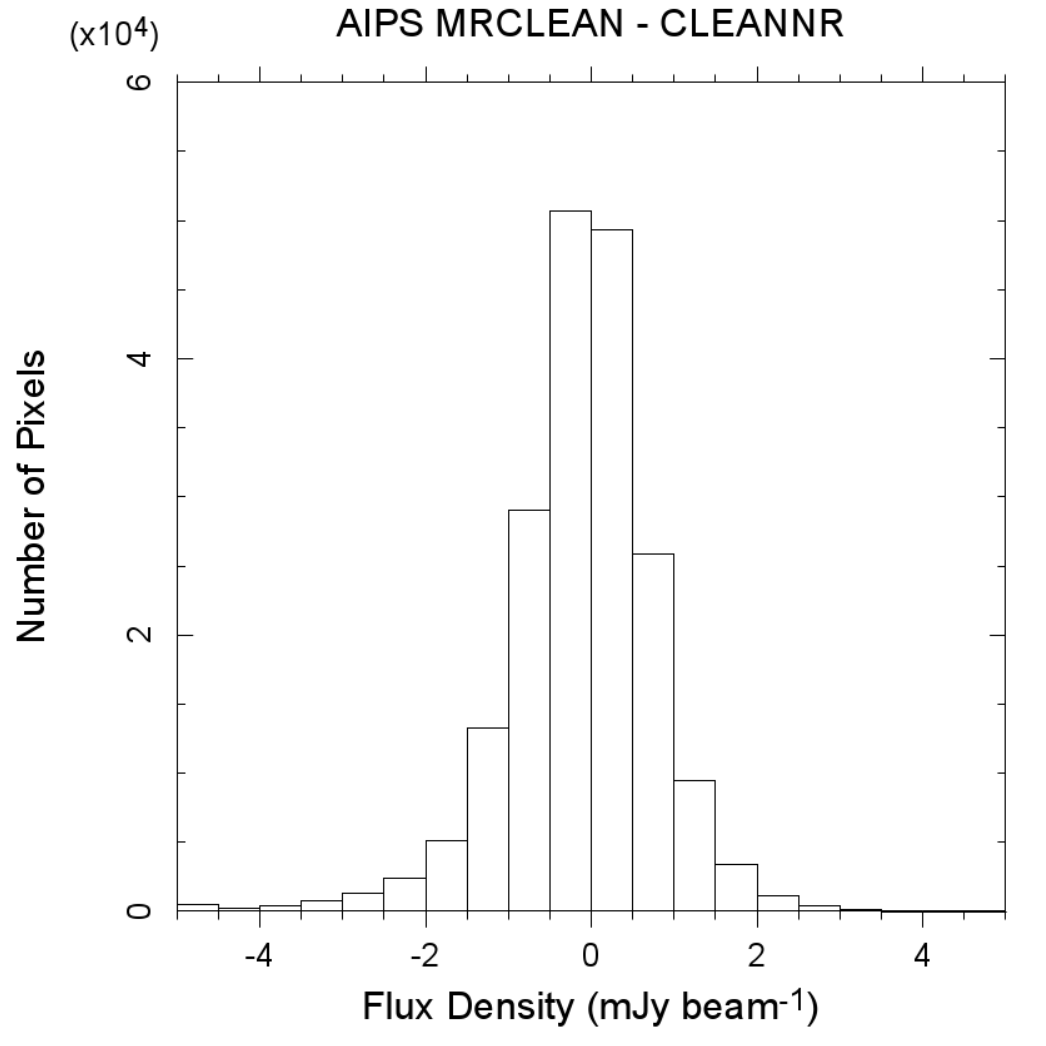}&
      \includegraphics[width=0.2\textwidth]{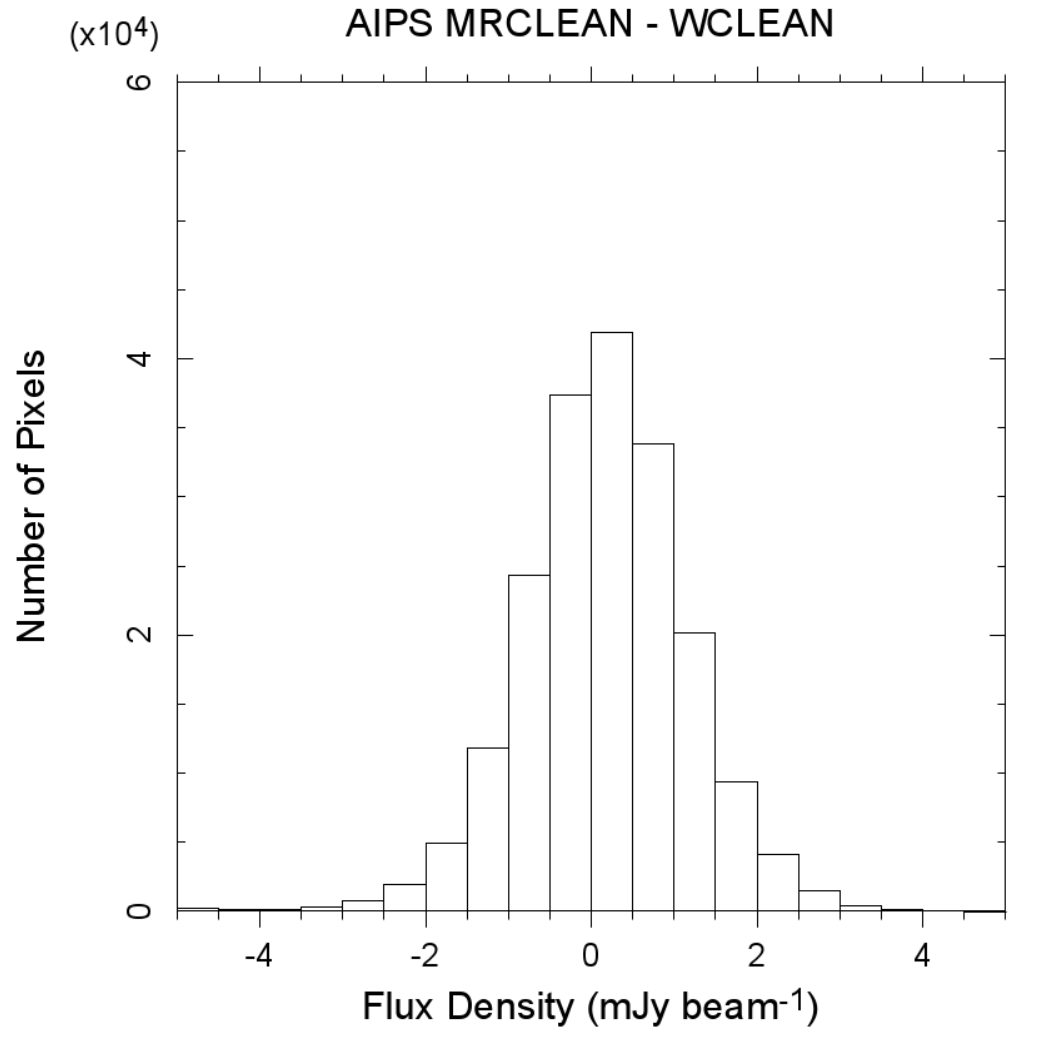}\\
    \end{tabular}
    \caption{Difference images (top row) of \mrclean\ minus
      residual-scaled \clean, non residual-scaled \clean\ and
      residual-scaled \wclean\ for Holmberg~II.  The original images were
      corrected for the primary-beam. For residual-scaled \clean\ all flux
      below $2\sigma$ of the \text{rms} noise in the \msclean\ restored
      image has been masked. For \wclean, only the difference within the
      \clean\ window used is shown.  Histograms (bottom row) of the flux
      values calculated within the same area of each image and equal to the
      area inside the window used for the \wclean\ are also shown.
      Gray-scale levels in the images run from $-5$ to
      $5$~\mjbeam.\label{fig:mr-minus-hol2}}
\end{sidewaysfigure}

The variations on \clean\ discussed here all use the same principles,
but different implementations to produce a final image.  It is
therefore interesting to see whether these different methods introduce
differences in the final images.  To that end, difference images of
\msclean\ and \mrclean\ minus \clean/\wclean\ have been constructed.
A difference image of the two scale-sensitive algorithms, \msclean\
minus \mrclean, was also made to compare the relative difference
between each.  All of these images were generated by performing a
pixel-by-pixel subtraction of two restored images. The resulting
images are shown in Figures \ref{fig:ms-minus-hol2} and
\ref{fig:mr-minus-hol2}.  Subtractions for residual-scaled and non
residual-scaled \clean\ are shown.  The former has all flux below
$2\sigma$ masked, as the comparison is only valid in regions where
there is source emission.  Similarly for \wclean, only the difference
within the window is shown.  The differences in the noise are shown
for the non-residual scaled \clean\ subtractions.  Also shown are
histograms of the flux values in each difference image.  The
histograms were calculated within the same region of each image, which
was equal to the window used for \wclean.

% \placefigure{fig:ms-minus-hol2}
% \placefigure{fig:mr-minus-hol2}

The difference between the \clean\ and \wclean\ subtractions for
either \msclean\ or \mrclean\ is minimal.  This is expected, as the
only difference is the application of a \clean\ window. The only major
differences in subtractions of \msclean\ and \mrclean\ minus classical
\clean\ are directly around and at the location of the source emission.
Negative peaks in these difference images correspond to locations of
maximum source emission in the actual images.  The \mrclean\ minus
\clean/\wclean\ difference images in Figure~\ref{fig:mr-minus-hol2}
also show additional positive difference flux around the negative
peaks.  

Looking at the difference image of \msclean\ minus \mrclean\ in
Figure~\ref{fig:ms-minus-hol2}, these locations show an average zero
difference flux. Overall, this difference image shows little
correlation with the source structure.  The small peaks of negative
difference flux visible in the image do appear to be at locations of
strong source emission as is the case for each scale-sensitive
algorithm minus classical \clean, but the extent of such differences
is much smaller than in the latter difference images.  Both
scale-sensitive algorithms are therefore performing a similar clean
process (\ie\ choosing components of similar strength size and
location).  But where there is compact, high peak flux emission
present, the scale-sensitive algorithms and classical \clean\ diverge
in their processing techniques.  This difference leads to greater
total flux recovery for scale-sensitive algorithms, as shown in the
previous sections.

%%%%####################################################################
\clearpage
\section{Applications of \texorpdfstring{\bmsclean}{MSCLEAN}}
\label{sec:applications}

Having performed and discussed the comparison of \msclean\ against
\clean\ and other algorithms the remainder of this paper is on some
real-world applications of \msclean.  We focus on two particular
applications that relate to small-scale extended structure in the \HI\
gas of galaxies.  In Section~\ref{sec:apps-holes}, the application of
\msclean\ to helping the search for \HI\ holes in the disk of galaxies
is discussed, while in Section~\ref{sec:apps-anomgas}, and initial
attempt is made to apply the increased contrast provided by \msclean\
to the study of faint, anomalous \HI\ gas structures outside of the
galactic disk.

\subsection{\texorpdfstring{\HIfat}{HI} Holes}
\label{sec:apps-holes}

High resolution \HI\ imaging has revealed a deeper level of structure
in the inter-stellar medium that forms an intricate tapestry of holes,
shells and bubbles that permeate and define a complex set of tunnels,
networks and cavities of under-densities through the tenuous
interstellar medium
\citep[\eg][]{bz_cbb,heiles_milk1,mcs_vim,bb_m31,ddh_m33,kim_lmc,stani_smc,wb_ic2574}.
Such small-scale structures provide an excellent environment in which
to compare \msclean\ to \clean. Here we compare the results of a
search for holes in the THINGS and \msclean\ cubes of the same galaxy.

To inspect the cubes and integrated moment maps, the KARMA software
suite of visualization tools was used \citep{karma}. In particular,
the \textsc{kvis} and \textsc{kpvslice} tools were extensively used
for finding the holes. These two tools provide two different ways of
looking at the \HI\ cube. The former provides views of the spatial and
spectral planes in the data cube. The latter allows arbitrary
position-velocity slices through the cube to be made, providing an
alternative view of a part of the disk. To find a hole, a search was
first conducted in \textsc{kvis} for under-densities in sequential
velocity channels.  Then, using \textsc{kpvslice}, a position-velocity
($pV$) cut through the hole was taken and the hole was classified as
one of three different types by observing its profile.

The qualities that define a hole from an under-density in the \HI\ gas
are somewhat subjective and depend on the signal-to-noise of their
feature. In this study, low signal-to-noise features that were of the
order of the size of the beam were generally classified as just
noise. A structure had to exist in more than four velocity channels as
well before considering it. To verify the reliability of the method of
identification genuine, a double-blind test was conducted. This
involved searching for holes in a single galaxy using the same
technique for both the \clean\ and \msclean\ data, and by two
different people. Only the basic observed properties of the hole
candidates were cataloged in the test, being the position, size and
shape of each hole. Also a `quality rating' between one and ten was
assigned. The quality rating defined the confidence that the observer
thought a candidate hole was a genuine structure. So a quality rating
of $7$ meant that the observer was $70\%$ confident that the candidate
was real. The end result of the double blind test was four catalogs,
two for each observer and imaging algorithm.

\begin{figure}[htbp]
  \centering
  \begin{tabular}{ccc}
    \includegraphics[angle=270,width=0.25\textwidth]{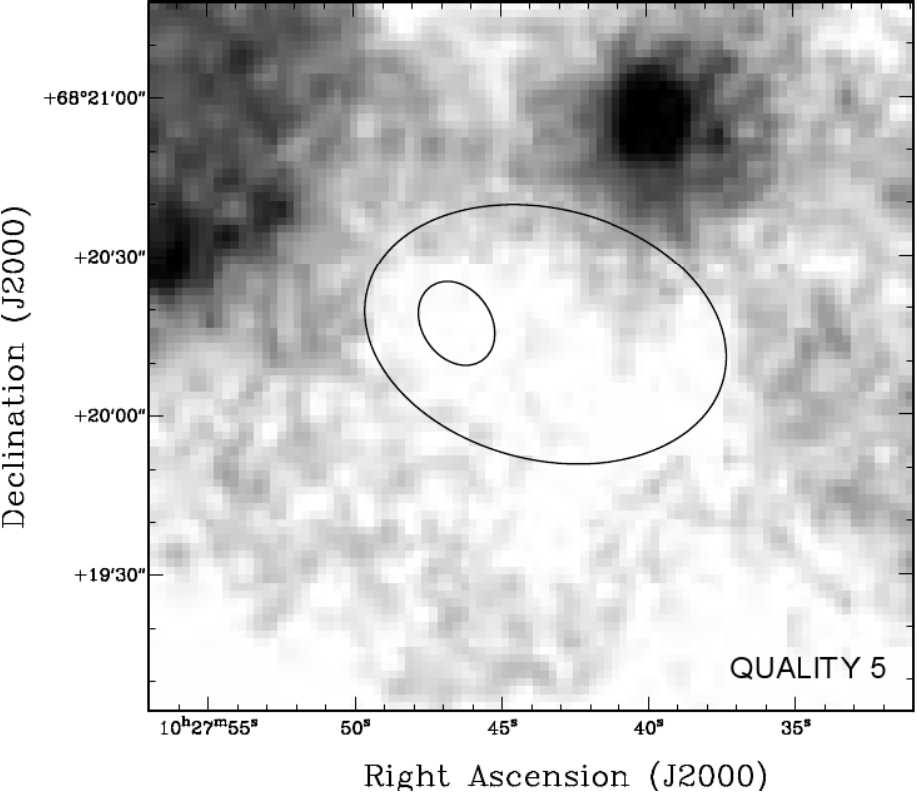}&
    \includegraphics[angle=270,width=0.25\textwidth]{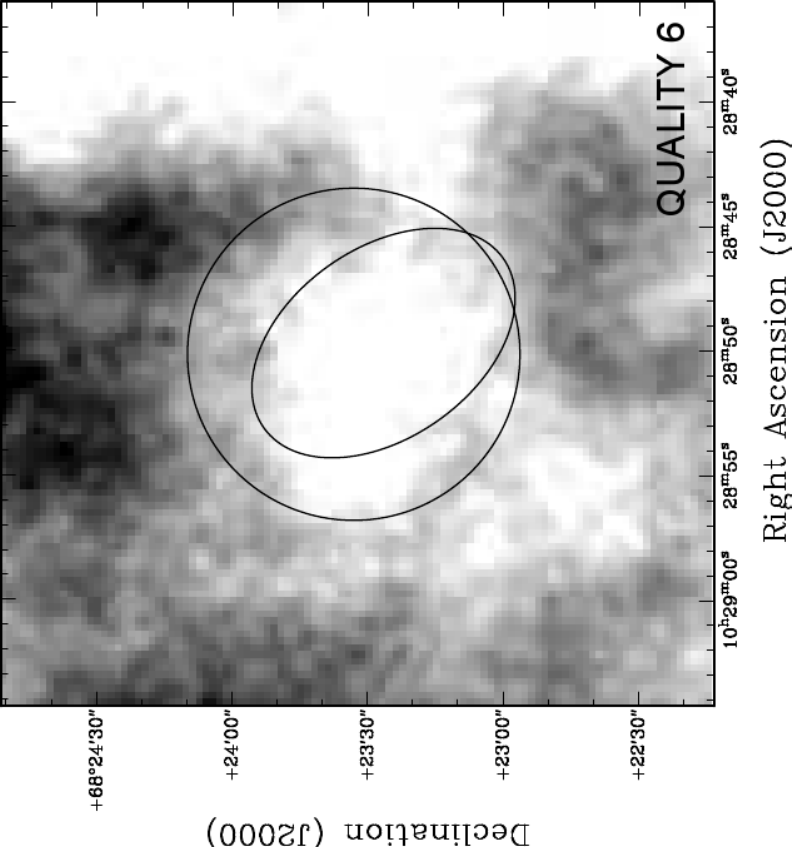}&
    \includegraphics[angle=270,width=0.25\textwidth]{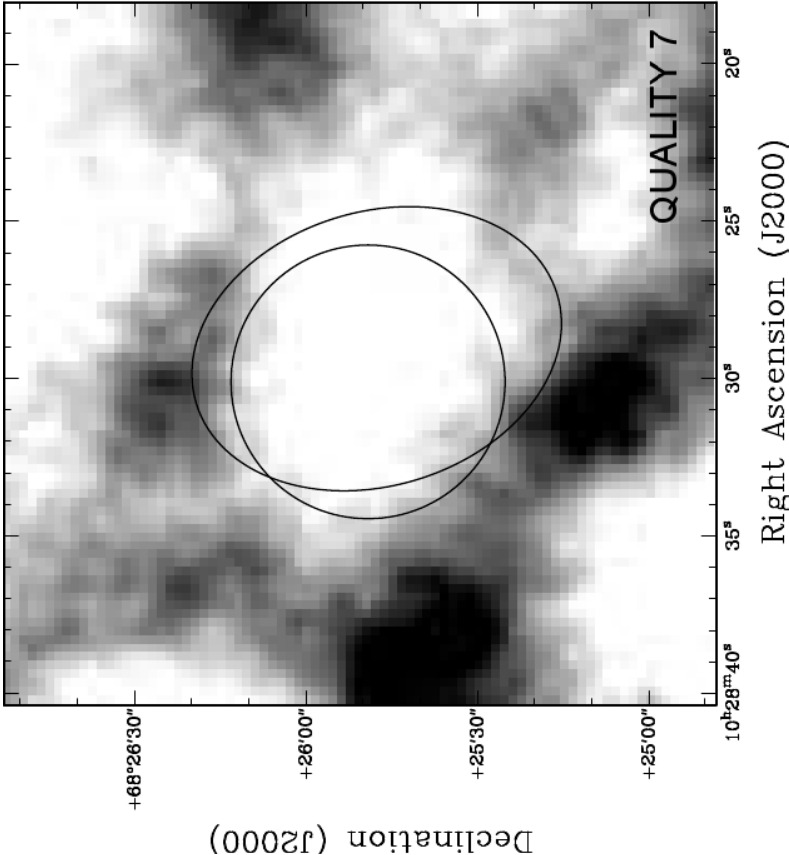}\\
  \end{tabular}
  \caption{Comparison of the agreement bewteen hole sizes and shapes
    with progressively higher quality ratings for the \msclean\
    catalogs created in the double-blind test.  The quality ratings
    increase from $5$ to $7$ from left to right.  The
    gray-scale levels run from $15$ to $260$
    \mjbeam~\kms.\label{fig:hole-quality}}
\end{figure}

There were significant variations between the catalogs of the two
observers in both the \clean\ and \msclean\ data. However, the
agreement between the catalogs, both as regards to size and shape of
the holes increased with increasing quality rating. This is shown in
Figure~\ref{fig:hole-quality} where the ellipses fitted to various
holes are shown for progressively higher quality ratings. It was
found, and as shown in the figure, that at around a quality value of
$6$, there was good agreement between the observers. That is, with the
search method and criteria used in this paper, all candidates found
with a quality rating of $6$ or above are believed to be real holes.

\begin{figure}[htbp]
  \centering
  \includegraphics[angle=270,width=0.6\textwidth]{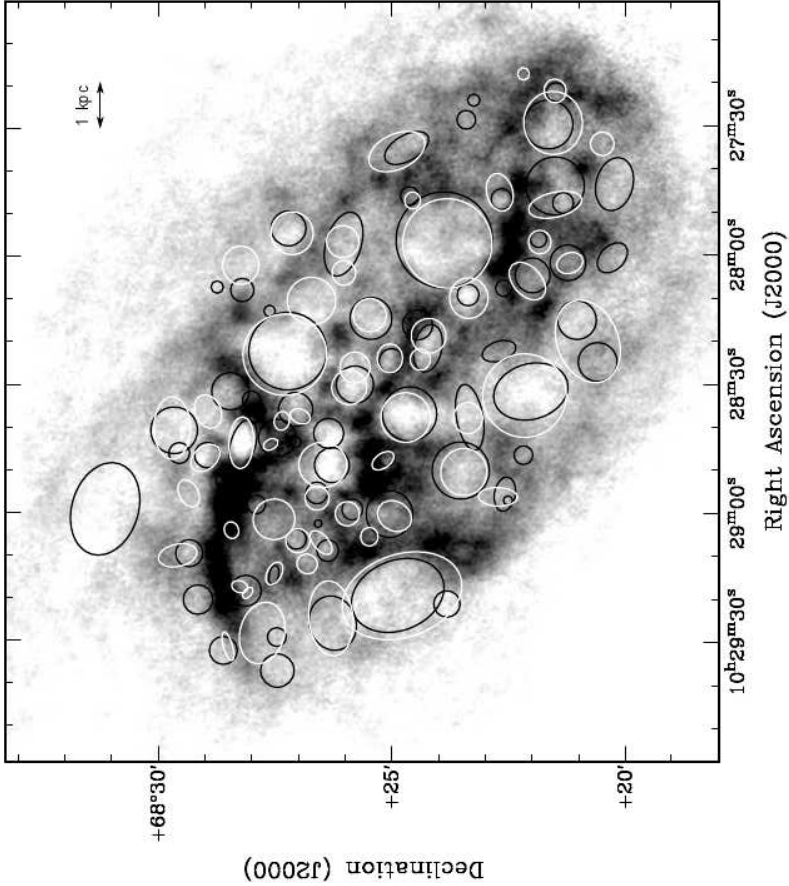}
  \caption{\msclean\ integrated \HI\ map with hole catalog overlays
    created from \msclean\ (black) and \clean\ (white) data.  The
    gray-scale levels run from $0$ to $251$
    \mjbeam~\kms.\label{fig:holecomp}} %min -13
\end{figure}

\begin{figure}[htbp]
  \centering
  \includegraphics[angle=270,width=0.6\textwidth,angle=270]{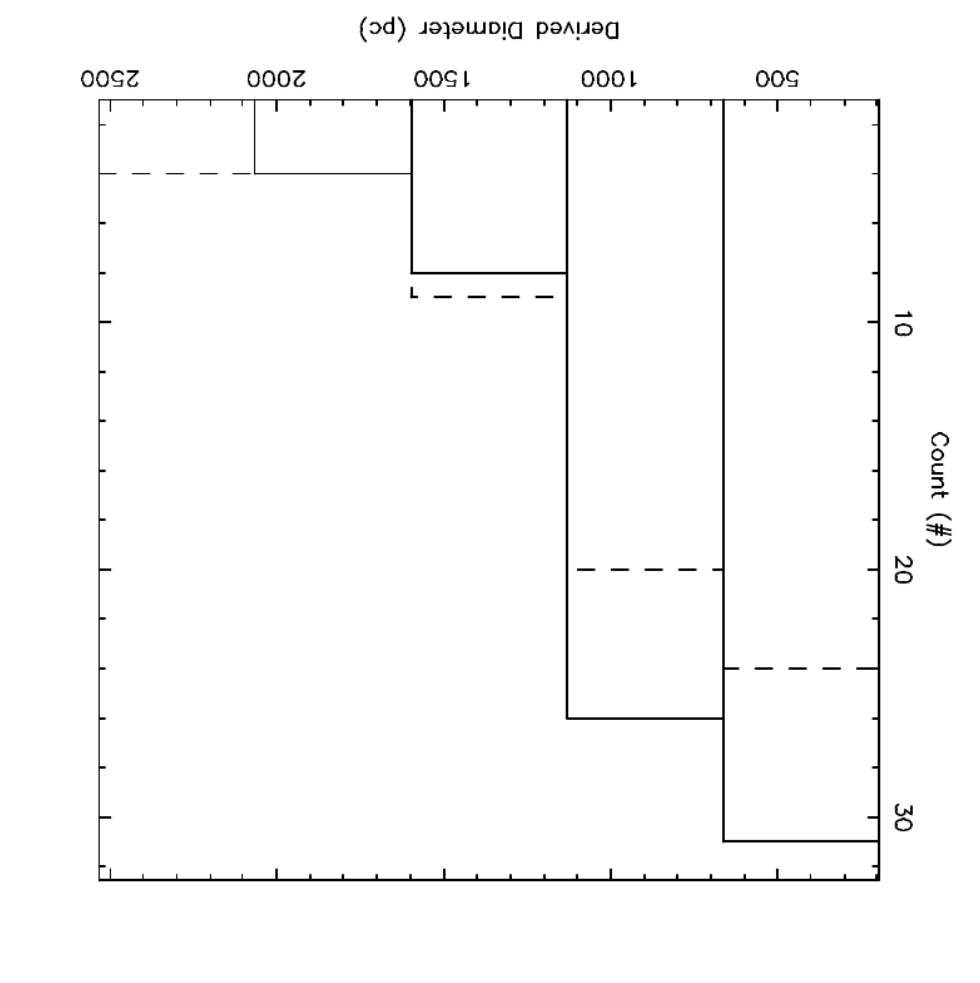}
  \caption{Histogram of the derived diameters of holes in the
    \clean\ (dashed histogram) and \msclean\ (solid histogram) hole
    catalogs of the galaxy IC~2574.\label{fig:ic2574-holecomphisto-diam}}
\end{figure}

The final hole catalog overlays for the \clean\ and \msclean\ data on
an \msclean\ natural-weighted integrated intensity map is shown in
Figure \ref{fig:holecomp}. Comparison of the overlays shows the
physical implications of the results discussed in the previous
sections.  Inspection of the \msclean\ hole overlay appears to show an
excess of small holes, especially in the inner
disk. Figure~\ref{fig:ic2574-holecomphisto-diam} shows a histogram of
the derived diameters of holes in both the \clean\ (dashed) and
\msclean\ (solid) catalogs. The difference is indicative of the extra
contrast provided by \msclean. Without the smoother background of the
\clean\ residual pedestal, structures are better defined and with a
higher contrast. Whether this extra population of small holes is
related to, \eg\ star-formation events, or whether we are observing
small scale turbulence, or as yet unrecognized systematic effects is
beyond the scope of this paper.

\clearpage

\subsection{Anomalous Gas}
\label{sec:apps-anomgas}

Increased resolution in radio images has also led to interesting
results beyond the high surface brightness disk of galaxies.
`Anomalous' neutral gas structures have been detected in a few nearby
galaxies such as NGC~891 \citep{swaters_n891_anomgas} and including
NGC~2403 \citep{frat_n2403_anomgas}. It has only been in recent, deep
\HI\ observations that these structures have been revealed as being
linked to a cold gaseous halo well above the main disk
\citep{oosterloo_n891_halogas}. Like holes, such structures require
significant angular resolution for detection. And unlike holes, their
location is mostly within a region where \clean\ characteristics may
play an important role in the interpretation of the data.

\begin{figure}[htbp]
  \centering
  \begin{tabular}{cc}
    \includegraphics[width=0.3\textwidth]{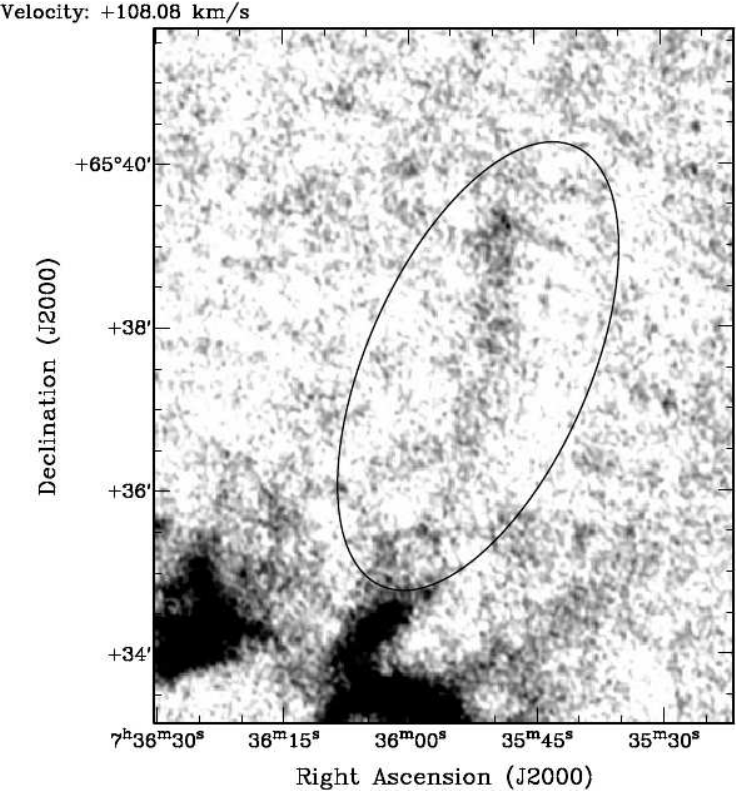}&
    \includegraphics[width=0.3\textwidth]{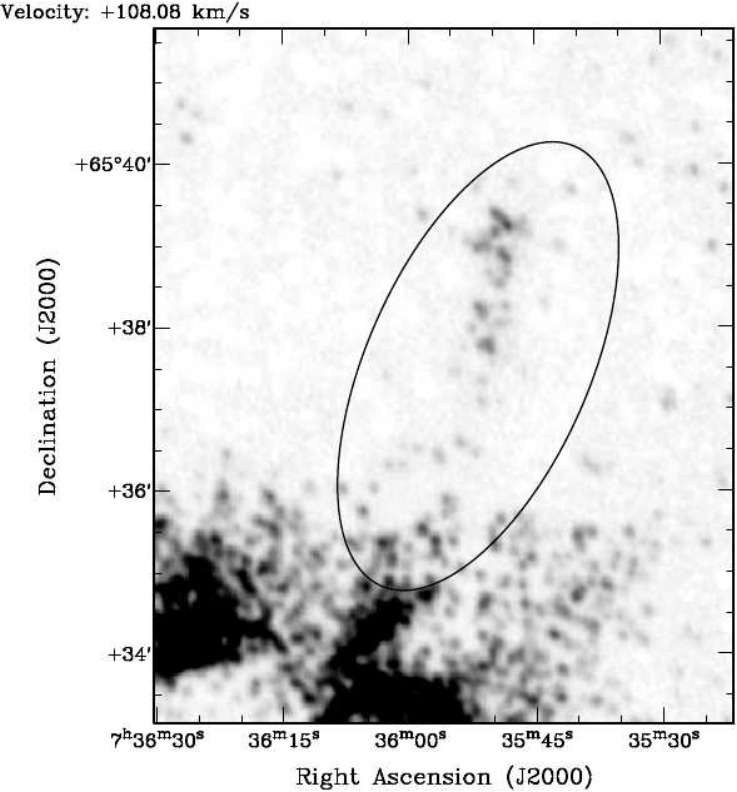}\\
  \end{tabular}
  \caption{Zoomed in channel map for NGC~2403 for both \msclean\
    (left) and \clean\ (right) data cubes.  Shown within the ellipse is
    the `anomalous' \HI\ structure as discovered by
    \cite{frat_n2403_anomgas}. The gray-scale levels run from $0$ to $15$
    \mjbeam.\label{fig:ngc2403-anomgas-chanmap-comp}} %min -3
\end{figure}

\begin{figure}[htbp]
  \centering
  \begin{tabular}{c}
    \includegraphics[angle=270,width=0.4\textwidth]{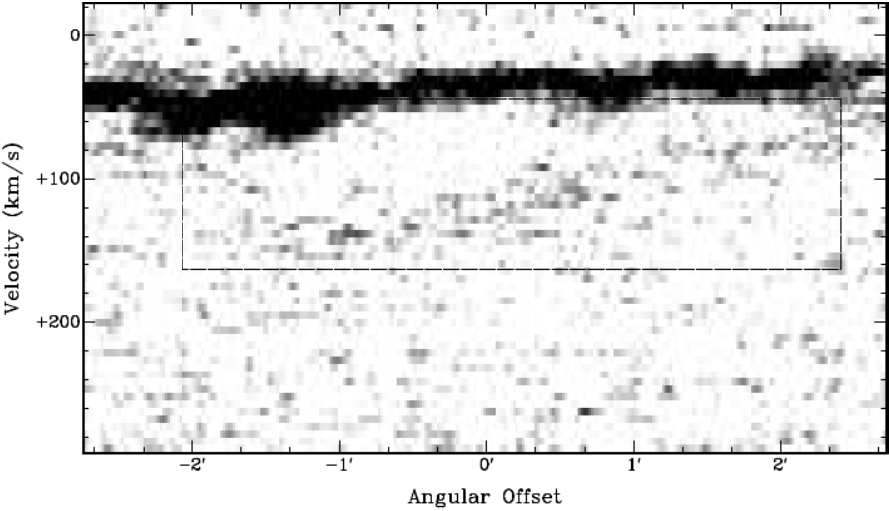}\\
    \includegraphics[angle=270,width=0.4\textwidth]{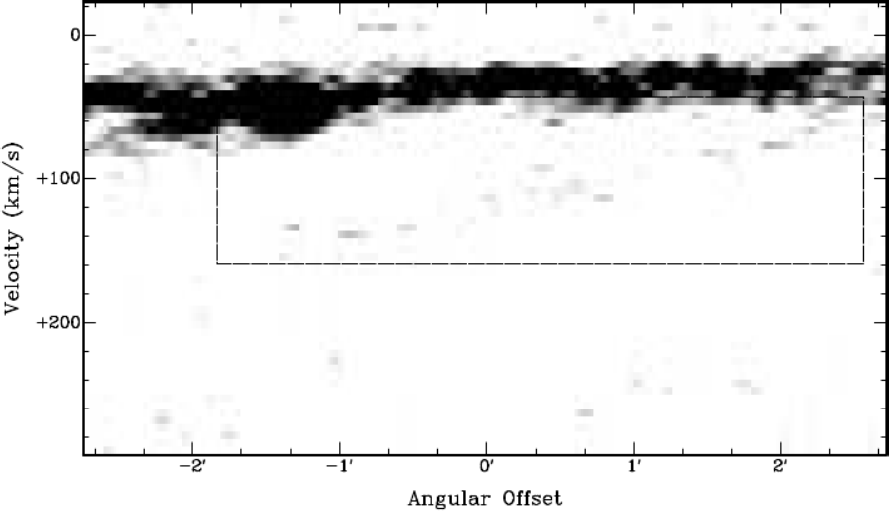}\\
  \end{tabular}
  \caption{Position-velocity ($pV$) cuts through the major axis of
    the ellipse around the structure observed in Figure
    \ref{fig:ngc2403-anomgas-chanmap-comp} in the \msclean\ (top) and
    \clean\ (bottom) natural-weighted data cubes. The gray-scale
    levels run from $0.2$ to $2$ \mjbeam.\label{fig:ngc2403-anomgas-pv-comp}}
\end{figure}

In this study, a comparison of the structures described in
\cite{frat_n2403_anomgas} was made in the \clean\ and \msclean\ data
sets. The KARMA software package was employed to make $pV$ slices and
channel map images as outlined in Section~\ref{sec:apps-holes}. The
structure located at $\sim108$~\kms\ by \cite{frat_n2403_anomgas} is
easily observed in both data-sets as shown in
Figure~\ref{fig:ngc2403-anomgas-chanmap-comp}. The effects of the
\clean\ bowl are also clearly visible in the \clean\ image as an
apparent decrease in the noise level. This has a negative impact on
the visibility of fine structure, however, in a $pV$ slice taken
across the major axis of the ellipse as shown in
Figure~\ref{fig:ngc2403-anomgas-pv-comp}. A lot of the structure is
hidden by the bowl, and much extra analysis would be needed to
quantify structures over the entire cube to the same quantitative
levels. In contrast, the \msclean\ image has a flat background and
more features of the structure are visible. It is clear that a more
detailed search will need to be conducted in NGC~2403. But as
indicated, \msclean\ greatly enhances the visibility of structures
near the disk.

%%%%####################################################################

\clearpage
\section{Summary}
\label{sec:summary}

This paper presents a demonstration of the Multi-Scale \clean\
algorithm on a sample of galaxies from THINGS. \msclean\ is an
extension of \clean\ that attempts to overcome some of the
short-comings of the latter algorithm, particularly relating to extended
sources such as galaxies. The THINGS data are well-suited to test and
compare the performance of \clean\ and \msclean. A summary of the
important results follows.

1. \msclean\ (which we here take to explicitly include both the CASA
as well as the AIPS Multi-Resolution \clean\ implementations) uses fewer
iterations than classical \clean\ and can \clean\ down to a low flux
threshold much more efficiently than the latter. The algorithm itself
is also relatively simple and easy to implement and is currently
available in the CASA and AIPS software packages.

2. \msclean\ removes the bowl and pedestal characteristics that occur
when \clean\ is applied to extended sources. In the \clean\ data the
extended wings of the dirty beam that are left uncleaned in the
pedestal smooth out a lot of the low-level fine-scale
structure. \msclean, by significantly reducing the pedestal, reveals
this structure more clearly. This leads to better contrast and higher
resolution at the lowest flux levels. The \clean\ bowl lowers the
apparent flux of all structures within it, due to the negative
background it creates. Extended structure usually has a low column
density compared to compact structure and so the effect is more
pronounced for extended structure. \msclean\ can greatly reduce this
bowl effect, thereby better recovering large-scale structure. The
noise characteristics of the final cubes remain the same.

3. \msclean\ illustrates the importance for residual flux scaling when
using classical \clean. This need for residual flux scaling arises
because classical \clean\ cleans down to a set flux level, leaving a
residual map which is characterized by the dirty beam.  The flux in
the residuals has to be rescaled by the ratio of the clean beam over
the dirty beam area before restoring the \clean\ components back on
the residual map.  \msclean\ does a better job in modeling the
emission, leaving virtually no residual and hence removing the need
for such scaling.

4. The added contrast provided by \msclean\ is useful for a number of
astronomical problems involving small-scale structure. This paper has
highlighted two such problems; \HI\ holes in the disk and anomalous
gas in the halo of nearby galaxies. Comparison of hole catalogs made
for the \clean\ and \msclean\ data-sets shows that \msclean\ reveals
finer structure in the central disk by removing the artificial
smoothness associated with \clean. A comparison of a known anomalous
gas structure in NGC~2403 demonstrated that the removal of the \clean\
bowl by \msclean\ better reveals structure outside of the disk as
well.

\acknowledgments

The work of WJGdB is based upon research supported by the South
African Research Chairs Initiative of the Department of Science and
Technology and National Research Foundation. EB gratefully
acknowledges financial support through an EU Marie Curie International
Reintegration Grant (Contract No. MIRG-CT-6-2005-013556). This
research has made use of the NASA/IPAC Extragalactic Database (NED)
which is operated by the Jet Propulsion Laboratory, California
Institute of Technology, under contract with the National Aeronautics
and Space Administration. Portions of the analysis presented here made
use of the Perl Data Language (PDL) developed by K. Glazebrook,
J. Brinchmann, J. Cerney, C. DeForest, D. Hunt, T. Jenness, T. Luka,
R. Schwebel, and C. Soeller and can be obtained from
\url{http://pdl.perl.org}. PDL provides a high-level numerical
functionality for the Perl scripting language \citep{pdlref}.  Some of
the figures in this paper were created and edited with the open-source
Inkscape vector graphics editor.  Inkscape can be obtained freely from
\url{http://www.inkscape.org}. We thank the anonymous referee for
useful and constructive comments which have significantly improved
several aspects of this paper.

Facilities: \facility{VLA}

%%%%####################################################################

\bibliographystyle{hapj}
\bibliography{apj-jour,paper}

\begin{thebibliography}{32}
\expandafter\ifx\csname natexlab\endcsname\relax\def\natexlab#1{#1}\fi

\bibitem[{{Bhatnagar} \& {Cornwell}(2003)}]{bc_msc}
{Bhatnagar}, S., \& {Cornwell}, T.~J. 2003, in Presented at the Society of
  Photo-Optical Instrumentation Engineers (SPIE) Conference, Vol. 5169,
  Astronomical Adaptive Optics Systems and Applications. Edited by Tyson,
  Robert K.; Lloyd-Hart, Michael. Proceedings of the SPIE, Volume 5169, pp.
  331-340 (2003)., ed. R.~K. {Tyson} \& M.~{Lloyd-Hart}, 331--340

\bibitem[{{Bhatnagar} \& {Cornwell}(2004)}]{bc_ssd}
{Bhatnagar}, S., \& {Cornwell}, T.~J. 2004, \aap, 426, 747,
  arXiv:astro-ph/0407225

\bibitem[{{Brand} \& {Zealey}(1975)}]{bz_cbb}
{Brand}, P.~W.~J.~L., \& {Zealey}, W.~J. 1975, \aap, 38, 363

\bibitem[{{Briggs}(1995)}]{briggs_robust}
{Briggs}, D.~S. 1995, in Bulletin of the American Astronomical Society,
  Vol.~27, Bulletin of the American Astronomical Society, 1444

\bibitem[{{Brinks} \& {Bajaja}(1986)}]{bb_m31}
{Brinks}, E., \& {Bajaja}, E. 1986, \aap, 169, 14

\bibitem[{{Brinks} \& {Shane}(1984)}]{bs_m31}
{Brinks}, E., \& {Shane}, W.~W. 1984, \aaps, 55, 179

\bibitem[{{Clark}(1980)}]{clark_clean}
{Clark}, B.~G. 1980, \aap, 89, 377

\bibitem[{{Cornwell} {et~al.}(1999){Cornwell}, {Braun}, \&
  {Briggs}}]{cornwell_deconv}
{Cornwell}, T., {Braun}, R., \& {Briggs}, D.~S. 1999, in Astronomical Society
  of the Pacific Conference Series, Vol. 180, Synthesis Imaging in Radio
  Astronomy II, ed. G.~B. {Taylor}, C.~L. {Carilli}, \& R.~A. {Perley}, 151

\bibitem[{{Cornwell}(2008)}]{msclean}
{Cornwell}, T.~J. 2008, ArXiv e-prints, 806, arXiv:astro-ph/0806.2228

\bibitem[{{Deul} \& {den Hartog}(1990)}]{ddh_m33}
{Deul}, E.~R., \& {den Hartog}, R.~H. 1990, \aap, 229, 362

\bibitem[{{Fraternali} {et~al.}(2002){Fraternali}, {van Moorsel}, {Sancisi}, \&
  {Oosterloo}}]{frat_n2403_anomgas}
{Fraternali}, F., {van Moorsel}, G., {Sancisi}, R., \& {Oosterloo}, T. 2002,
  \aj, 123, 3124, astro-ph/0203405

\bibitem[{{Glazebrook} \& {Economou}(1997)}]{pdlref}
{Glazebrook}, K., \& {Economou}, F. 1997, The Perl Journal, 5

\bibitem[{{Gooch}(1996)}]{karma}
{Gooch}, R. 1996, in ASP Conf. Ser. 101: Astronomical Data Analysis Software
  and Systems V, ed. G.~H. {Jacoby} \& J.~{Barnes}, 80

\bibitem[{{Heiles}(1979)}]{heiles_milk1}
{Heiles}, C. 1979, \apj, 229, 533

\bibitem[{{H{\"o}gbom}(1974)}]{hogbom_clean}
{H{\"o}gbom}, J.~A. 1974, \aaps, 15, 417

\bibitem[{{J{\"o}rs{\"a}ter} \& {van Moorsel}(1995)}]{jvm_residflux}
{J{\"o}rs{\"a}ter}, S., \& {van Moorsel}, G.~A. 1995, \aj, 110, 2037

\bibitem[{{Kim} {et~al.}(1999){Kim}, {Dopita}, {Staveley-Smith}, \&
  {Bessell}}]{kim_lmc}
{Kim}, S., {Dopita}, M.~A., {Staveley-Smith}, L., \& {Bessell}, M.~S. 1999,
  \aj, 118, 2797

\bibitem[{{McCray} \& {Snow}(1979)}]{mcs_vim}
{McCray}, R., \& {Snow}, Jr., T.~P. 1979, \araa, 17, 213

\bibitem[{{Muller} {et~al.}(2004){Muller}, {Stanimirovi{\'c}}, {Rosolowsky}, \&
  {Staveley-Smith}}]{msrs_stat}
{Muller}, E., {Stanimirovi{\'c}}, S., {Rosolowsky}, E., \& {Staveley-Smith}, L.
  2004, \apj, 616, 845, arXiv:astro-ph/0408259

\bibitem[{{Oosterloo} {et~al.}(2007){Oosterloo}, {Fraternali}, \&
  {Sancisi}}]{oosterloo_n891_halogas}
{Oosterloo}, T., {Fraternali}, F., \& {Sancisi}, R. 2007, Astronomical Journal,
  134, 1019, arXiv:0705.4034

\bibitem[{{Rots}(1980)}]{rots_singledish}
{Rots}, A.~H. 1980, \aaps, 41, 189

\bibitem[{{Schwab}(1984)}]{schwab_clean}
{Schwab}, F.~R. 1984, \aj, 89, 1076

\bibitem[{{Schwarz}(1978)}]{schwarz_cleananal}
{Schwarz}, U.~J. 1978, \aap, 65, 345

\bibitem[{{Stanimirovic} {et~al.}(1999){Stanimirovic}, {Staveley-Smith},
  {Dickey}, {Sault}, \& {Snowden}}]{stani_smc}
{Stanimirovic}, S., {Staveley-Smith}, L., {Dickey}, J.~M., {Sault}, R.~J., \&
  {Snowden}, S.~L. 1999, \mnras, 302, 417

\bibitem[{{Starck} {et~al.}(1994){Starck}, {Bijaoui}, {Lopez}, \&
  {Perrier}}]{sblp_wavelet}
{Starck}, J.-L., {Bijaoui}, A., {Lopez}, B., \& {Perrier}, C. 1994, \aap, 283,
  349

\bibitem[{{Starck} {et~al.}(2002){Starck}, {Pantin}, \&
  {Murtagh}}]{deconv_review}
{Starck}, J.~L., {Pantin}, E., \& {Murtagh}, F. 2002, \pasp, 114, 1051

\bibitem[{{Swaters} {et~al.}(1997){Swaters}, {Sancisi}, \& {van der
  Hulst}}]{swaters_n891_anomgas}
{Swaters}, R.~A., {Sancisi}, R., \& {van der Hulst}, J.~M. 1997, \apj, 491,
  140, arXiv:astro-ph/9707150

\bibitem[{{Tan}(1986)}]{tan_cleananal}
{Tan}, S.~M. 1986, \mnras, 220, 971

\bibitem[{{Thompson} {et~al.}(2001){Thompson}, {Moran}, \&
  {Swenson}}]{tms_radioint}
{Thompson}, A.~R., {Moran}, J.~M., \& {Swenson}, Jr., G.~W. 2001,
  Interferometry and Synthesis in Radio Astronomy, 2nd Edition (Interferometry
  and synthesis in radio astronomy by A.~Richard Thompson, James M.~Moran, and
  George W.~Swenson, Jr.~2nd ed.~ New York : Wiley, c2001.xxiii, 692 p.~:
  ill.~; 25 cm.~''A Wiley-Interscience publication.'' Includes bibliographical
  references and indexes.~ISBN : 0471254924)

\bibitem[{{Wakker} \& {Schwarz}(1988)}]{ws_mrclean}
{Wakker}, B.~P., \& {Schwarz}, U.~J. 1988, \aap, 200, 312

\bibitem[{{Walter} \& {Brinks}(1999)}]{wb_ic2574}
{Walter}, F., \& {Brinks}, E. 1999, \aj, 118, 273

\bibitem[{{Walter} {et~al.}(2008){Walter}, {Brinks}, {de Blok}, {Bigiel},
  {Kennicutt}, {Jr.}, {Thornley}, \& {Leroy}}]{THINGS}
{Walter}, F., {Brinks}, E., {de Blok}, W.~J.~G., {Bigiel}, F., {Kennicutt},
  R.~C., {Jr.}, {Thornley}, M.~D., \& {Leroy}, A.~K. 2008, \aj,
  arXiv:astro-ph/0810.2125

\end{thebibliography}

% \include{figures}
% \include{tables}

%%%%####################################################################

\end{document}